\documentclass[a4paper,11pt]{article}
\pdfoutput=1 
\usepackage{comment}
\usepackage{jheppub} 
\usepackage[normalem]{ulem}
\usepackage[T1]{fontenc} 
\usepackage{lipsum}
\usepackage{xcolor}
\usepackage{amsmath}
\usepackage{multirow}
\usepackage{tikz}
\usepackage{tikz-feynman}
\usepackage{caption}
\usepackage{subcaption}
\usepackage{slashed}
\usepackage{graphicx}

\title{\boldmath 
Connecting $B\to K^{(\ast)}\nu\nu$ and $K\to \pi\nu\nu$ decays with neutrino physics
}

\author[a,b]{Asmaa~Abada,\,}
\author[a]{Claire~Chevallier,\,}
\author[c]{Luighi~P.~S.~Leal,\,}
\author[a]{Anna~Llauradó,\,}
\author[a]{Olcyr~Sumensari,\,}
\author[c]{Renata~Zukanovich Funchal\,}

\affiliation[a]{Université Paris-Saclay, CNRS/IN2P3, IJCLab, 91405 Orsay, France}
\affiliation[b]{Institut Universitaire de France (IUF), Paris, France}
\affiliation[c]{Departamento de Física Matemática, Instituto de Física\\
Universidade de São Paulo, 05315-970 São Paulo, Brazil}

\emailAdd{asmaa.abada@ijclab.in2p3.fr}
\emailAdd{anna.filella-llaurado@ijclab.in2p3.fr}
\emailAdd{claire.chevallier@ijclab.in2p3.fr}
\emailAdd{luighi.leal@usp.br}
\emailAdd{olcyr.sumensari@ijclab.in2p3.fr}
\emailAdd{zukanov@if.usp.br}

\abstract{We investigate the contributions of Lepton Number Violating (LNV) effective operators to the rare decays $B\to K^{(\ast)}\nu\nu$ and $K\to \pi\nu\nu$. Such operators can modify the kinematic distributions of these processes, providing distinctive probes of physics beyond the Standard Model. Through a renormalization-group analysis, we show that the Standard Model Effective Field Theory (SMEFT) operators responsible for these effects are subject to stringent indirect constraints from neutrino physics. In particular, we find that the mild excess reported by Belle-II in the $B^+\to K^+\nu\nu$ channel cannot be explained by LNV SMEFT operators without introducing significant fine-tuning in neutrino masses. We then show that these constraints can be evaded in the SMEFT minimally extended by a light right-handed neutrino, allowing for sizable effects in rare meson decays. Finally, we explore the implications of this viable scenario for low-energy processes, including neutrinoless double-beta decays.}

\begin{document} 
\maketitle
\flushbottom

\section{Introduction}
\label{sec:intro}

Flavor Changing Neutral Currents (FCNCs) are among the most sensitive probes of New Physics effects in low-energy observables. Recently, the Belle-II experiment has provided the first evidence for the $B^+ \to K^+\nu\bar{\nu}$ process~\cite{Belle-II:2023esi} and NA62 has clearly observed the $K^+\to \pi^+ \nu\bar{\nu}$ decay~\cite{Chang:2026vvx}. These determinations represent important milestones in precision tests of the Standard Model (SM) flavor dynamics. These processes are of particular interest as they are dominated by short-distance effects~\cite{Buras:2014fpa}, which makes them theoretically cleaner than analogous processes involving charged leptons. In particular, the $B^+ \to K^+\nu\bar{\nu}$ measurement reported by Belle-II exhibits a mild discrepancy with respect to the SM prediction, which has prompted various interpretations based on the Effective Field Theory (EFT) approach.

Assuming that the scale of New Physics is much larger than the electroweak scale, the Standard Model Effective Field Theory (SMEFT) provides the most suitable framework for describing low-energy observables~\cite{Buchmuller:1985jz}. Within the SMEFT, the leading contributions to the $d_i\to d_j \nu\bar{\nu}$ transition (with $i\neq j$) arise from dimension-six operators~\cite{Buras:2014fpa}. In particular, this scenario can accommodate Belle-II results through effective coefficients with third-generation leptons~\cite{Allwicher:2023xba,Bause:2023mfe,Allwicher:2024ncl,Marzocca:2024hua}. However, the particularity of these experimental searches is that neutrinos are not directly observed. Therefore, these searches are also sensitive to light and invisible New Physics particles that could also be produced on-shell in these decays~\cite{Altmannshofer:2023hkn,Kamenik:2011vy,DiLuzio:2025qkc,He:2022ljo,McKeen:2023uzo}. Similarly, extended EFT scenarios in the neutrino sector could also be tested by the same measurements~\cite{Felkl:2023ayn,Gorbahn:2023juq,Buras:2024ewl,Rosauro-Alcaraz:2024mvx,Endo:2026qof}. In both cases, a clear prediction is a modification of the kinematic distributions of the $B^+\to K^+ + \mathrm{inv}$ or $K^+\to \pi^+ + \mathrm{inv}$ decays, which are not affected by dimension-six SMEFT operators~\cite{Becirevic:2023aov}, thus providing a clear experimental handle to disentangle these scenarios~\cite{Belle-II:2025lfq}.

In this study, we will investigate extended EFT frameworks in the neutrino sector that contribute to these processes, going beyond the dimension-six SMEFT operators. We will explore the flavor phenomenology of these scenarios, with a particular emphasis on their
interplay with constraints from neutrino physics. Without introducing additional light fields, the only possibility is to consider dimension-seven SMEFT operators, which violate lepton number ($L$) and induce  $\Delta L=2$ transitions such as $d_i\to d_j\nu\nu$~\cite{Buras:2024ewl}. However, this scheme is subject to various theoretical and phenomenological considerations. These include the very specific ultraviolet (UV) completions needed to suppress dimension-six contributions to rare decays, while keeping 
sizable dimension-seven effects. 
Furthermore, as recently argued in Ref.~\cite{Endo:2026qof}, the same operators contributing to $s\to d\nu\nu$ and $b\to s\nu\nu$ transitions will necessarily generate large 
loop-induced
contributions to active neutrino masses and to neutrinoless double-beta ($0\nu\beta\beta$) decays (see also Ref.~\cite{Cirigliano:2017djv}).

The above-mentioned constraints on dimension-seven operators can be captured in a Renormalization-Group (RG) analysis, based on the one-loop anomalous dimensions computed in Ref.~\cite{Zhang:2023kvw}. In particular, the flavor-changing effects in quark currents arise from the misalignment of up- and down-type quark Yukawa couplings, as encoded in the Cabibbo-Kobayashi-Maskawa (CKM) matrix. In this paper, we will demonstrate that requiring the absence of fine-tuning in neutrino masses is sufficient to preclude observable dimension-seven contributions to $B^+ \to K^+ + \rm inv$ decays. In contrast, large contributions to kaon observables such as 
$K^+ \to \pi^+ + \rm inv$ remain in principle possible, in agreement with Ref.~\cite{Endo:2026qof}. However, as we will argue, the dimension-seven operators can be dominant in kaon decays only in very specific UV-complete scenarios, with a peculiar flavor pattern needed to suppress dimension-six operators. A similar naturalness argument for neutrino masses has been discussed in the context of $0\nu\beta\beta$ decays~\cite{Cirigliano:2017djv} and the neutrino magnetic-moment~\cite{Bell:2005kz}.  

Finally, we will show that a minor extension of the SMEFT with the inclusion of a light (SM singlet) 
Majorana neutrino $N$ is sufficient to circumvent the above-mentioned limitations~\cite{Rosauro-Alcaraz:2024mvx}, allowing us to predict measurable effects in $B\to K^{(\ast)}+\mathrm{inv}$ decays in a natural framework. In this scenario, it is possible to write dimension-six operators that contribute to the $d_i\to d_j N N$ and $d_i\to d_j \nu N$ transitions, which can arise from minimal UV completions. While RG evolutions also have an impact on neutrino phenomenology by inducing neutrino Yukawa-like operators~\cite{Ardu:2024tzb}, we will demonstrate that these effects are much less severe than in the previous case, thanks to a seesaw-like suppression in the expressions for the light-neutrino masses, as well as in the induced mixing between the active and sterile neutrinos.

The remainder of this paper is organized as follows. In Sec.~\ref{sec:left}, we study the low-energy EFT Lagrangian with SM neutrinos (LEFT), as well as the EFT obtained with the addition of a Majorana sterile neutrino in the spectrum ($\nu$LEFT), confronting the effective coefficients to low-energy data from FCNC processes in both cases.
In Secs.~\ref{sec:pheno-smeft} and \ref{sec:pheno-nusmeft}, we formulate the corresponding EFTs above the electroweak scale -- namely, the SMEFT and the $\nu$SMEFT, respectively -- both of which respect the $SU(3)_c \times SU(2)_L \times U(1)_Y$ gauge symmetry, and we derive the phenomenological constraints induced by RG evolution. Our main findings are summarized in Sec.~\ref{sec:conclusion}.

\section{Low-energy EFT}
\label{sec:left}

In this Section, we formulate the EFT approach to describe the $b\to s\nu\nu$ and $s\to d\nu\nu$ transitions, and we derive constraints on the relevant effective coefficients using rare decays of kaons and $B$-mesons. Two effective scenarios will be considered: (i) operators with only SM neutrinos, $\nu_L$; and (ii) operators built with an additional light sterile neutrino, $N~\sim({\bf 1},\,{\bf 1},\,0)$, i.e.~transforming as a singlet under the SM group $SU(3)_c\times SU(2)_L \times U(1)_Y$.~\footnote{We consider the EFT built with only one sterile neutrino. The generalization to a higher number of sterile neutrinos is straightforward. } In the latter case, $N$ is assumed to be sufficiently light, allowing it to be produced on-shell in meson decays. While sharing common features at low energies, these two scenarios have distinct UV realizations and phenomenological features that we will explore in Sec.~\ref{sec:pheno-smeft} and \ref{sec:pheno-nusmeft}, respectively.

\subsection{Which is the underlying EFT?}

In the following, we define the low-energy effective Lagrangian and identify the leading operators contributing to the $d_i\to d_j \nu\nu$ transition (with $i\neq j$) in the two scenarios that we consider.

\subsubsection{Left-handed neutrinos}

Under the minimal assumption that only SM fields are present at low-energy scales, the corresponding EFT is the so-called LEFT, which can be written at leading order in the expansion as follows,
\begin{align}
\label{eq:left}
    \mathcal{L}_{\mathrm{LEFT}} \supset \dfrac{1}{v^2}\sum_{X=L,R} C_{V_{XL}}\,{O}_{V_{XL}}  + \dfrac{1}{v^2}\bigg{[}\sum_{X=L,R}C_{S_{XL}}\,{O}_{S_{XL}} + C_{T_L}\, O_{T_L} + \mathrm{h.c.}\bigg{]}\,,
\end{align}

\noindent where $v=(\sqrt{2}G_F)^{-1/2}$ is the electroweak vacuum expectation value, $G_F$ denotes the Fermi constant, and $C_I$ stand for the effective coefficients of the $O_I$ operators, which are invariant under $SU(3)_c \times U(1)_{\mathrm{em}}$. The leading operators with SM neutrinos are~\cite{Buras:2014fpa}
\begin{align}
    \label{eq:left-lnc}
    {O}_{\substack{V_{LL} \\ ij \alpha\beta }}&=\big{(}\bar{d}_{Li} \gamma^\mu d_{Lj}\big{)}\big{(}\bar{\nu}_{L\alpha}\gamma_\mu \nu_{L\beta}\big{)} \,, & O_{\substack{V_{RL}\\ij\alpha\beta}} & =\big{(}\bar{d}_{Ri} \gamma^\mu d_{Rj}\big{)}\big{(}\bar{\nu}_{L\alpha}\gamma_\mu \nu_{L\beta}\big{)} \,, 
\end{align}

\noindent where quark flavor indices are denoted by Latin symbols, with $i\neq j$. Lepton flavor indices are denoted by Greek symbols. Within the SM, the only nonzero coefficient  is $C_{V_{LL}}$, with neutrino flavor diagonal and universal interactions, 
\begin{align}
C_{\substack{V_{LL}\\ij\alpha\beta}}\equiv\delta_{\alpha\beta}\,C_{\substack{V_{LL}\\ij\alpha\alpha}}^{\mathrm{SM}}\,,
\end{align}

\noindent which are given by
\begin{align}
C_{\substack{V_{LL}\\sb\alpha\alpha}}^{\mathrm{SM}}&=-\,\dfrac{\alpha_{\mathrm{em}}}{\pi s^2_W } \lambda^t_{sb}\, X_t\,,\qquad\qquad
C_{\substack{V_{LL}\\ds\alpha\alpha}}^{\mathrm{SM}}=-\,\dfrac{\alpha_{\mathrm{em}}}{\pi s^2_W } \Big{(} \lambda^t_{ds}\, X_t +\lambda_{ds}^c\, X_c^{\alpha} \Big{)}\,,
\end{align}

\noindent with the definitions of $\lambda_{ij}^q\equiv V_{qi}^\ast V_{qj}$ and $s^2_W \equiv \sin^2 \theta_W$, where $V$ denotes the CKM matrix, $\theta_W$ is the weak mixing angle, and $\alpha_{\mathrm{em}}$ stands for the fine-structure constant. The Inami-Lim functions $X_t$ and $X_c^\alpha$ capture the short-distance effects~\cite{Inami:1980fz}, which have been precisely computed in Ref.~\cite{Buchalla:1993bv}.

Operators with scalar and tensor currents belong to a different category from the vector ones introduced above as they violate Lepton number ($L$)~\cite{Buras:2024ewl},
\begin{align}
    \label{eq:left-lnv}
    {O}_{\substack{S_{LL}\\ij\alpha\beta}}&=\big{(}\bar{d}_{Ri} d_{Lj}\big{)}\big{(}\overline{\nu^C_{L\alpha}}\nu_{L\beta} \big{)} \,, & {O}_{\substack{S_{RL}\\ij\alpha\beta}}&=\big{(}\bar{d}_{Li} d_{Rj}\big{)}\big{(}\overline{\nu^C_{L\alpha}}\nu_{L\beta} \big{)} \,, \\[0.35em]
    {O}_{\substack{T_L\\ ij \alpha\beta}}&= \big{(}\bar{d}_{Ri} \sigma^{\mu\nu} d_{Lj}\big{)}\big{(}\overline{\nu^C_{L\alpha}}\sigma_{\mu\nu}\nu_{L\beta}\big{)}\,.\nonumber
\end{align}

\noindent where fermion charge-conjugation is defined as $\psi^C \equiv C \bar{\psi}^T$, with 
$C=i\gamma_2\gamma_0$. Note, in particular, that these operators are renormalized by QCD, with the anomalous dimensions known up to four-loop order~\cite{Vermaseren:1997fq}. Moreover, they are constrained by the properties of the charge-conjugation matrix, $\overline{\psi^C} \psi^\prime = + \overline{\psi^{\prime C}} \psi$ and $\overline{\psi^C} \sigma^{\mu\nu}\psi^\prime = -\overline{\psi^{\prime C}} \sigma^{\mu\nu}\psi$, where $\psi^{(\prime)}$ are generic fermion fields. Therefore, scalar and tensor effective coefficients are symmetric and anti-symmetric in neutrino flavor space, respectively.

Finally, we note that the operators in Eq.~\eqref{eq:left-lnc} appear at dimension six once the full SM gauge symmetry is imposed (i.e., in the SMEFT), whereas those in Eq.~\eqref{eq:left-lnv} appear only at dimension seven~\cite{Buras:2024ewl}. Therefore, the latter can only be dominant if a mechanism suppresses the dimension-six operators. The realization of such a scenario relies on very specific UV conditions, as will be discussed in Sec.~\ref{sec:pheno-smeft}. 

\subsubsection{Right-handed neutrinos}

In the presence of a light right-handed neutrino $N$, also known as a sterile neutrino, there are additional operators that can be written down in the EFT formulated below the electroweak scale, which will be dubbed as $\nu$LEFT in the following,
\begin{align}
\label{eq:nuleft}
    \mathcal{L}_{\nu\mathrm{LEFT}} \supset \dfrac{1}{v^2}\sum_{X=L,R} \bar{C}_{V_{XR}}\, \bar{O}_{V_{XR}} + \dfrac{1}{v^2}\bigg{[}\sum_{X=L,R} \bar{C}_{S_{XL}}\,\bar{O}_{S_{XL}} + \bar{C}_{T_L}\, \bar{O}_{T_L} + \mathrm{h.c.}\bigg{]},
\end{align}

\noindent where $\bar{C}$ are the effective coefficients of the operators $\bar{O}$  that contain the light field $N$,
\begin{align}
    \bar{O}_{\substack{V_{LR}\\ ij NN}}&=\big{(}\bar{d}_{Li} \gamma^\mu d_{Lj}\big{)}\big{(}\bar{N}\gamma_\mu N\big{)} \,, & \bar{O}_{\substack{V_{RR}\\ij NN}} & =\big{(}\bar{d}_{Ri} \gamma^\mu d_{Rj}\big{)}\big{(}\bar{N}\gamma_\mu N\big{)} \,,\\[0.35em]
    \bar{O}_{\substack{S_{LL}\\ijN\beta}}&=\big{(}\bar{d}_{Ri} d_{Lj}\big{)}\big{(}\bar{N}\nu_{L\beta} \big{)} \,,   & \bar{O}_{\substack{S_{RL}\\ijN\beta}}&=\big{(}\bar{d}_{Li} d_{Rj}\big{)}\big{(}\bar{N}\nu_{L\beta} \big{)} \,, \nonumber \\[0.35em]
    \bar{O}_{\substack{T_L\\ ij N\beta}}&= \big{(}\bar{d}_{Ri} \sigma^{\mu\nu} d_{Lj}\big{)}\big{(}\bar{N}\sigma_{\mu\nu}\nu_{L\beta}\big{)}\,. \nonumber
\end{align}

\noindent These operators contribute to processes such as $B\to K\nu N$ and $B\to K N N$ provided they are kinematically allowed, i.e.~$m_N < m_B-m_{K}$ and $2m_N < m_B-m_{K}$, respectively, $m_N$ being the mass of the sterile neutrino. In this EFT scenario, the scalar and tensor structures become allowed already at dimension six when the SM gauge symmetry is imposed~\cite{Rosauro-Alcaraz:2024mvx}, as will be discussed in Sec.~\ref{sec:pheno-nusmeft}. Notice, also, that these operators are subject to the same QCD running as those described in Sec.~\ref{sec:left}.

Before moving into the discussion on the phenomenological bounds, we notice that a different basis is considered in Ref.~\cite{Felkl:2023ayn}, with $N^C$ being identified with a fourth-neutrino generation. 
In this way, compact expressions are provided for the relevant rare decays, which apply to both scenarios discussed above. 
In our study, we opt to write down two different bases to better distinguish them, as they are realized by different constructions above the electroweak scale, with very different phenomenological features, as will be discussed in Sec.~\ref{sec:pheno-smeft} and \ref{sec:pheno-nusmeft}. We note, however, that we found full agreement between the expressions from Ref.~\cite{Felkl:2023ayn} and our results, which are collected in Appendix~\ref{app:formulas}.

\subsection{$B\to K \nu\nu$ and $B\to K^\ast \nu\nu$}
\label{ssec:bsnunu}

We now discuss the constraints from exclusive decays based on the $b\to s\nu\nu$ transition on the effective couplings introduced in Eq.~\eqref{eq:left} and \eqref{eq:nuleft}. These processes have attracted considerable interest in recent years, motivated by the first observation of the $B^+\to K^+\nu\nu$ channel by Belle-II~\cite{Belle-II:2023esi}, 
\begin{align}
\label{eq:BK-exp}
    \mathcal{B}(B^+ \to K^+ +\rm{inv})^{\mathrm{exp}} &= \big{(}2.3 \pm 0.5^{+0.5}_{-0.4}\big{)}\times 10^{-5}\,, 
\end{align}

\noindent which lies about $3\sigma$ above its SM prediction~\cite{Becirevic:2023aov},~\footnote{Long-distance contributions from $B^+\to \tau^+(\to K^+\bar{\nu})\nu$ are not included in the SM prediction given above~\cite{Kamenik:2009kc}, as they are subtracted experimentally in Ref.~\cite{Belle-II:2023esi}.}
\begin{align}
    \mathcal{B}(B^+ \to K^+ \nu \nu)^{\mathrm{SM}} &= (4.44\pm0.14_{\text{FF}}\pm0.27_{\text{par}})\times 10^{-6}\,,  
\end{align}

\noindent where we break down the form-factor uncertainty from the parametric one, which is dominated by the CKM matrix-elements. This prediction is based on the combined fit of $B\to K$ form factors computed by means of Lattice QCD (LQCD) simulations~\cite{Bailey:2015dka,Parrott:2022zte}, cf.~Ref.~\cite{Becirevic:2023aov}. For definiteness, we consider the $|\lambda_{sb}^{t}| \equiv |V_{tb} V_{ts}^\ast|= (39.3\pm 1.0)\times 10^{-2}$ value obtained through CKM unitarity from $|V_{cb}| = (40.0\pm 1.0)\times 10^{-3}$~\cite{FlavourLatticeAveragingGroupFLAG:2024oxs}, which is extracted from $B\to D \ell\bar{\nu}$ data by using the $B\to D$ form factors from LQCD~\cite{FermilabLattice:2015ilb}.~\footnote{Notice that using the inclusive value instead, namely $|V_{cb}|=(42.2 \pm  0.8)\times 10^{-3}$~\cite{HFLAV:2022esi}, would lead to $\mathcal{B}(B^+ \to K^+\nu\nu)^{\mathrm{SM}} = (4.94\pm 0.15_{\text{FF}}\pm0.22_{\text{par}})\times 10^{-6}$\,.}  Notice that the long-standing discrepancy between exclusive and inclusive determinations of $|V_{cb}|$ introduces an important ambiguity in the above predictions~\cite{Buras:2021nns,Becirevic:2023aov}. Constraints from $\Delta F=2$ processes seem to favor a value of $|V_{cb}|$ closer to the inclusive one, although excellent control of the meson-mixing matrix element is needed for these observables~\cite{FlavourLatticeAveragingGroupFLAG:2024oxs}. Currently, the experimental uncertainty is largely dominant, so that the specific choice of $|V_{cb}|$ has no significant impact on the interpretation of $b\to s\nu\bar{\nu}$ data. However, it will be fundamental to address this issue in the future to match the foreseen experimental sensitivity of Belle-II with $50~\mathrm{ab}^{-1}$~\cite{Belle-II:2018jsg} and of proposed experiments such as FCC-ee~\cite{Amhis:2023mpj}.

Another useful constraint comes from the $B\to K^\ast \nu\nu$ decay modes, for which there is only an upper limit thus far~\cite{Belle:2017oht},
\begin{align}
    \mathcal{B}(B^0 \to K^{\ast 0}+ {\rm inv})^{\mathrm{exp}} &< 2.7\times 10^{-5}~(90\%~\mathrm{CL})\,,
\end{align}
which is to be compared to the SM prediction~\cite{Becirevic:2023aov},
\begin{align}
    \mathcal{B}(B^0 \to K^{\ast 0}\nu\nu)^{\mathrm{SM}} &= (9.00\pm0.83_{\text{FF}}\pm 0.49_{\text{par}})\times 10^{-6}\,,  
\end{align}

\noindent which is obtained by using the $B\to K^\ast$ form-factors from Ref.~\cite{Bharucha:2015bzk}, combined with the same CKM inputs as given above. The $B\to K^\ast$ branching fraction, as well as the longitudinal polarization fraction ($F_L$), can provide useful tests at Belle-II of the mild excess observed in $B^+\to K^+ +{\rm inv}$ decays~\cite{Buras:2014fpa,Allwicher:2023xba}. Finally, we note that the experimental limits on the fully invisible decay mode, $\mathcal{B}(B_s\to {\rm inv})<5.6\times 10^{-4}$ (90$\%$ CL), obtained from a reinterpretation of LEP data~\cite{Alonso-Alvarez:2023mgc}, are, in principle, sensitive to pseudoscalar operators, but turn out to be less constraining than the observables discussed above.

\begin{figure}[!t]
    \begin{center}
    \includegraphics[width=0.499\textwidth]{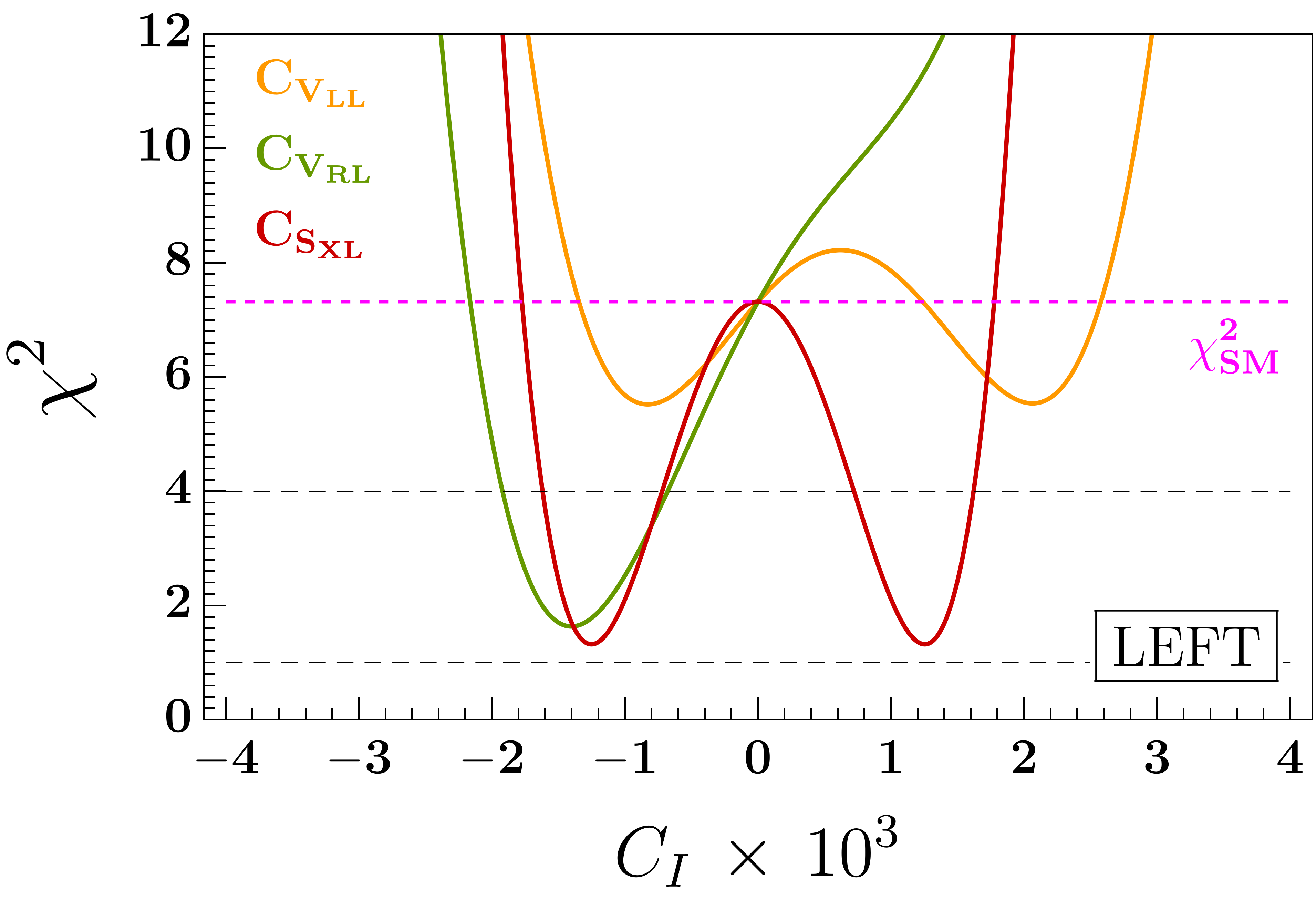}~\includegraphics[width=0.499\textwidth]{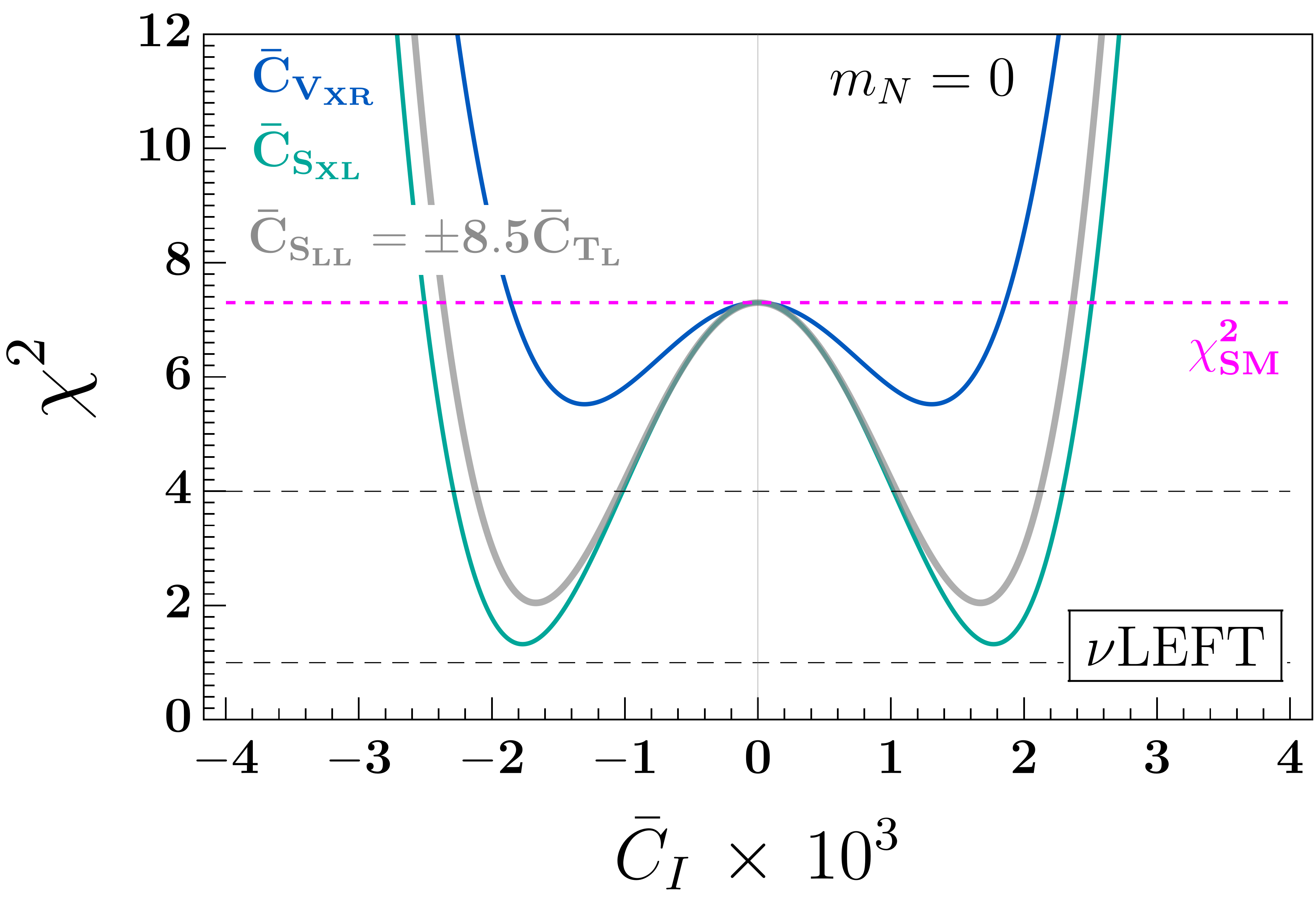}
    \end{center}
    \caption{\label{fig:chisq-left-B}\small \sl Value of the $\chi^2$ for individual Wilson coefficients fitted for the experimental result of $\mathcal{B}(B^+ \to K^+ + \mathrm{inv})$ and consistent with the limit on $\mathcal{B}(B^0 \to K^{\ast 0} + \mathrm{inv})$, compared to the SM value $\chi^2_{\mathrm{SM}}\simeq 7.3$. In the left panel, we consider the LEFT scenario for $C_{V_{LL}}$, $C_{V_{RL}}$ and $C_{S_{XL}}$, with $X=L,R$, cf.~Eq.~\eqref{eq:left}. In the right panel, we consider the $\nu$LEFT scenario for $\bar{C}_{V_{XR}}$, $\bar{C}_{S_{XL}}$, with $X=L,R$, as well as the combination $\bar C_{S_{LL}} \simeq \pm 8.5 \, \bar{C}_{T_{L}}$, cf.~Eq.~\eqref{eq:nuleft}.
    For simplicity, we set $m_N =0$ in the right panel. In both cases, all coefficients are  given at $\mu=m_b$, 
    quark flavor indices are fixed to $(i,j)=(2,3)$, and we consider a single active-neutrino flavor.   
    }
\end{figure}

\subsubsection*{Numerical analysis} Several EFT reinterpretations of Belle-II results have been performed, considering the leading operators in the SMEFT~\cite{Allwicher:2023xba,Bause:2023mfe,Allwicher:2024ncl} and the sub-leading ones that violate $L$~\cite{Buras:2024ewl}, as well as EFT scenarios with a sterile neutrino~\cite{Rosauro-Alcaraz:2024mvx}. We perform a $\chi^2$ analysis in Fig.~\ref{fig:chisq-left-B} to constrain the Wilson coefficients $C_I$ $(\bar{C}_I)$ using the experimental results on $B^+\to K^+ + {\rm inv}$ and $B^0\to K^{\ast 0} + {\rm inv}$. 
We distinguish the LEFT scenario with active neutrinos (left panel) from the $\nu$LEFT scenario with an additional sterile neutrino (right panel), as introduced in Eq.~\eqref{eq:left} and \eqref{eq:nuleft}, respectively. For simplicity, in the latter case, we assume that the sterile neutrino mass $m_N$ is negligible with respect to $m_B$. In Fig.~\ref{fig:chisq-left-B}, we consider combinations of Wilson coefficients that arise in UV concrete models, as will be discussed in Sec.~\ref{ssec:smeft-uv-completions} and \ref{ssec:nusmeft-uv-completions}, with a single flavor of active neutrinos. More specifically, in the left panel, we consider 
the fit for the individual coefficients
$C_{V_{XL}}$ and $C_{S_{XL}}$ (with $X=L,R$), which are defined in Eqs.~\eqref{eq:left-lnc} and \eqref{eq:left-lnv}.~\footnote{Notice that $C_{T_L}$ is anti-symmetric in neutrino flavors, as defined in Eq.~\eqref{eq:left-lnv}, thus vanishing for a single neutrino flavor. } In the right panel, we consider the fit for
the individual coefficients $\bar{C}_{V_{XR}}$ and $\bar{C}_{S_{XL}}$ (with $X=L,R$
), in addition to the combination $\bar{C}_{S_{LL}} \simeq \pm 8.5\, \bar{C}_{T_L}$, which is predicted in scalar leptoquark models~\cite{Becirevic:2016yqi,Dorsner:2016wpm}.
In all cases, we accounted for the QCD running from the $\Lambda \simeq 1~\mathrm{TeV}$ scale down to $\mu=m_b$. Our main findings, in agreement with previous studies, are summarized below:
\begin{itemize}
    \item[$\bullet$] The lepton-number conserving operators ${O}_{V_{LL}}$ and ${O}_{V_{RL}}$ can both accommodate Belle-II data, with a mild preference for the latter, which is less affected by the $B^0\to K^{\ast 0}+ {\rm inv}$ constraints~\cite{Bause:2023mfe,Allwicher:2023xba}. These operators could appear in the SMEFT at dimension six, with the following effective scales,
    \begin{align}
        \Lambda_{V_L}^{(6)}\Big|_{\text{best fit}} &\equiv  \dfrac{v}{\sqrt{|C_{V_{LL}}|}} 
        \simeq 9~\mathrm{TeV}\,, \\[0.4em] 
        \Lambda_{V_R}^{(6)}\Big|_{\text{best fit}} &\equiv  \dfrac{v}{\sqrt{|C_{V_{RL}}|}} 
        \simeq 7~\mathrm{TeV}\,, \nonumber
    \end{align}

    \noindent where flavor indices are omitted, and where we only keep the solution from Fig.~\ref{fig:chisq-left-B} with the smallest Wilson coefficient for $C_{V_{LL}}$.

    \item[$\bullet$] The scalar operators in the first EFT scenario can also provide a good description of current data~\cite{Buras:2024ewl}. Since these operators necessarily arise at dimension seven above the electroweak scale, they amount to a lower value of the effective  scale,
    \begin{align}
        \Lambda_S^{(7)}\Big|_{\text{best fit}} &\equiv  \dfrac{v}{\sqrt[3]{|C_{S_{XL}}|}} 
        \simeq 2~\mathrm{TeV}\,, 
    \end{align}
    which is due to the different scaling on $\Lambda$ and since this operator does not interfere with the SM one.
    
    \item[$\bullet$] Finally, the $\nu$LEFT operators with a sterile neutrino in the right panel of Fig.~\ref{fig:chisq-left-B} can also accommodate current data~\cite{Rosauro-Alcaraz:2024mvx}. The scenario with vector operators $\bar{O}_{V_{XR}}$ (with $X=L,R$) leads to the effective scale,
    \begin{align}
        \bar{\Lambda}^{(6)}_V &\Big|_{\text{best fit}} \equiv  \dfrac{v}{\sqrt{|\bar{C}_{V_{XR}}|}} 
        \simeq 7~\mathrm{TeV}\,, 
    \end{align}
    where we recall that we use an overbar to distinguish the $\nu$LEFT operators. Similarly, in the scalar case, the coefficients $\bar{C}_{S_{XL}}$ lead to
    \begin{align}
        \label{eq:lambda-nusmeft-B-scalar}
        \bar{\Lambda}^{(6)}_S\Big|_{\text{best fit}} &\equiv  \dfrac{v}{\sqrt{|\bar{C}_{S_{XL}}|}} 
        \simeq 6~\mathrm{TeV}\,. 
    \end{align}
    
    \noindent This value remains almost unaffected in the particular scenario with both scalar and tensor couplings, as predicted in leptoquark models, cf.~Sec.~\ref{ssec:nusmeft-uv-completions}.
\end{itemize}

\noindent The viable cases identified above will be used as benchmarks for the RG analysis in the following Sections. 

One important caveat of the numerical analysis presented in Fig.~\ref{fig:chisq-left-B} is that the experimental constraints from $B\to K^{(\ast)}+\mathrm{inv}$ searches depend on the theoretical priors used in the experimental searches. In particular, scalar operators can significantly modify the kinematic distributions of these processes, thereby possibly affecting the resulting bounds. Recently, the Belle-II collaboration has provided a model-agnostic likelihood for the $B\to K+\mathrm{inv}$ constraints~\cite{Belle-II:2025lfq,Gartner:2024muk}, allowing for a consistent reinterpretation of the experimental constraints in the presence of New Physics. We have explicitly verified that the $2\sigma$ constraint on the scalar Wilson coefficients, $C_{S_{XL}}\in{(0.7,1.6)\times 10^{-3}}$, obtained in our analysis in Fig.~\ref{fig:chisq-left-B}, would be modified to $C_{S_{XL}}\in{(1.5,3.6)\times 10^{-3}}$ when this reinterpretation is employed.~\footnote{Note that the $C_{S_{XL}}$ range provided in Ref.~\cite{Belle-II:2025lfq} has been obtained after marginalizing over the full set of operators. Instead, we perform a fit with a single effective coefficient, as motivated by minimalistic New Physics scenarios.} In the following, we nevertheless employ the constraints obtained in our naive analysis, as the differences arising from the reinterpretation do not affect our main conclusions, and since this reinterpretation is not available for the scenario with a massive sterile neutrino, as will be considered in Sec.~\ref{sec:pheno-nusmeft}. Whenever relevant, we comment on their potential impact.

\subsection{$K\to \pi \nu\nu$ }
\label{ssec:sdnunu}

We now discuss the rare kaon decays $K^+ \to \pi^+\nu\nu$, which are also theoretically clean observables and which are very sensitive to New Physics effects. Recently, the NA62 collaboration reported the most precise determination of this process~\cite{Chang:2026vvx},
\begin{align}
    \label{eq:Kpinunu-exp}
    \mathcal{B}(K^+ \to \pi^+ + {\rm inv})^{\mathrm{exp}} &= \big{(}9.6^{+1.9}_{-1.8}\big{)}\times 10^{-11}\,,
\end{align}
which is compatible with the SM prediction,~\cite{Brod:2021hsj}
\begin{align}
    \label{eq:Kpinunu-sm}
    \mathcal{B}(K^+ \to \pi^+\nu\nu)^{\mathrm{SM}} &= \big{(}7.75 \pm 0.16_{\text{SD}} \pm 0.24_{\text{LD}}\pm 0.59_{\text{par}}\big{)}\times 10^{-11}\,,
\end{align}
where we split the uncertainty into short-distance (SD) and long-distance (LD) contributions, and the parametric uncertainties (par), respectively. The $K\to \pi$ vector form-factor $(f_+)$ shape is taken from Ref.~\cite{Carrasco:2016kpy}, which is combined with the FLAG average for its overall normalization, $f_+(0)=0.9698\pm 0.0017$~\cite{FlavourLatticeAveragingGroupFLAG:2024oxs}.~\footnote{Alternatively, if we had considered the form-factors determined from $K\to\pi\ell\nu$ experimental data~\cite{Mescia:2007kn}, we would have obtained instead, $\mathcal{B}(K^+\to\pi^+\nu\nu)^{\mathrm{SM}} = \big{(}7.55 \pm 0.16_{\text{SD}} \pm 0.24_{\text{LD}}\pm 0.55_{\text{par}}\big{)}\times 10^{-11}$, using the same inputs for the other parameters. This value is smaller than the one quoted in Eq.~\eqref{eq:Kpinunu-sm}, but fully compatible within uncertainties.} The parametric uncertainties are dominated by the CKM inputs~\cite{Brod:2021hsj}. To determine $|\lambda_{ds}^{t}| \equiv |V_{ts} V_{td}^\ast|$, we use once again CKM unitarity, combined with the value for the Cabibbo angle $\lambda \equiv |V_{us}|=0.2252\pm 0.0006$ extracted from $K_{\mu 2}/\pi_{\mu 2}$ in Ref.~\cite{Eboli:2025vks} using the theoretical inputs from Ref.~\cite{DiCarlo:2019thl,Cirigliano:2022yyo}, $|V_{cb}|\equiv A \lambda^2= (40.0\pm 1.0)\times 10^{-3}$ from $B\to D\ell \bar{\nu}$ data~\cite{FlavourLatticeAveragingGroupFLAG:2024oxs}, and $\bar{\rho}=0.157\pm 0.021$ and $\bar{\eta}=0.376\pm 0.035$ from the New Physics fit from UTfit~\cite{UTfit:2022hsi}. These inputs allow us to determine $|\lambda_{sd}^{t}| =(3.3\pm 0.2)\times 10^{-4} $, with the dominant uncertainty arising from $|V_{cb}|$. We stress, once again, that the SM predictions for rare decays would differ if the inclusive value of $|V_{cb}|$ were used instead~\cite{Buras:2021nns}, but these uncertainties remain subdominant relative to the experimental ones.~\footnote{More specifically, the inclusive value $|V_{cb}|=(42.2 \pm  0.8)\times 10^{-3}$~\cite{HFLAV:2022esi}, combined with the lattice QCD form-factors~\cite{Carrasco:2016kpy}, would lead to $\mathcal{B}(K^+ \to \pi^+\nu\nu)^{\mathrm{SM}} = \big{(}8.96 \pm 0.19_{\text{SD}} \pm 0.27_{\text{LD}}\pm 0.44_{\text{par}}\big{)}\times 10^{-11}$.} For the other theoretical inputs, we closely follow Ref.~\cite{Brod:2021hsj}, including the long-distance contributions estimated in Ref.~\cite{Isidori:2005xm}.

Complementarity constraints can be obtained from the $K_L\to \pi^0\nu\nu$ decay, which is sensitive to CP-violating contributions~\cite{Grossman:1997sk}. Currently, the most stringent constraints are given by KOTO, which has set the upper limit $\mathcal{B}(K_L\to \pi^0 + \rm inv)<2.2\times 10^{-9}$ ($90\%$ CL)~\cite{KOTO:2024zbl}, which is about two orders of magnitude above its SM prediction~\cite{Brod:2021hsj}. In the presence of right-handed neutrinos, it is possible to enhance the $K_{L(S)}\to \nu\nu$ decays as well~\cite{Gninenko:2014sxa,Abada:2016plb}. However, for all neutral kaon decays, the current experimental sensitivity is not sufficient to test the scenarios that we consider~\cite{BESIII:2025kjj}. In the future, the KOTO-II experiment is expected to greatly enhance the experimental sensitivity on the $K_L\to\pi^0+ \rm inv$ channel~\cite{KOTO:2025gvq}, providing strong constraints on New Physics contributions.

\subsubsection*{Numerical analysis} 

\begin{figure}[!t]
    \begin{center}
    \includegraphics[width=0.5\textwidth]{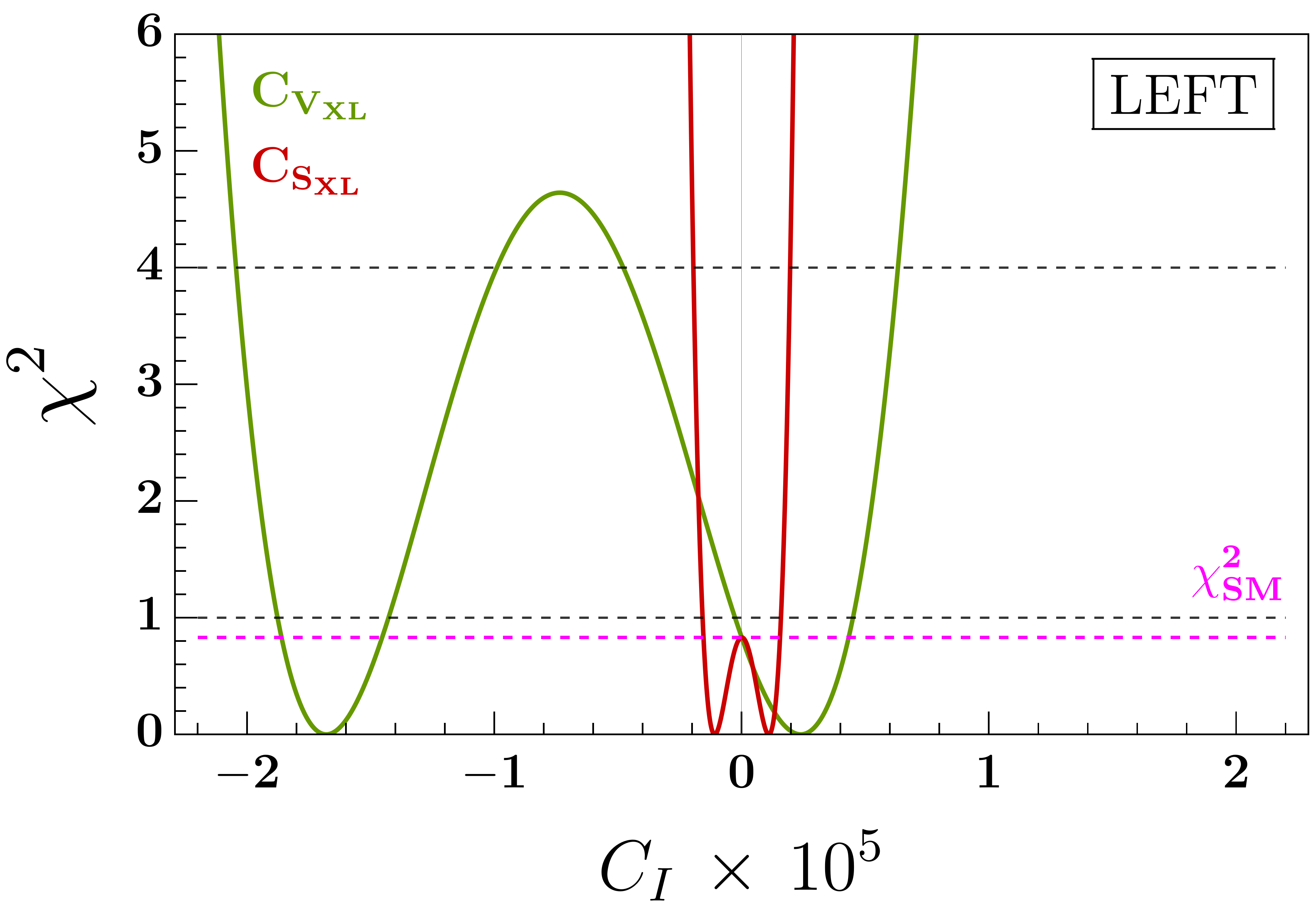}~\includegraphics[width=0.5\textwidth]{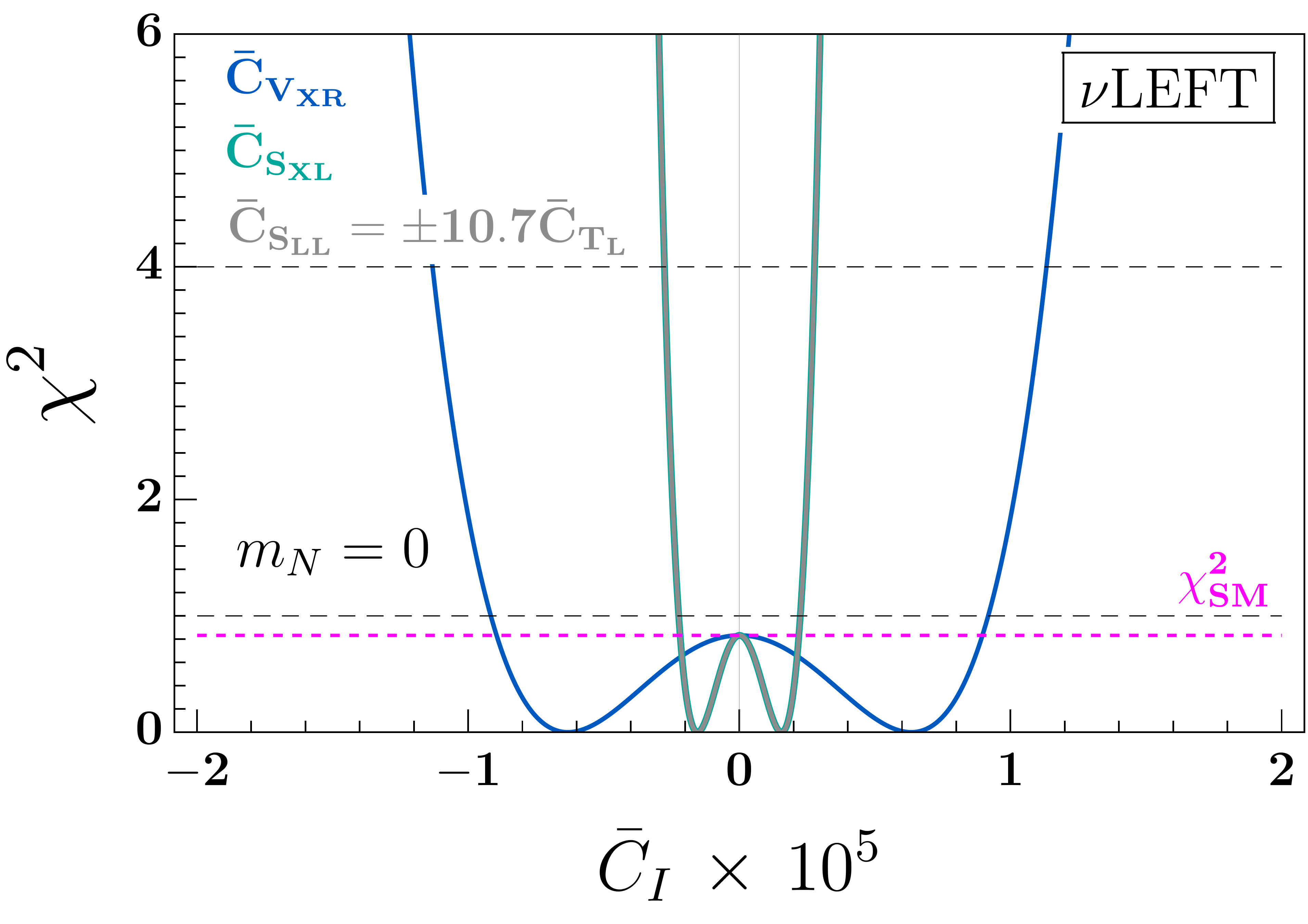}
    \caption{\small \sl \label{fig:chisq-left-K}Value of the $\chi^2$ for individual Wilson coefficients fitted for the experimental result of $\mathcal{B}(K^+ \to \pi^+ + \mathrm{inv})$  compared to the SM value $\chi^2_{\mathrm{SM}}\simeq 0.84$. In the left panel, we consider the LEFT scenario for $C_{V_{XL}}$ and $C_{S_{XL}}$, with $X=L,R$, cf.~Eq.~\eqref{eq:left}. In the right panel, we consider the $\nu$LEFT scenario for $\bar{C}_{V_{XR}}$, $\bar{C}_{S_{XL}}$, with $X=L,R$, and the combination $\bar{C}_{S_{LL}} \simeq \pm 10.7 \,\bar{C}_{T_{L}}$, cf.~Eq.~\eqref{eq:nuleft}. For simplicity, we set $m_N=0$ in the right panel. In both cases, quark flavor indices are fixed to $(i,j)=(1,2)$, and we consider a single active-neutrino flavor.   }
    \end{center}
\end{figure}

In Fig.~\ref{fig:chisq-left-K}, we perform a similar $\chi^2$ analysis of kaon data within the LEFT (left panel), and the $\nu$LEFT (right panel). In the latter case, we assume that the sterile neutrino has a negligible mass in comparison to $m_K$. We consider the same scenarios as in Sec.~\ref{ssec:bsnunu}, namely the individual effective coefficients, as well as the combination of scalar and tensor coefficients appearing in leptoquark models. More specifically, in the latter scenario, the Wilson coefficients are related via $\bar{C}_{S_{LL}}\simeq \pm 10.7\,\bar{C}_{T_L}$ at the scale $\mu=2~\mathrm{GeV}$, after accounting for the QCD running from the $\Lambda \simeq 1$ TeV scale. Since $K^+\to \pi^+ +\rm inv$ data agrees with its SM prediction, we can derive lower bounds on the New Physics scale for the different scenarios: 
\begin{itemize}
    \item[$\bullet$] The simplest scenario is the one with vector coefficients $C_{V_{XL}}$ (with $X=L,R$), as before, which are generated by dimension-six operators in the SMEFT. Current data allow us to set the following lower bound on the effective scale of New Physics,
    \begin{align}
        {\Lambda}^{(6)}_{V_X} \Big|_{\text{best fit}}&\equiv  \dfrac{v}{\sqrt{|{C}_{V_{XL}}|}} 
        \gtrsim 160~\mathrm{TeV}\,, 
    \end{align}
    
    \noindent which is given to $2\sigma$ accuracy. Notice, in particular, that the effects of these operators are dominated by interference terms, as depicted by the asymmetric intervals in Fig.~\ref{fig:chisq-left-K}.
    
    \item[$\bullet$] The scalar operators $C_{S_{XL}}$ (with $X=L,R$) can also accommodate current kaon data provided that
    \begin{align}
        {\Lambda}^{(7)}_{S}\Big|_{\text{best fit}} &\equiv  \dfrac{v}{\sqrt[3]{|{C}_{S_{XL}}|}} 
        \gtrsim 20~\mathrm{TeV}\,, 
    \end{align}

    \noindent where we consider, once again, a different scaling of the LEFT coefficients on the EFT cutoff, as these operators only appear at dimension seven above the electroweak scale.
    
    \item[$\bullet$] Finally, assuming that sterile neutrinos appear with mass $m_N<m_K-m_\pi$, the operators in the right panel of Fig.~\ref{fig:chisq-left-K} can also describe these processes. For instance, the vector $\bar{C}_{V_{XR}}$ and scalar $\bar{C}_{S_{XL}}$ coefficients can be probed with an effective scale,
    \begin{align}
        \label{eq:lambda-nusmeft-K-scalar}
        \bar{\Lambda}^{(6)}_{V}\Big|_{\text{best fit}} &\equiv  \dfrac{v}{\sqrt{|{C}_{V_{XR}}|}}
        \gtrsim 98~\mathrm{TeV}\,, \\[0.3em]
        \bar{\Lambda}^{(6)}_{S}\Big|_{\text{best fit}} &\equiv  \dfrac{v}{\sqrt{|{C}_{S_{XL}}|}}
        \gtrsim 198~\mathrm{TeV}\,, \nonumber
    \end{align}

    where we have neglected $m_N$, for simplicity.
\end{itemize}

\noindent As expected, FCNC kaon decays can generally probe much higher New Physics scales than the $B$-meson decays discussed in Sec.~\ref{ssec:bsnunu}, as they are more strongly suppressed within the SM. Nevertheless, these different effective scales can be easily reconciled if a hierarchical flavor structure is considered in the quark sector~\cite{Allwicher:2024ncl}. 

\

In Secs.~\ref{sec:pheno-smeft} and~\ref{sec:pheno-nusmeft}, we discuss the formulation of the gauge-invariant EFTs that lead to the above low-energy operators, both with only the  SM fields (namely, within the SMEFT) and with additional sterile neutrinos (i.e., in the $\nu$SMEFT), respectively.

\section{SMEFT phenomenology}
\label{sec:pheno-smeft}

In this section, we study the interplay between the $d_i\to d_j\nu\nu$ transition ($i\neq j$) and neutrino-physics observables, assuming that these processes are affected by dimension-seven Lepton Number Violating (LNV) operators in the SMEFT. In Sec.~\ref{ssec:smeft}, we remind the relevant operators that contribute to these processes in the SMEFT. In Secs.~\ref{ssec:smeft-nu-masses} and \ref{ssec:smeft-0nubb}, we explore the connection between these observables induced by RG evolution. In Sec.~\ref{ssec:smeft-numerical}, we perform our numerical analysis. Finally, in Sec.~\ref{ssec:smeft-uv-completions}, we summarize our conclusions for this scenario and discuss possible UV realizations.

\subsection{Operator basis}
\label{ssec:smeft}
We begin by writing the SMEFT operators relevant for the
$d_i \to d_j \nu\nu$ transition, with $i\neq j$, using the conventions that are summarized in Appendix~\ref{app:conventions}.
We denote the lepton and quark doublets by $l$ and $q$, respectively, whereas the weak singlet fermions are denoted by $u$, $d$ and $e$. Lepton flavor indices are denoted by Greek symbols, as before, whereas lower Latin symbols are used for quark flavors. Moreover, we explicitly show $SU(2)_L$ indices using upper Latin symbols.  In the flavor sector, we write the Yukawa Lagrangian as follows,
\begin{align}
\mathcal{L}_{\mathrm{yuk}}=-y_d\,\bar{q}{H}d-y_u\,\bar{q} \widetilde{H} u-y_\ell\,\bar{l}{H}e + \mathrm{h.c.}\,,
\end{align}

\noindent where flavor indices are omitted, the conjugate Higgs-doublet is defined by $\widetilde{H}=i \tau^2 H^\ast$, and the quark and lepton Yukawa matrices are assumed to be diagonal,
\begin{align}
\label{eq:yukawa-convention}
y_d = \mathrm{diag}(y_d,y_s,y_b)\,,\qquad y_u = V^\dagger\,\mathrm{diag}(y_u,y_c,y_t)\,, \qquad y_\ell = \mathrm{diag}(y_e,y_\mu,y_\tau)\,,
\end{align}

\noindent where $y_{\psi_i}=\sqrt{2} \, m_{\psi_i}/v$ (with $\psi\in \lbrace d,u,\ell \rbrace$) are the Yukawa couplings and $m_{\psi_i}$ their corresponding masses. With the above conventions, the SMEFT Lagrangian can be  written as
\begin{align}
    \label{eq:smeft}
    \mathcal{L}_{\mathrm{SMEFT}} = \mathcal{L}_{\mathrm{SM}} &+ \dfrac{1}{2\Lambda} \,\left(\mathcal{C}^{(5)}_{LH}\,\mathcal{O}^{(5)}_{LH}+\text{h.c}\right)\\[0.35em]
    &+ \dfrac{1}{\Lambda^2} \sum_I\,\mathcal{C}^{(6)}_I\,\mathcal{O}^{(6)}_I + \dfrac{1}{\Lambda^3} \sum_I\,\left(\mathcal{C}^{(7)}_I\,\mathcal{O}^{(7)}_I +\text{h.c}\right)+\dots\, \nonumber
\end{align}

\noindent At dimension five, only the Weinberg operator appears~\cite{Weinberg:1979sa}, 
\begin{align}
\label{eq:Weinberg}
\mathcal{O}^{(5)}_{\substack{LH\\ \alpha\beta}} =  \epsilon^{ij} \epsilon^{mn} \big{(}l_{\alpha}^{i\,T} C l_{\beta}^m \big{)} H^j H^n\, ,
\end{align}
which is responsible for generating neutrino masses, where we remind that $C=i\gamma_2\gamma_0$ denotes fermion charge-conjugation.~\footnote{Our convention for the Levi-Civita symbol is $\epsilon_{12}=-\epsilon_{21}=+1$.} The dimension-six operators relevant to our study are~\footnote{For the sake of simplicity, and whenever it does not lead to ambiguity, we omit the superscript on the Wilson coefficients denoting the operator dimension.} 
\begin{align}
\label{eq:smeft-d6}
\mathcal{O}_{\substack{lq\\ \alpha\beta ij}}^{(1)}&=\left(\bar{l}_
\alpha\gamma_\mu l_\beta \right)\left(\bar{q}_i\gamma^\mu q_j\right) \,, &
\mathcal{O}_{\substack{lq\\ \alpha\beta ij}}^{(3)}&=\left(\bar{l}_\alpha\gamma_\mu \tau^I l_\beta\right)\left(\bar{q}_i\gamma^\mu\tau^I q_j\right)\,,\\[0.35em]
\mathcal{O}_{\substack{ld\\ \alpha\beta ij}}&=\left(\bar{l}_\alpha\gamma_\mu l_\beta\right)\left(\bar{d}_i\gamma^\mu d_j\right) \,,\nonumber
\end{align}

\noindent where $\tau^I$ ($I=1,2,3$) denote the Pauli matrices. These operators conserve lepton number, thus contributing to the LEFT  operators in Eq.~\eqref{eq:left-lnc} below the electroweak scale. 
Lepton-number-violating contributions to these processes only appear at dimension seven, through a single gauge-invariant operator~\cite{Lehman:2014jma},
\begin{align}
\label{eq:smeft-d7}
\mathcal{O}_{\substack{\bar{d} l q l H 1\\ i\alpha j\beta }} &=\epsilon^{a b} \epsilon^{d e}\left(\overline{d_{i}} \,l_{\alpha }^a\right)\left(q_{j }^{b\,T} C \,l_{\beta}^d\right) H^e\,,
\end{align}
which generates scalar and tensor contributions via Fierz rearrangement, cf.~below.

\subsubsection*{Tree-level matching}

The above operators can be matched onto Eq.~\eqref{eq:left} at the electroweak scale $\mu_{\mathrm{ew}}\simeq m_Z$,
\begin{align}
\label{eq:smeft-match}
C_{\substack{V_{LL}\\ ij\alpha\beta }}(\mu_{\mathrm{ew}})&=\frac{v^2}{\Lambda^2}\,\mathcal{C}_{\substack{lq\\ \alpha\beta ij}}^{(1-3)}(\mu_{\mathrm{ew}})\,, & C_{\substack{V_{RL}\\ ij\alpha\beta }}(\mu_{\mathrm{ew}}) &=\frac{v^2}{\Lambda^2} \, \mathcal{C}_{\substack{ld\\ \alpha\beta ij}}(\mu_{\mathrm{ew}})\,, \\[0.45em]
C_{\substack{S_{LL}\\ij\alpha\beta}}(\mu_{\mathrm{ew}}) &= -\frac{1}{2\sqrt{2}}\frac{v^3}{\Lambda^3} \, \mathcal{C}_{\substack{\bar{d}l q l H1\\ i \beta j\alpha}}^{[S]}(\mu_{\mathrm{ew}})\,, & C_{\substack{S_{RL}\\ij\alpha\beta}}(\mu_{\mathrm{ew}}) &= 0\,, \nonumber \\[0.45em]
C_{\substack{T_{L}\\ij\alpha\beta}}(\mu_{\mathrm{ew}})  &= -\frac{1}{8\sqrt{2}}\frac{v^3}{\Lambda^3} \, \mathcal{C}_{\substack{\bar{d}l q l H1\\i \beta j\alpha}}^{[A]}(\mu_{\mathrm{ew}}) \,, \nonumber
\end{align}

\noindent where, for convenience, we define $\mathcal{C}_{lq}^{1\pm 3} \equiv \mathcal{C}_{lq}^{1}\pm \mathcal{C}_{lq}^{3}$, as well as the symmetric and anti-symmetric combinations of Wilson coefficients in lepton flavors,
\begin{equation}
\label{eq:sym-antisym-flavor}
\mathcal{C}_{\substack{\bar{d}l ql H1\\i\alpha j\beta}}^{[S]} \equiv \frac{1}{2}\bigg{(}\mathcal{C}_{\substack{\bar{d}l ql H1\\ i\alpha j\beta}}+\mathcal{C}_{\substack{\bar{d}l ql H1\\ i\beta j\alpha}}\bigg{)}\,, \qquad\quad
\mathcal{C}_{\substack{\bar{d}l ql H1\\i\alpha j\beta}}^{[A]} \equiv \frac{1}{2}\bigg{(}\mathcal{C}_{\substack{\bar{d}l ql H1\\i\alpha j\beta}}-\mathcal{C}_{\substack{\bar{d}l ql H1\\ i\beta j\alpha}}\bigg{)}\,,
\end{equation}

\noindent and, similarly, for the other dimension-seven operators that will be introduced later. Notice, in particular, that the coefficient $C_{S_{RL}}$ is only induced at higher order in the SMEFT expansion. 

Clearly, from the EFT power counting, dimension-six operators will provide the dominant contribution to these processes in most concrete scenarios, unless they are absent or suppressed by powers of $v/\Lambda$ relative to those appearing at higher dimensions. To achieve this hierarchy in the effective coefficients, the UV completions of these operators must satisfy very restrictive conditions, which will be explored in Sec.~\ref{ssec:smeft-uv-completions}. 

Finally, we stress once again that the dimension-seven operators in Eq.~\eqref{eq:smeft-d7} lead to modifications of the kinematic distributions of $K\to \pi \nu\nu$ and $B\to K\nu\nu$ decays~\cite{Buras:2024ewl}. Such effects cannot be induced by dimension-six operators,
since the operators in Eq.~\eqref{eq:smeft-d6} have the same Lorentz structure as the SM contributions, thus only rescaling the full di-neutrino spectrum~\cite{Becirevic:2023aov}. However, dimension-seven operators suffer from stringent constraints from neutrino observables, as we explore in the following. More specifically, this operator mixes through RG evolution into the dimension-five and seven Weinberg operators~\cite{Zhang:2023kvw}, which contribute to neutrino masses~\cite{Chala:2021juk}. Furthermore, through RG effects, this operator can also potentially induce new effects in $0\nu\beta\beta$ decays~\cite{deVries:2026ujz}.

\subsection{Neutrino masses and operator mixing}
\label{ssec:smeft-nu-masses}

We now discuss the impact of dimension-seven operators introduced in Eq.~\eqref{eq:smeft-d7} on generating neutrino masses through RG effects \cite{Chala:2021juk}. We assume that only the coefficient $\mathcal{C}_{\substack{\bar{d}l q l H1}}^{(7)}$ is present in the UV, with quark-flavor indices $i\neq j$, which are needed to describe the $3\to 2$ and $2\to 1$ transitions, which enter $b\to s$ and $s\to d$ transitions at tree level, respectively. Majorana neutrino masses receive tree-level contributions from the Weinberg operator $\smash{\mathcal{O}_{LH}^{(5)}}$ defined in Eq.~\eqref{eq:Weinberg} and from its dimension-seven analog,
\begin{align}
\mathcal{O}_{\substack{LH\\ \alpha\beta}}^{(7)}\,= \epsilon^{ab} \epsilon^{cd} \big{(}l_{\alpha}^{a\,T} C l_{\beta}^c \big{)} H^b H^d\,\big{(}H^\dagger H\big{)}\,,
\end{align}

\noindent which amounts to the following neutrino-mass matrix at tree-level, 
\begin{align}
\label{eq:nu-masses}
m_\nu^{(0)} = -\dfrac{v^2}{2\Lambda}\bigg{[}\mathcal{C}_{LH}^{(5)}+\mathcal{C}_{LH}^{(7)} \,\dfrac{v^2}{\Lambda^2}\bigg{]}\,, 
\end{align}

\noindent where flavor indices are omitted and the Wilson coefficients are evaluated at $\mu=\mu_{\mathrm{ew}}$. Although suppressed by additional powers of $\mu_H^2/\Lambda^2$, 
where $\mu_H^2$ is the quadratic term in the Higgs potential,
the RG-induced contributions to neutrino masses from dimension-five and seven Weinberg operators, turn out to be of the same order in the EFT scenario that we consider, as we will demonstrate in the following.~\footnote{Our convention for the Higgs potential reads $\mathcal{L}_{\mathrm{Higgs}} \supset - \mu_H^2\, H^\dagger H-\lambda \,(H^\dagger H)^2$, cf.~Appendix~\ref{app:conventions}.} 

\subsubsection{RG evolution}

The one-loop RG equations can be generically written as 
\begin{align}
\dot{\mathcal{C}}_I \equiv 16\pi^2 \dfrac{\mathrm{d}\mathcal{C}_I}{\mathrm{d}\log \mu} = \hat{\gamma}_{JI}(\mu)\, \mathcal{C}_J\,,
\end{align}

\noindent where $\mathcal{C}_I$ is a generic Wilson coefficient and $\hat{\gamma}$ denotes the one-loop anomalous dimension matrix. In our scenario, the dominant contributions to the running of the Weinberg operators are given by~\cite{Zhang:2023kvw},~\footnote{Notice that our definition of $\mathcal{O}_{LH}^{(5)}$ is the Hermitian conjugate of the one used in Ref.~\cite{Zhang:2023kvw}.}
\begin{align}
\dot{\mathcal{C}}_{\substack{LH\\ \alpha\beta}}^{(7)} &= -{2N_c}\, \bigg{[} \lambda \,\big{(}y_d\big{)}_{rp} - \big{(}y_d\,y_d^\dagger\,y_d\big{)}_{rp} \bigg{]}\, \mathcal{C}_{\substack{\bar{d}l q l H1\\ p\alpha r\beta}}^{[S]} +4 N_c\,\bigg{[} \lambda \,\big{(}y_u^\dagger\big{)}_{rp} - \big{(}y_u^\dagger\,y_u\,y_u^\dagger \big{)}_{rp} \bigg{]}\mathcal{C}_{\substack{\bar{q}ullH\\ pr\alpha \beta}}^{[S]}+\dots \,, \nonumber     \\[0.45em]
\label{eq:rg-smeft}
\dot{\mathcal{C}}_{\substack{LH\\ \alpha\beta}}^{(5)} &= -\dfrac{2\mu_H^2}{\Lambda^2} N_c  \bigg{[} \big{(}y_d\big{)}_{rp} \, \mathcal{C}_{\substack{\bar{d}l q l H1\\ p\alpha r\beta}}^{[S]} - 2\big{(}y_u\big{)}_{rp}\,\mathcal{C}_{\substack{\bar{q}ullH\\ pr\alpha \beta}}^{[S]} \bigg{]} \, +\dots \,,
\end{align}

\begin{figure}[t]
\centering
\includegraphics[width=0.95\linewidth]{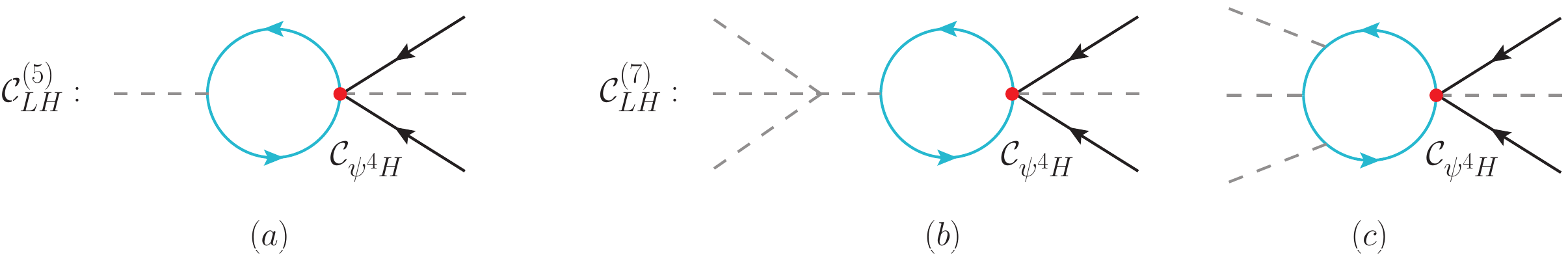}
\caption{\small \sl Relevant one-loop diagrams contributing to the mixing of the $\psi^4H$ operators $\mathcal{O}_{\bar{d}lqlH1}$ and $\mathcal{O}_{\bar{q}ullH}$ into the dimension-five and seven Weinberg operators, $\smash{\mathcal{O}_{LH}^{(5)}}$ and $\smash{\mathcal{O}_{LH}^{(7)}}$, respectively. Quark lines are depicted in blue.}
\label{fig:diags-mass-smeft}
\end{figure}
\begin{figure}[t]
\centering
\includegraphics[width=0.65\linewidth]{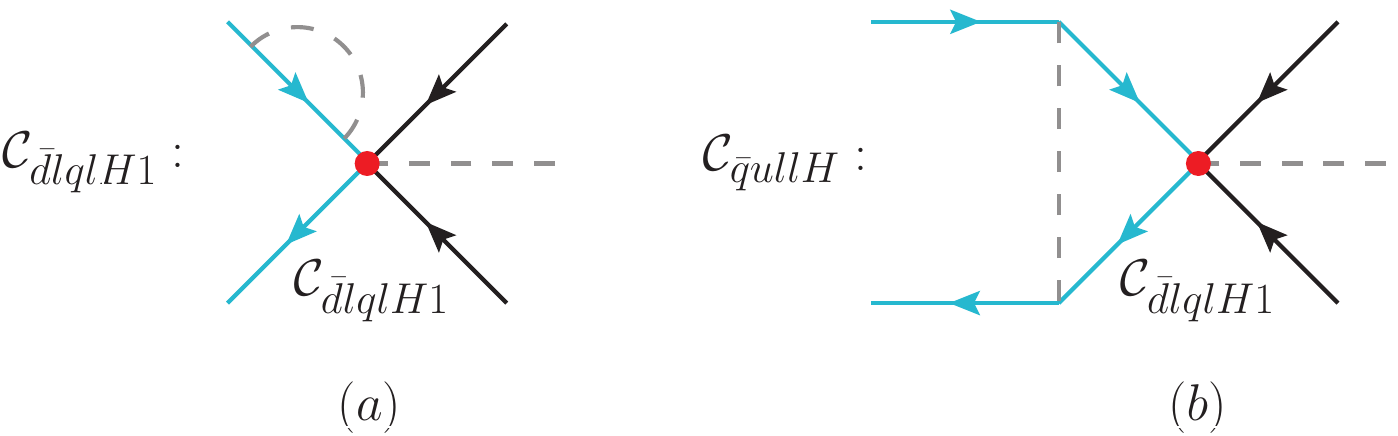}
\caption{\small \sl One-loop diagram introducing the self-mixing of operators of the type $\mathcal{O}_{\bar{d}lqlH1}$ (left panel) and the mixing of $\mathcal{O}_{\bar{d}lqlH1}$ into $\mathcal{O}_{\bar{q}ullH}$ through Yukawa interactions (right panel). In both cases, the insertion of Yukawas can change the quark flavor of the operators. Quark lines are depicted in light blue. }
\label{fig:diags-mass-smeft-bis}
\end{figure}


\noindent where $N_c=3$ denotes the number of colors, the summation over repeated quark-flavor indices is omitted, and we remind that the upper index $[S]$ denotes symmetric lepton-flavor indices, cf.~Eq.~\eqref{eq:sym-antisym-flavor}. The ellipses represent the other contributions that are not relevant for our study, including, e.g., the diagonal terms of the anomalous-dimension matrices. The terms on the right-hand side are illustrated in Fig.~\ref{fig:diags-mass-smeft}, and they comprise the direct running of $\mathcal{C}_{\substack{\bar{d}l q l H1}}$ into the Weinberg operator --  provided that flavor combinations do not vanish -- as well as the effects from $\mathcal{C}_{\substack{\bar{q}ullH}}$ defined as follows, 
\begin{align}
\mathcal{O}_{\substack{\bar{q}ullH\\ pr \alpha \beta}} = \epsilon^{ab}\big{(}\bar{q}_{p} u_{r}\big{)}\big{(}l_\alpha^T C l_\beta^a\big{)}H^b\,,
\end{align}

\noindent which can enter as intermediate contribution to neutrino masses in our scenario, via a two-step running, i.e.~$\mathcal{C}_{\substack{\bar{d}l q l H1}} \to \mathcal{C}_{\substack{\bar{q}ullH}} \to \mathcal{C}_{LH}^{(5,7)}$. Indeed, it is necessary to keep this intermediate operator, as the operators entering FCNC processes at tree level (i.e., with $p\neq r$) will not contribute to the right-hand side of Eq.~\eqref{eq:rg-smeft} in the down-aligned flavor basis. However, these operators can mix through the Yukawa interactions~\cite{Zhang:2023kvw}, as depicted in Fig.~\ref{fig:diags-mass-smeft-bis},
\begin{align}
\label{eq:rg-smeft-bis}
\dot{\mathcal{C}}_{\substack{\bar{d}l q l H1\\ p\alpha r\beta}} &= \dfrac{1}{2}\big{(}y_u y_u^\dagger\big{)}_{tr}\,{\mathcal{C}}_{\substack{\bar{d}l q l H1\\ p\alpha t\beta}}+\dots\,,\\[0.45em]
\dot{\mathcal{C}}_{\substack{\bar{q}ullH\\ pr \alpha \beta}} &= \big{(}y_d\big{)}_{ps}\big{(}y_u\big{)}_{tr}\,{\mathcal{C}}_{\substack{\bar{d}l ql H1\\ s\alpha t\beta}}+\dots\,, \nonumber
\end{align}

\noindent where we have explicitly written on the right-hand side only the RG contributions that can change flavors. Therefore, the $s\to d\nu\nu$ and $b\to s\nu\nu$ transitions are connected through RG evolution to the Weinberg operator at the ``one-loop squared'' order. Indeed, we remind that $y_u = V^\dagger \,\mathrm{diag}\big{(}y_u,y_c,y_t\big{)}$ carries the CKM matrix in our flavor basis, cf.~Eq.~\eqref{eq:yukawa-convention}.

\subsubsection*{Leading-logarithm solution}

We can now solve Eqs.~\eqref{eq:rg-smeft} and \eqref{eq:rg-smeft-bis} to determine the RG-contribution to the neutrino masses $m_\nu\equiv m_\nu^{(0)}+\delta m_\nu$, where $m_{\nu}^{(0)}$ denotes the neutrino mass computed at tree-level in Eq.~\eqref{eq:nu-masses} and $\delta m_\nu$ describes the RG-induced contributions,
\begin{align}
\label{eq:mass-smeft-ij}
({\delta m_\nu})_{\alpha\beta} =\dfrac{N_c}{(16\pi^2)^2} \dfrac{v^4}{\Lambda^3}\log^2 \Big{(}\dfrac{\mu_{\mathrm{ew}}}{\Lambda}\Big{)}\,\Big{[}(y_u y_u^\dagger)^2 \,y_d -\dfrac{1}{4} y_u y_u^\dagger y_d y_d^\dagger y_d\Big{]}_{ji}\,\mathcal{C}_{\substack{\bar{d}l q l H1\\
i\alpha j\beta}}^{[S]}(\Lambda)+\dots \,,
\end{align}
\noindent 
where we have neglected the running of SM couplings to first approximation~\cite{Buras:2018gto}. In the above equation, $(y_u y_u^\dagger)^2 \,y_d = V^\dagger \hat{y}_u^4 V \hat{y}_d$ and $y_u y_u^\dagger y_d y_d^\dagger y_d = V^\dagger \hat{y}_u^2 V \hat{y}_d^3$ are non-diagonal, where $\hat{y}_q \equiv \mathrm{diag}(y_{q_1},\,y_{q_2},\,y_{q_3})$  denotes the Yukawa matrices (with $q=u,d$) after diagonalization. We stress that only contributions with $i\neq j$ are kept on the right-hand side of Eq.~\eqref{eq:mass-smeft-ij}, as these are the operators relevant for FCNC transitions. Therefore, in these scenarios, the leading contributions to $m_\nu$ appear only at order $\log^2(\mu_{\mathrm{ew}}/\Lambda)$, i.e.~in the second leading-logarithm. 

From Eq.~\eqref{eq:mass-smeft-ij}, we notice that the leading contributions are proportional to five insertions of Yukawa couplings, $y_u^4\,y_d$ and $y_u^2\,y_d^3$. This is because the linear terms on the quark Yukawas in Eq.~\eqref{eq:rg-smeft}, which would lead to cubic contributions, do not contribute to neutrino masses, as they cancel between dimension-five and seven operators after replacing $\mu_H^2=-\lambda v^2$. Finally, we also identify that the first term in Eq.~\eqref{eq:mass-smeft-ij} comes from the two-step mixing with an intermediate operator $\mathcal{C}_{\substack{\bar{q}ullH}}$, which gives the dominant contribution after replacing the bottom and top-quark Yukawas in Eq.~\eqref{eq:mass-smeft-ij}. These effects will be proportional to $y_b\,y_t^4\,V_{ti} V_{tj}^\ast$, thus with a GIM-like suppression factor, which makes the contributions larger for the operators entering the $b\to s$ transition than for those relevant for the $s\to d$ transition.~\footnote{Notice, however, that the contributions proportional to the charm-quark mass are not negligible for the $s\to d$ transition and will be considered in our numerical analysis.}

\subsubsection{Numerical significance} 

Numerically, we find that the operators entering the $b\to s\nu\nu$ transition amount to the following contribution to the neutrino-mass matrix,
 \begin{align}
    (\delta m_\nu)_{\alpha\beta} \simeq\mathcal{C}_{\substack{\bar{d}lqlH1\\3\alpha 2\beta}}^{[S]}(\Lambda)\times \cfrac{2.4\,\mathrm{keV}}{(\Lambda/1\,\text{TeV})^3}\,+\,\mathcal{C}_{\substack{\bar{d}lqlH1\\2\alpha 3\beta}}^{[S]}(\Lambda)\times \cfrac{54\,\mathrm{eV}}{(\Lambda/1\,\text{TeV})^3}\,+\,\dots\,,
\end{align}
where we have replaced $\Lambda=1~\mathrm{TeV}$ in the argument of logarithm in~Eq.~\eqref{eq:mass-smeft-ij} and factored out $\mathcal{C}^{(7)}/\Lambda^3=\mathcal{O}(\mathrm{TeV}^{-3})$, as currently probed by Belle-II results, cf.~Sec.~\ref{sec:left}. The above results imply that a strong fine-tuning in $(\delta m_\nu)_{\alpha\beta}$ is needed
to cancel the above contributions to neutrino masses, which exceed by orders of magnitude the cosmological bounds on the sum of neutrino masses, namely $m_\nu \lesssim 0.1~\mathrm{eV}$~\cite{Planck:2018vyg}. Such a cancellation could arise, e.g., from direct contributions to $\mathcal{C}_{LH}^{(5)}(\Lambda)$ and  $\mathcal{C}_{LH}^{(7)}(\Lambda)$. Alternatively, the RG-induced contributions to $\delta m_\nu$ could be suppressed by considering operators with fully anti-symmetric lepton flavors. However, both constructions are highly model-dependent and would require substantial fine-tuning of the UV interactions to accommodate viable active-neutrino masses. 

In the kaon sector, the analogous expression reads
\begin{align}\label{eq:deltamnuRG}
    (\delta m_\nu)_{\alpha\beta} \simeq\mathcal{C}_{\substack{\bar{d}lqlH1\\2\alpha 1\beta}}^{[S]}(\Lambda)\times \cfrac{0.46\,\mathrm{eV}}{(\Lambda/1\,\text{TeV})^3}\,+\,\mathcal{C}_{\substack{\bar{d}lqlH1\\1\alpha 2\beta}}^{[S]}(\Lambda)\times \cfrac{0.02\,\mathrm{eV}}{(\Lambda/1\,\text{TeV})^3}\,+\,\dots\,,
\end{align}
for which there is no fine-tuning issue, given the stronger CKM suppression of the RG-induced contributions and the smaller values of $\mathcal{C}^{(7)}/\Lambda^3$ that are probed experimentally. Yet, in this case, the question of whether it is possible to write a UV completion of this EFT scenario that does not generate dimension-six operators for the $s\to d$ transition remains, as will be explored in Sec.~\ref{ssec:smeft-uv-completions}.

Before closing this section, we notice that the interplay between FCNC processes and neutrino masses has been recently discussed in Ref.~\cite{Endo:2026qof}, in which a fixed-order two-loop calculation of neutrino masses has been performed. Their contributions to neutrino masses turn out to be comparable in size, but smaller than the RG effects computed above, suggesting that the logarithmic contributions are dominant.

\subsection{$0\nu\beta\beta$ decays}
\label{ssec:smeft-0nubb}

The EFT operators involving electron neutrinos are potentially subject to additional bounds from experimental limits on the $0\nu\beta\beta$ decay half-life~\cite{KamLAND-Zen:2024eml}. These processes do not receive tree-level contributions from operators with different quark flavors, such as those entering, e.g., the $b\to s \nu_e\nu_e$ transition.~\footnote{Notice that $\mathcal{O}_{\bar{d}lqlH1}$ with quark flavor indices $d_1$ and $q_2$ contributes to $K\to \pi \nu_e\nu_e$ and it generates tree-level contributions to $0\nu\beta\beta$ via the CKM misalignment in left-handed quark doublets, as will be discussed in Sec.~\ref{ssec:smeft-numerical}.} However, they can receive non-negligible contributions from radiative corrections~\cite{deVries:2026ujz}. In particular, the Yukawa-induced contributions to the RG equations can mix operators with different quark flavors through the misalignment of down- and up-type quark Yukawas, similarly to the discussion in Sec.~\ref{ssec:smeft-nu-masses}.

\subsubsection{RG evolution}

In the following, we consider the $\mathcal{O}_{\substack{\bar{d}l q l H1}}$ operator, with the flavor indices relevant for the $s\to d\nu\nu$ and $b\to s\nu\nu$ transitions, and estimate the corresponding RG-induced constraints from $0\nu\beta\beta$ decays.~\footnote{See Ref.~\cite{Endo:2026qof} for a similar discussion, in which a fixed-order one-loop calculation is performed.} By using the anomalous-dimension matrix computed in Ref.~\cite{Zhang:2023kvw}, we find that the first leading-logarithm contribution from these operators to those entering $0\nu\beta\beta$ decays at tree level is suppressed. Indeed, the leading effects to $0\nu\beta\beta$ decays arise from the two-step running,
$\mathcal{C}_{\substack{\bar{d}l q l H1}} \to \mathcal{C}_{\substack{\bar{d}l ql H2}} \to \mathcal{C}_{l HW}$, where
\begin{align}
    \mathcal{O}_{\substack{\bar{d} l q l H 2\\ i\alpha j\beta }} &=\epsilon^{a d} \epsilon^{b e}\left(\overline{d_{i}} l_{\alpha }^a\right)\left(q_{j }^b C l_{\beta}^d\right) H^e\,,\\[0.4em]
    \mathcal{O}_{\substack{lHW\\ \alpha \beta}} &= \epsilon^{ab}(\epsilon \tau^I)^{de}\big{(}l_\alpha^ aC \sigma^{\mu\nu} l_\beta^d\big{)}H^b H^e W^{I}_{\mu\nu}\,, \nonumber
\end{align} 
\noindent which can be matched to dimension-six and dimension-nine LEFT operators that contribute to $0\nu\beta\beta$ decays for lepton flavor indices $\alpha=\beta=1$~\cite{Cirigliano:2017djv}. 
The relevant operator mixing contributions read~\cite{Zhang:2023kvw},
\begin{align}
\dot{\mathcal{C}}_{\substack{\bar{d}l ql H2\\ i\alpha k\beta}} &= \big{(}y_u y_u^\dagger\big{)}_{jk}\,\mathcal{C}_{\substack{\bar{d} l ql H1\\ i\alpha j\beta}}+\dots\,,\\[0.35em]
\dot{\mathcal{C}}_{\substack{l HW\\ \alpha \beta}} &= -\frac{N_c}{2}g_2 \,\big{(}y_d\big{)}_{ki} \,\mathcal{C}_{\substack{\bar{d} l q l H2 \\ i \alpha k \beta}}^{[S]}  +\dots\,, \nonumber
\end{align}

\noindent where we remind that $y_uy_u^\dagger = V^\dagger \hat{y}_u^2 V$, which allows for non-diagonal flavor contributions through the CKM matrix, and the ellipses represent additional terms that are not relevant to our analysis. By performing a logarithmic expansion, the relevant coefficient at the electroweak scale reads,
\begin{align}
    \label{eq:clHW}
    \mathcal{C}_{\substack{l HW\\ 11}} (\mu_{\mathrm{ew}}) = -\dfrac{N_c\,g_2}{4(16\pi^2)^2}\log^2 \Big{(}\dfrac{\mu_{\mathrm{ew}}}{\Lambda}\Big{)} \,\big{(}y_u y_u^\dagger y_d\big{)}_{ji}\,\mathcal{C}_{\substack{\bar{d} l ql H1\\ i1 j1}}(\Lambda) + \dots\,,
\end{align}

\noindent where we have set $\alpha=\beta=1$, and the ellipses comprise sub-leading contributions for the operators that we consider. Such operators contribute to $0\nu\beta\beta$ decays via contact interactions, after integrating out the $W$-boson propagators~\cite{Cirigliano:2018yza}.

\subsubsection{Numerical significance}

In our numerical analysis, we consider the experimental limit on the $0\nu\beta\beta$ half-life obtained by KamLAND-Zen for $^{136}\mathrm{Xe}$, namely $\smash{T_{1/2}^{0\nu}>3.8
\times 10^{26}~\mathrm{yr}}$ ($90\%$~CL)~\cite{KamLAND-Zen:2024eml}. We match the SMEFT operators to the LEFT, which can then be matched to the Chiral Lagrangian~\cite{Cirigliano:2017djv,Cirigliano:2017tvr}, using the low-energy  constants (LECs) determined in Ref.~\cite{Bhattacharya:2016zcn,ParticleDataGroup:2024cfk}. By combining these Wilson coefficients with the nuclear matrix elements (NMEs) from Ref.~\cite{Menendez:2017fdf} (see also Ref.~\cite{Hyvarinen:2015bda}), and using the phase-space factors reported in Ref.~\cite{Horoi:2017gmj}, it is possible to compute the half-life for this process through the master formula described in Ref.~\cite{Cirigliano:2018yza}. 

By assuming that the contribution from the operator $\mathcal{O}_{\bar{d}lqlH1}$ dominates over those arising from Majorana neutrino masses, and setting the unknown LECs to the order of magnitude of their absolute values derived from naive dimensional analysis~\cite{Cirigliano:2018yza}, we obtain the following limits for the operators relevant for $B\to K^{(\ast)} \nu\nu$ decays, namely those with quark-flavor indices $(i,j)=(3,2)$ and $(2,3)$,
\begin{align}
    &\dfrac{1}{\Lambda^3}\,\mathcal{C}_{\substack{\bar{d} l ql H1\\ 3121}}(\Lambda)  < (222\,{\rm GeV})^{-3}\,, & \dfrac{1}{\Lambda^3}\,{\mathcal{C}_{\substack{\bar{d} l ql H1\\ 2131}}(\Lambda)}< (63\,{\rm GeV})^{-3} \,,
\end{align}

\noindent where we have set the renormalization scale to $\Lambda=1~\mathrm{TeV}$ in the logarithm. It is worth noting that the hierarchy between these two constraints is given by $m_b/m_s$, due to the different down-quark flavors entering Eq.~\eqref{eq:clHW}. These constraints turn out to be rather weak due to the loop and CKM suppression, and thus are not competitive with those derived from the fine-tuning considerations for neutrino masses.

A similar analysis can be performed for the operators entering $K\to \pi \nu\nu$ decays, namely those with quark-flavor indices $(i,j)=(2,1)$ or $(1,2)$, where $i$ and $j$ denote, as before, the flavor indices of the right-handed down-quark and left-handed quark doublet, respectively. In particular, for the first operator, tree-level contributions are possible in our basis, since $q_2= (V^\dagger u_L)_2 = V_{us}^\ast\,u_L + V_{cs}^\ast \, c_L + V_{ts}^\ast\,t_L$ contains the up-quark with a (mild) $V_{us}$ suppression. For this reason, $0\nu\beta\beta$ decays will lead to very stringent bounds for this specific operator,
\begin{align}
    &\dfrac{1}{\Lambda^3}\,{\mathcal{C}_{\substack{\bar{d} l ql H1\\ 1121}}(\Lambda)}< (187 \,{\rm TeV})^{-3}\,,
\end{align}

\noindent which is competitive to the sensitivity from $K\to \pi + \mathrm{inv}$ searches, cf.~Sec.~\ref{sec:left}. For the second operator, tree-level contributions are absent. Therefore, the leading contributions would come from RG effects in this case. However, the CKM and quark-mass suppression in Eq.~\eqref{eq:clHW} would be even stronger than for the operators entering $B$-meson decays, thus not leading to useful constraints.

\begin{figure}[!t]
\centering
    \centering
    \includegraphics[width=0.62\textwidth]{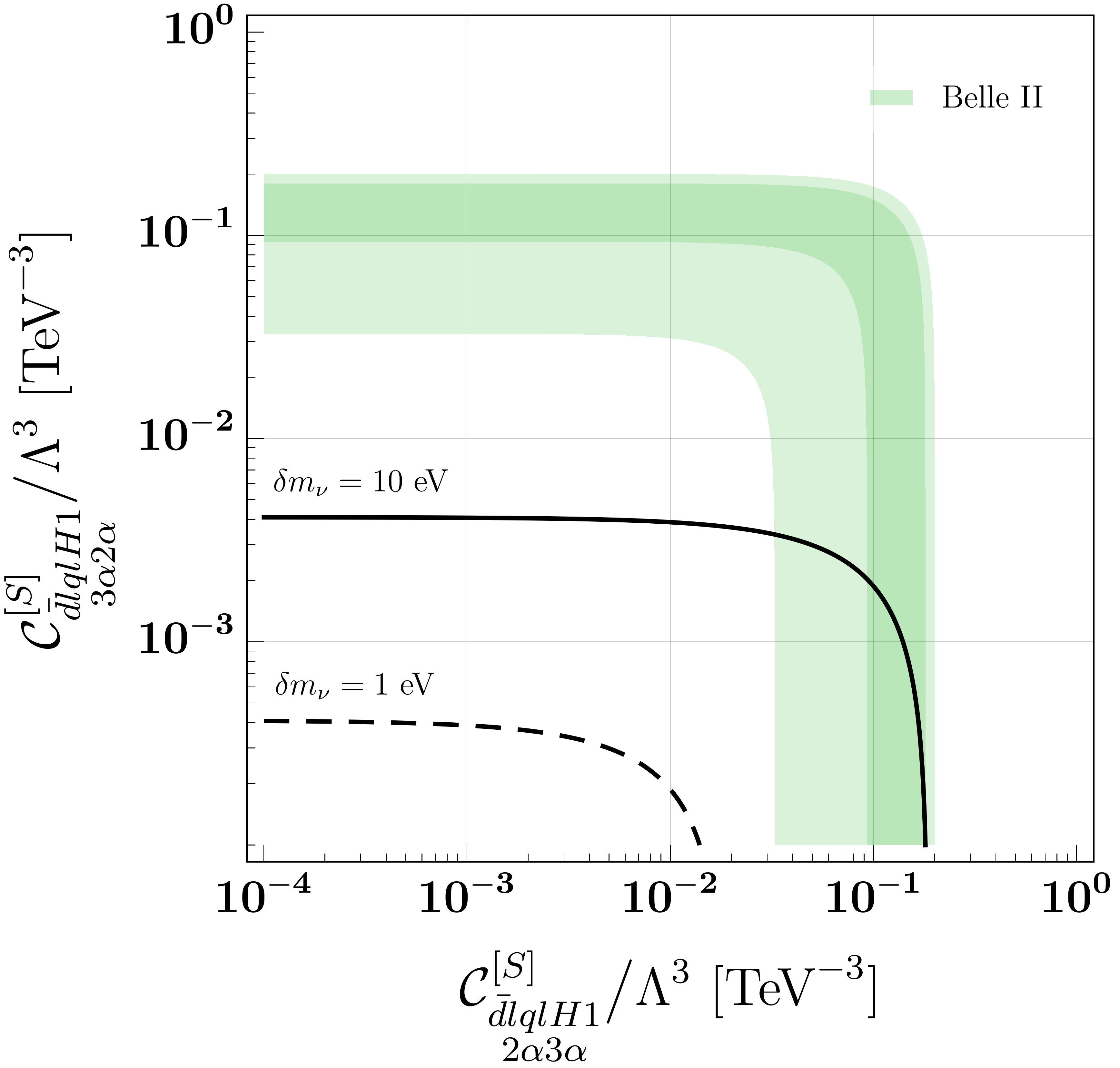}
\caption{\small \sl   
Constraints from $B\to K\nu\nu$ decays~\cite{Belle-II:2023esi} on the $\mathcal{C}_{\bar{d}l q l H1}/\Lambda^{3}$ effective coefficient, evaluated at the scale $\Lambda \simeq 1~\mathrm{TeV}$, with flavor indices $2\alpha 3\alpha$ and $3\alpha 2\alpha$, are depicted by the light (dark) green regions to $1\sigma$ ($2\sigma$) accuracy. These constraints are confronted with 
the requirement that the RG contributions to neutrino masses $\delta m_\nu$ are smaller than $10~\mathrm{eV}$ (black solid line) and $1~\mathrm{eV}$ (black dashed line). 
}
\label{fig:num-smeft-BK}
\end{figure}

\subsection{Phenomenological analysis}
\label{ssec:smeft-numerical}

In the following, we confront the $B\to K\nu\nu$ and $K\to \pi\nu\nu$ decays discussed in Sec.~\ref{sec:left} with the indirect constraints from the smallness of neutrino masses and the non-observation of $0\nu\beta\beta$ decays, as discussed above. We will discuss the implications for $B$-meson and kaon observables separately.

\subsubsection*{\underline{$B\to K^{(\ast)}\nu\nu$}\,: }

In Fig.~\ref{fig:num-smeft-BK}, we depict the allowed regions from $B^+\to K^+ \nu\nu$~\cite{Belle-II:2023esi} and $B^0\to K^{\ast 0} \nu\nu$~\cite{Belle:2017oht} decays in the two-dimensional plane of $\mathcal{C}_{\bar{d}l q l H1}/\Lambda^{3}$ evaluated at the scale $\Lambda \simeq 1~\mathrm{TeV}$, with quark flavor indices $(i,j)=(2,3)$ and $(3,2)$, i.e.,~the Wilson coefficients that contribute at tree-level to these processes. The requirement that RG-induced contributions to neutrino masses $\delta m_\nu$ are smaller than $1~\mathrm{eV}$ and $10~\mathrm{eV}$ is depicted by the black lines on the same plot, cf.~Sec.~\ref{ssec:smeft-nu-masses}. As already anticipated in the above discussion, the sensitivity of Belle-II is still not sufficient to probe EFT scenarios, with dimension-seven operators, that would not suffer from fine-tuning issues in neutrino masses.~\footnote{The discussion of whether dimension-seven operators can induce the dominant effects in these decays is postponed to Sec.~\ref{ssec:smeft-uv-completions}.}   Notice, also, that constraints from $0\nu\beta\beta$ decays are not displayed, as they are not competitive with meson decays for this transition, cf.~Sec.~\ref{ssec:smeft-0nubb}.

\begin{figure}[!t]
\centering
    \centering
    \includegraphics[width=0.62\textwidth]{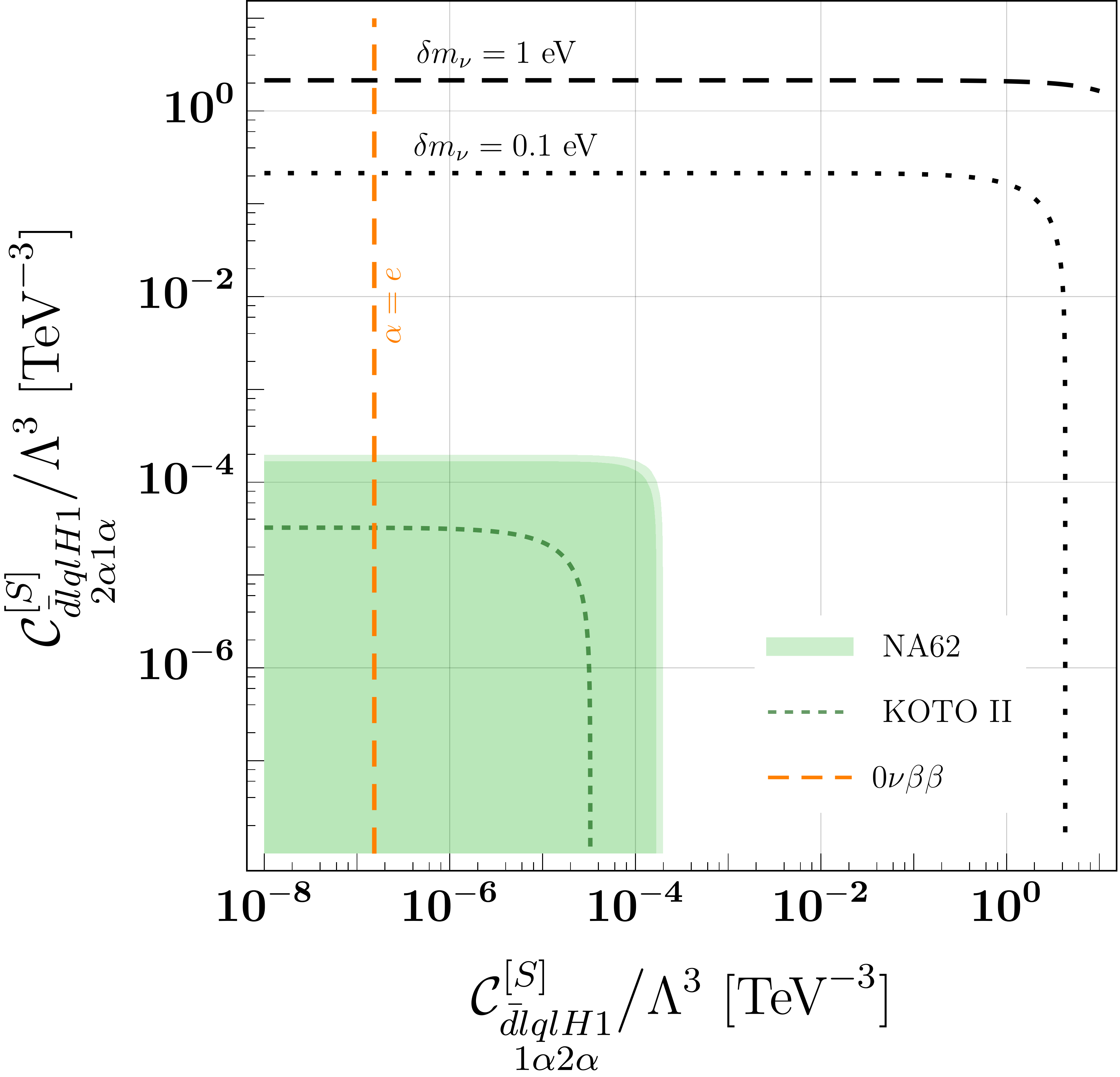}
\caption{\small \sl  Constraints from $K^+\to \pi^+\nu\nu$ decays~\cite{Chang:2026vvx} on the   $\mathcal{C}_{\bar{d}l q l H1}/\Lambda^{3}$ effective coefficient, evaluated at the scale $\Lambda \simeq 1~\mathrm{TeV}$,  
with flavor indices $1\alpha 2\alpha$ and $2\alpha 1\alpha$, are depicted by the light-green region. Future prospects for the KOTO-II experiment~\cite{KOTO:2025gvq}, assuming SM-like central values, are depicted by the green line. These constraints are confronted to
the requirement that the RG contributions to neutrino masses $\delta m_\nu$ are smaller than $1~\mathrm{eV}$ (black dashed line) and $0.1~\mathrm{eV}$ (black dotted line). Furthermore, constraints from $0\nu\beta\beta$ decays are represented by the orange line.
}
\label{fig:num-smeft-K}
\end{figure}

In conclusion, the EFT scenarios that generate measurable lepton-number-violating contributions to the $b\to s\nu\nu$ transition in the SMEFT are associated with a sizable fine-tuning in neutrino masses. While this is not a strict exclusion of these scenarios, it suggests that possible UV completions of these operators would generate too large masses for left-handed neutrinos, unless a specific cancellation is engineered at the UV scale.

\subsubsection*{\underline{$K\to \pi\nu\nu$}\,: }

In Fig.~\ref{fig:num-smeft-K}, we perform a similar analysis for the dimension-seven effective coefficients that enter the $K^+\to \pi^+\nu\nu$~\cite{Chang:2026vvx} decay, namely, the effective coefficient $\mathcal{C}_{\bar{d}l q l H1}/\Lambda^{3}$, with quark flavor indices $(i,j)=(1,2)$ and $(2,1)$, which is introduced at the scale $\Lambda \simeq 1~\mathrm{TeV}$. Following the discussion in Sec.~\ref{ssec:smeft-nu-masses}, we find that the indirect constraints from neutrino masses are not competitive with rare kaon decays.  The two main reasons for this difference with respect to $B$-meson decays are: (i) the two stronger CKM-suppression of the RG-contributions to neutrino masses encoded in $\delta m_\nu$, see Eq.~\eqref{eq:deltamnuRG}; and (ii) the smaller values of $\mathcal{C}^{(7)}/\Lambda^3$ probed by $K^+\to\pi^+\nu\nu$.

In agreement with Sec.~\ref{ssec:smeft-0nubb}, we find that $0\nu\beta\beta$ decays provide meaningful bounds for the operator with a second-generation quark doublet ($q_2$), since it contains an upper component $(V^\dagger u_L)_2 = V_{us}^\ast\,u_L + V_{cs}^\ast \, c_L + V_{ts}^\ast\,t_L$ via the CKM matrix. Therefore, this operator contributes to $0\nu\beta\beta$ decays at tree-level with a mild $V_{us}$ suppression. These constraints are weaker for the operator made with $q_3$, as it does not induce tree-level contributions to the transitions with valence quarks. Notice, also, that the current constraints on $K_L\to \pi^0 \nu\nu$ from KOTO~\cite{KOTO:2024zbl} are not displayed in Fig.~\ref{fig:num-smeft-K}, as they are not yet competitive with the charged-kaon mode. However, in the future, the proposed KOTO-II experiment~\cite{KOTO:2025gvq} will have enough sensitivity to be competitive with the $K^+ \to \pi^+ \nu\nu$ constraints, as depicted by the dark-green dashed lines in Fig.~\ref{fig:num-smeft-K}.



\subsection{Which UV completions?}
\label{ssec:smeft-uv-completions}

We have shown that the current sensitivity on $\mathcal{B}(B\to K^{(\ast)}+\mathrm{inv})$ at Belle-II is not sufficient to probe dimension-seven coefficients that would not be associated with large contributions to neutrino masses, as shown in Fig.~\ref{fig:num-smeft-BK}. However, in the kaon sector, the experimental precision of NA62 on $\mathcal{B}(K^+\to \pi^+ + \mathrm{inv})$ can probe similar EFT contributions, without facing fine-tuning issues for neutrino masses, thanks to the larger CKM suppression of the corresponding RG contributions, as depicted in Fig.~\ref{fig:num-smeft-K}. The remaining question is whether the working assumption of this section, 
namely that the leading contribution to these processes arises \emph{only} at \emph{dimension seven}, can indeed be realized in minimalistic concrete models of New Physics. We stress that our goal is not to construct full-fledged UV completions, but rather to investigate whether such a hierarchy can be realized in minimal concrete models with new heavy bosons and/or fermions above the TeV scale, which would be embedded in a more complete theory.

To address the above question, we consider the classification of possible weakly-coupled and renormalizable UV completions of dimension-seven SMEFT operators that was made in Ref.~\cite{Li:2023cwy}. We are interested in scenarios that can generate the $\mathcal{O}_{\bar{d}lqlH1}$ operator, with flavor indices that contribute to the $s\to d \nu\nu$ transition, but without generating contributions to this transition already at dimension six, i.e.,~to the operators $\mathcal{O}_{lq}^{(1-3)}$ and $\mathcal{O}_{ld}$, which would be dominant if proportional to the same couplings. Another relevant constraint is that the scenarios should not generate the Weinberg operator at tree-level, otherwise, they would run into a similar problem with neutrino masses as the scenarios explored above.

There are only two possible topologies that can generate the $\mathcal{O}_{\bar{d}lqlH1}$ operator at tree level in weakly-coupled renormalizable scenarios, as depicted in Fig.~\ref{fig:dim7-topo}.~\footnote{Notice that two mediators are needed to generate contributions that are not suppressed by the SM Yukawas and that violate quark flavor at tree level.} For the topology (a), two scalars $\Delta_1$ and $\Delta_2$ must be introduced above the electroweak scale. SM gauge invariance constrains 
the possible mediators to be: (i) a scalar triplet, $\Xi \sim ({\bf 1}, {\bf 3}, 1)$, and a doublet, $\varphi \sim ({\bf 1}, {\bf 2}, 1/2)$; or (ii) a pair of scalar leptoquarks, namely $\widetilde{R}_2 \sim ({\bf 3}, {\bf 2}, 1/6)$, combined with either $S_1\sim ({\bf \bar{3}}, {\bf 1}, 1/3)$ or $S_3\sim ({\bf \bar{3}}, {\bf 3}, 1/3)$:

\begin{figure}
    \centering
    \includegraphics[width=0.75\linewidth]{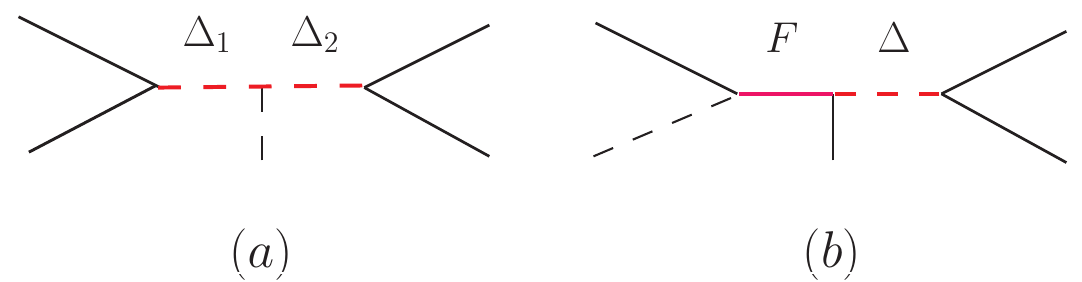}
    \caption{\label{fig:dim7-topo} \small \sl Possible topologies of tree-level UV completions of the dimension-seven operator $\mathcal{O}_{\bar{d}lqlH1}$. The red lines denote a new heavy scalar ($\Delta$) or fermion ($F$) field.}
\end{figure}

\begin{itemize}
    \item[$\bullet$] In the first case, the dimension-seven coefficients will depend on the $(l^C\,l)\,\Xi$ and $\Xi^\dagger\, H \varphi$ couplings, where flavor and $SU(2)_L$ contractions are omitted, for simplicity. However, it is impossible to forbid the couplings $\Xi^\dagger \, H^2$ and $|\Xi|^2\, |H|^2$ with a symmetry principle, while still keeping the contributions to the dimension-seven coefficient that we are interested in. Therefore, this scenario would unavoidably generate tree-level contributions to the Weinberg operator, as in the type-II seesaw mechanism~\cite{Magg:1980ut}.
    \item[$\bullet$] In the scenarios with pairs of scalar leptoquarks, contributions to neutrino masses only arise
    at loop level~\cite{Fajfer:2024uut}. However, each of the scalar leptoquarks that contribute to the dimension-seven operator $\mathcal{C}_{\bar{d}lqlH1}$ would also induce tree-level contributions to the dimension-six coefficients entering the $d_i\to d_j\nu\nu$ transitions, namely $\smash{\mathcal{C}_{lq}^{(1-3)}}$ and $\mathcal{C}_{ld}$. For instance, in the scenario with $\widetilde{R}_2$ and $S_1$ leptoquarks~\cite{Dorsner:2022twk},
\begin{align}
\mathcal{L} \supset\;& \big{(}D_{\mu}S_1\big{)}^\dagger \, \big{(}D^{\mu}S_1\big{)} + \big{(}D_{\mu} \widetilde{R}_2\big{)}^\dagger \, \big{(}D^{\mu}\widetilde{R}_2\big{)} - m_{S_1}^2 \, S_1^\dagger S_1 
 - m_{\tilde{R}_2}^2 \widetilde{R}_2^\dagger \widetilde{R}_2  \\[0.35em]
&+ \Big( y_{1L}^{i\alpha}\,\overline{q^C_i}\,\epsilon\,l_{\alpha}\,S_1 
+ y_{2L}^{i\alpha}\,\bar{d}_i\,l_{\alpha}\,\epsilon\,\widetilde{R}_2 
- \lambda_{RS}\,\widetilde{R}_2^\dagger\,H\,S_1^\ast + \mathrm{h.c.} \Big)\,, \nonumber
\end{align}

\noindent where $y_{1L}^{i\alpha}$ and $y_{2L}^{i\alpha}$ denotes the leptoquark Yukawa couplings, $m_{S_1}$ are  $m_{\tilde{R}_2}$ are their corresponding masses, where $\lambda_{RS}$ stands for the trilinear coupling that induces the breaking of $L$. The diagram in panel (a) of Fig.~\ref{fig:dim7-topo} leads to the following dimension-seven coefficient,
\begin{equation}
   \dfrac{1}{\Lambda^3}\, \mathcal{C}_{\substack{\bar{d}lqlH1\\ i\alpha j\beta}}=+\frac{\lambda_{RS}\,y_{1L}^{j\beta}\,y_{2L}^{i\alpha}}{m_{S_1}^2\,m_{\tilde{R_2}}^2}\,.
\end{equation}
The exchange of a single leptoquark gives rise to the dimension-six coefficients,
\begin{align}
  \label{eq:smeft-lq-d6}
  \dfrac{1}{\Lambda^2}\,\mathcal{C}_{\substack{lq\\ \alpha\beta ij}}^{(1)} &= -\dfrac{1}{\Lambda^2}\,\mathcal{C}_{\substack{lq\\ \alpha\beta ij}}^{(3)} = +\dfrac{y_{1L}^{j\beta} \,y_{1L}^{i\alpha\,\ast}}{4\,m_{S_1}^2}\,, &
  \dfrac{1}{\Lambda^2}\,\mathcal{C}_{\substack{ld\\ \alpha\beta ij}} &= -\dfrac{y_{2L}^{i\beta}\,y_{2L}^{j\alpha\,\ast}}{2\,m_{\widetilde{R}_2}^2}\,
\end{align}

\noindent which are correlated to the dimension-seven ones. Since $m_{S_1}\gg v$ and $m_{\widetilde{R}_2}\gg v$, given the limits from the direct searches at the LHC, it is clear that the dimension-six coefficients would lead to the dominant contribution to $d_i\to d_j\nu\nu$ observables. The only possibility to invert the hierarchy between the coefficients of dimension seven and six operators is to postulate a flavor structure that suppresses the latter. This could be achieved if each leptoquark couples predominantly to a specific quark generation -- for instance, a viable choice could be $y_{1L}^{s\alpha} \neq 0$ and $y_{2L}^{b\alpha} \neq 0$, with the other leptoquark couplings set to zero. In this case, the dimension-six coefficients would only be induced through CKM loops, with a GIM-like suppression, see~e.g.~Ref.~\cite{Fajfer:2018bfj}, thus potentially making them smaller than the dimension-seven ones. 
\end{itemize}
For topology (b), one fermion $F$ and a scalar $\Delta$ must be introduced above the electroweak scale. Depending on the quantum numbers of the scalar $\Delta$, two scenarios are possible:
\begin{itemize}
    \item The scalar $\Delta$ can have the same quantum numbers as the SM Higgs doublet, with the fermion field constrained to be either $N\sim ({\bf 1},{\bf 1},0)$ or $\Sigma\sim({\bf 1},{\bf 3},0)$. In these scenarios, the dimension-seven coefficients depend on the couplings $\bar{l}\widetilde{\Delta}N$ or $\bar{l}(\vec{\tau}\cdot\vec{\Sigma})\widetilde{\Delta}$, which unavoidably generate tree-level contributions to the Weinberg operator, as in the type-I and type-III seesaw mechanisms, respectively.

    \item The scalar $\Delta$ could be triplet $\Xi \sim ({\bf 1}, {\bf 3}, 1)$, or the scalar leptoquarks $S_1 \sim ({\bf 3}, {\bf 1}, {1/3})$, $\widetilde{R}_2 \sim ({\bf 3}, {\bf 2},{1/6})$ and $S_3 \sim ({\bf \bar{3}}, {\bf 3}, 1/3)$, which already appeared in the topology (a) in Fig.~\ref{fig:dim7-topo}.  As noted above, these scenarios would either generate contributions to the tree-level Weinberg operator or to the dimension-six coefficients entering the $d_i\to d_j\nu\nu$ transitions, unless a specific flavor ansatz is imposed.
\end{itemize}

\noindent In summary, possible UV completions of the dimension-seven operators that contribute to $d_i\to d_j\nu\nu$ transition either generate a large tree-level contribution to the Weinberg operator, or they simultaneously generate dimension-seven and dimension-six operators, with the latter being largely dominant. The only exceptions are the scenarios involving scalar leptoquarks, with a specific \emph{flavor ansatz} to suppress the combinations of Yukawa couplings that enter the dimension-six coefficients. Since the dominance of dimension-seven operators assumed in this section is thus not generic, we do not pursue a detailed phenomenological analysis of these leptoquark scenarios, the predictions of which would largely depend on the assumed flavor structure.

\section{$\nu$SMEFT phenomenology}
\label{sec:pheno-nusmeft}

In this Section, we explore an alternative to the previous scenario, which consists of introducing a light right-handed neutrino $N \sim ({\bf 1}, {\bf 1}, 0)$ to the SMEFT framework. We will demonstrate that such a scenario can generate measurable effects in $B\to K + \mathrm{inv}$ and $K\to \pi + \mathrm{inv}$ decays, potentially distorting their $q^2$-spectra, while avoiding a fine-tuning problem in neutrino masses. 

The remainder of this section is organized as follows. In Sec.~\ref{ssec:nu-smeft}, we introduce the EFT operators containing the right-handed neutrino $N$, which contribute to FCNC transitions. In Sec.~\ref{ssec:nusmeft-nu-masses}, we analyze the RG contributions to neutrino masses and mixing in this framework. In Sec.~\ref{ssec:nusmeft-0nubb}, we estimate the constraints from $0\nu\beta\beta$ decays to these operators. The implications of this scenario to neutrino and flavor searches are discussed in Sec.~\ref{ssec:nusmeft-numerical}.
Finally, we briefly explore the UV completions of the $\nu$SMEFT in Sec.~\ref{ssec:nusmeft-uv-completions}.

\subsection{Operator basis}
\label{ssec:nu-smeft}

The Lagrangian describing the SMEFT extended with a light right-handed neutrino $N\sim({\bf 1},\,{\bf 1},\,0)$ can be written as,
\begin{align}
    \label{eq:nu-smeft}
    \mathcal{L}_{\nu \mathrm{SMEFT}} & = \mathcal{L}_{\mathrm{SM}}+ \mathcal{L}_N+\sum_{d\geq 5}\dfrac{1}{\Lambda^{d-4}} \sum_I\,\bar{\mathcal{C}}^{(d)}_I\,\bar{\mathcal{O}}^{(d)}_I+\dots\,,
\end{align}

\noindent with 
\begin{align}
    \label{eq:nu-smeft-bis}
    \mathcal{L}_N = \bar{N} i \slashed{\partial} N - \Big{[}\frac{m_N}{2}\, \overline{N^C} N+\big{(}{y}_N\big{)}_{\alpha N}\,\bar{l}_\alpha  \,\widetilde{H} N+\mathrm{h.c.}\Big{]}\,,
\end{align}

\noindent where $m_N$ is a Majorana mass, and $y_{N}$ is the neutrino Yukawa matrix. We will consider a simplified scenario with a single sterile fermion field $N$, light enough to be produced on-shell in kaon and $B$-meson decays.~\footnote{Notice that at least two sterile neutrinos would be needed to reproduce realistic masses of active neutrinos. However, for the purposes of this work, it is sufficient to consider a single state, which captures the relevant phenomenology.} The main difference of this scenario from the one discussed above is that scalar and tensor operators in Eq.~\eqref{eq:nuleft} arise at dimension six~\cite{delAguila:2008ir},
\begin{align}
    \label{eq:nusmeft-d6}
    \bar{\mathcal{O}}_{\substack{N d\\NNij}}&= \left(\bar{N} \gamma^\mu N\right)\left(\bar{d}_i \gamma_\mu d_j\right)\,,&
  \bar{\mathcal{O}}_{\substack{N q\\NNij}} & =\left(\bar{N} \gamma^\mu N\right)\left(\bar{q}_i \gamma_\mu q_j\right)\,,\\[0.35em]
  \bar{\mathcal{O}}_{\substack{l N q d\\\alpha N i j}} & =(\bar{l}_\alpha N) \epsilon(\bar{q}_i d_j)\,, &
  \bar{\mathcal{O}}_{\substack{l d q N\\\alpha i j N}} & =(\bar{l}_\alpha d_i) \epsilon(\bar{q}_j N)\,.\nonumber
\end{align}

\noindent For convenience, we define the operators~\cite{Ardu:2024tzb}
\begin{align}
\label{eq:nu-smeft-redundant}
\bar{\mathcal{O}}_{\substack{lNqd\\ \alpha N ij}}^{(3)} &\equiv -4\,\bar{\mathcal{O}}_{\substack{lNqd\\ \alpha N ij}}-8 \, \bar{\mathcal{O}}_{\substack{ldqN\\ \alpha ji N}} \,, \\[0.35em]
\bar{\mathcal{O}}_{\substack{lNqd\\ \alpha N ij}}^{(1)} &\equiv \bar{\mathcal{O}}_{\substack{lNqd\\ \alpha N ij}}\,, \nonumber
\end{align}

\noindent which will be used in the following instead of $\mathcal{O}_{ldqN}$ and $\bar{\mathcal{O}}_{\substack{lNqd}}$ to make our phenomenological expressions more compact, as scalar and tensor contributions become then isolated. Furthermore, we introduce the Yukawa-like operator,
\begin{align}
\label{eq:nusmeft-yukawa-like}
\bar{\mathcal{O}}_{\substack{NH\\ \alpha N}}=\big{(}\bar{l}_\alpha\widetilde{H}N\big{)}\big{(}H^\dagger H\big{)}\,,
\end{align}

\noindent which will appear useful in the discussion of neutrino masses that follows.

\subsubsection*{Tree-level matching}

The matching of the above operators to Eq.~\eqref{eq:left-lnv} at the electroweak scale is given by
\begin{align}
\label{eq:nusmeft-matching}
\bar{C}_{\substack{V_{LR}\\ijNN}}(\mu_{\mathrm{ew}}) &=\frac{v^2}{\Lambda^2}\,\bar{\mathcal{C}}_{\substack{Nq\\NNij}}(\mu_{\mathrm{ew}})  \,,\qquad & \bar{C}_{\substack{V_{RR}\\ij NN}}(\mu_{\mathrm{ew}}) &=\frac{v^2}{\Lambda^2}\,\bar{\mathcal{C}}_{\substack{Nd\\NN ij}}(\mu_{\mathrm{ew}}) \,, \\[0.35em]
\bar{C}_{\substack{S_{LL}\\ijN\alpha}}(\mu_{\mathrm{ew}}) &=\frac{v^2}{\Lambda^2}\,\bar{\mathcal{C}}^{(1)\,\ast}_{\substack{\ell N qd\\\alpha N ji}}(\mu_{\mathrm{ew}})\,, 
& \bar{C}_{\substack{S_{RL}\\ij N \alpha}}(\mu_{\mathrm{ew}}) &=0\,,\nonumber\\[0.35em]
\bar{C}_{\substack{T\\ij N \alpha}}(\mu_{\mathrm{ew}}) &=\frac{v^2}{\Lambda^2}\,\bar{\mathcal{C}}^{(3)\,\ast}_{\substack{l Nqd\\\alpha N ji }}(\mu_{\mathrm{ew}})\,,\nonumber
\end{align}

\noindent where $\bar{C}_{S_{RL}}$ only appears at higher orders in the EFT expansion. Since operators with a different Lorentz structure appear already at dimension six in this scenario, it is more natural to expect them to receive measurable contributions compared to the previous case. However, the same issue of fine-tuning might arise in a different form, as the dimension-six operators in Eq.~\eqref{eq:nusmeft-d6} renormalize the neutrino Yukawa couplings, thus changing the neutrino spectrum and inducing a potentially large active-sterile neutrino mixing, depending on the values of $m_N$. 
In the following, we show that these constraints are significantly weaker than in the SMEFT case and therefore do not pose a problem.

\subsection{RG effects and neutrino Yukawa}
\label{ssec:nusmeft-nu-masses}

\begin{figure}[t]
\centering
\includegraphics[width=0.95\linewidth]{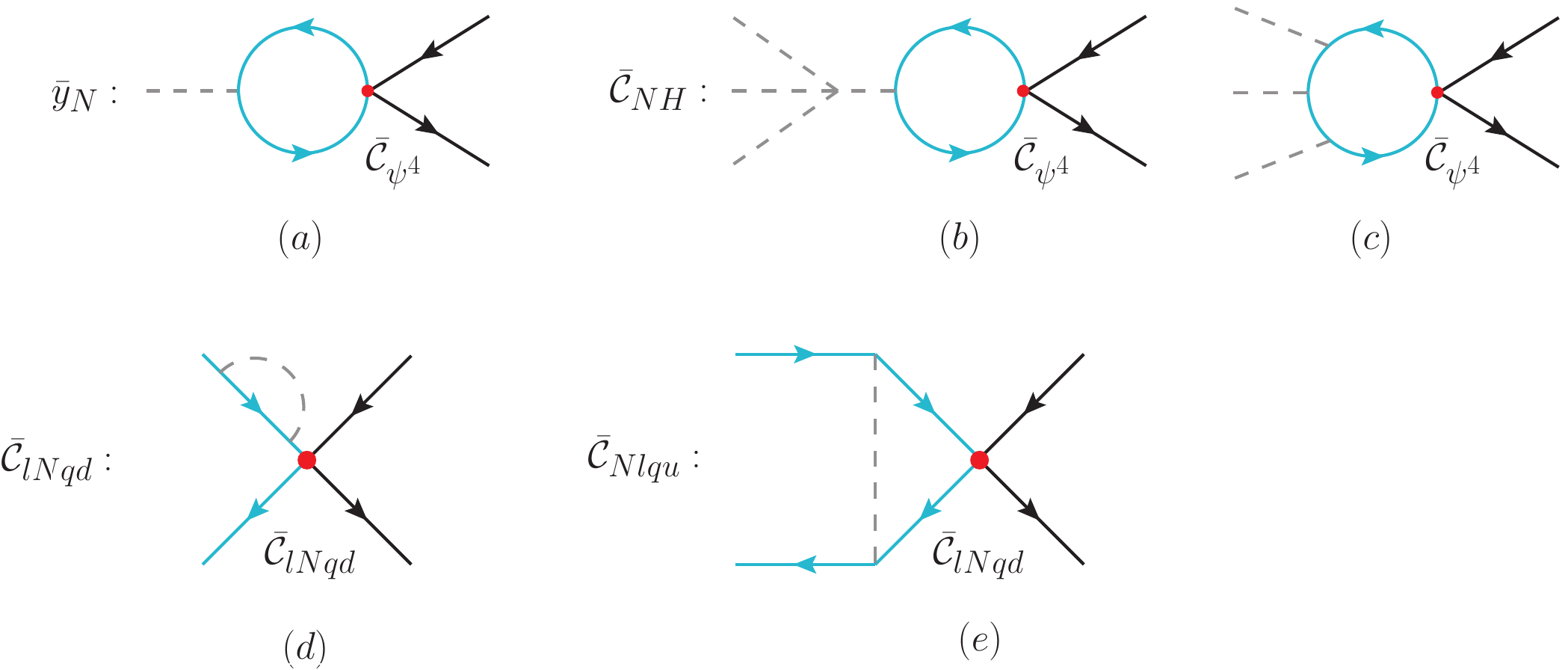}
\caption{\small \sl \label{fig:diags-mass-nusmeft} Relevant one-loop diagrams for operator mixing in the SMEFT extended with a right-handed neutrino $N\sim ({\bf 1}, {\bf 1}, 0)$. Quark lines are depicted in light blue.}
\end{figure}

We now discuss the impact of dimension-six operators introduced in Eq. \eqref{eq:nusmeft-d6} on active-neutrino masses through RG effects. In this scenario, the neutrino mass matrix ($M_\nu$) can be written up to dimension six as follows,
\begin{align}
\label{eq:nu-mass-matrix}
M_\nu =  \begin{pmatrix}
\;0\; & \;m_D\; \\[0.3em]
\;m_D^T \; & \;m_N\; 
\end{pmatrix}\,,
\end{align}

\noindent where the off-diagonal term $(m_D)$ is a lepton-number-conserving Dirac mass. The latter receives contributions after spontaneous electroweak symmetry breaking from the neutrino Yukawa coupling $y_{N}$ defined in Eq.~\eqref{eq:nu-smeft-bis}, as well as from the Yukawa-like operator defined in Eq.~\eqref{eq:nusmeft-yukawa-like},
\begin{align}
\label{eq:mD-tree}
    m_D=\frac{v}{\sqrt{2}}\left(y_N-\frac{v^2}{2\Lambda^2}\, \bar{\mathcal{C}}_{NH}\right)\,,
\end{align}

\noindent where flavor indices are omitted and the couplings are evaluated at the electroweak scale $\mu \simeq \mu_{\mathrm{ew}}$. We assume that the semileptonic operators introduced in Eq.~\eqref{eq:nusmeft-d6} are present at the UV scale $\Lambda$ and we estimate the RG contributions to the off-diagonal term $m_D$. From Fig.~\ref{fig:diags-mass-nusmeft}, we see that both the dimension-four and six operators receive one-loop contributions and, similarly to the discussion in Sec.~\ref{ssec:smeft-nu-masses}, they contribute to neutrino masses at the same order, in spite of the dimension-six ones being suppressed by $\mu_H^2/\Lambda^2$.

We consider the anomalous-dimension matrix reported in Ref.~\cite{Ardu:2024tzb} for the dimension-six operators in the SMEFT extended with a sterile neutrino, and we recast the results from Ref.~\cite{Jenkins:2013zja} for the running of the neutrino Yukawa coupling,
\begin{align}
\label{eq:nusmeft-rge}
     \dot{\bar{\mathcal{C}}}_{\substack{NH\\ \alpha N}} & =  4N_c \bigg{[}\lambda\,\big{(}y_d^\dagger\big{)}_{ij}-\big{(}y_d^\dagger y_d y_d^\dagger \big{)}_{ij}\bigg{]} \, \bar{\mathcal{C}}_{\substack{lNqd\\ \alpha N ji}}^{(1)} -4 N_c \bigg{[}\lambda\, \big{(}y_u\big{)}_{ij}-\big{(}y_u y_u^\dagger y_u \big{)}_{ij}\bigg{]}\,\bar{\mathcal{C}}_{\substack{Nlqu\\ N \alpha ij}}^{\ast}+ \dots \,, \\[0.4em]
     \dot{y}_{\substack{N\\ \alpha N}} & = -\dfrac{2\mu_H^2}{\Lambda^2}\,N_c \bigg{[}\big{(}y_d^\dagger\big{)}_{ij}\,\bar{\mathcal{C}}_{\substack{lNqd\\ \alpha N ji}}^{(1)} + \big{(}y_u\big{)}_{ij}\,\bar{\mathcal{C}}_{\substack{Nlqu\\ N \alpha ij}}^{\ast}\bigg{]} + \dots \,,\nonumber
\end{align}
where we remind that $\lambda$ is the quartic coupling in the Higgs potential, and the ellipses represent contributions that are not relevant to our study. Similarly to Sec.~\ref{ssec:smeft-nu-masses}, we also introduce the operator,
\begin{align}
    \bar{\mathcal{O}}_{\substack{Nlqu\\ N\alpha ij}} = \big{(}\bar{N}l_\alpha\big{)}\big{(}\bar{q}_i u_j\big{)} \,,
\end{align}

\noindent which contributes to the neutrino Yukawa evolution as an intermediate operator, which is essential since the first leading-logarithm contributions from the operators in Eq.~\eqref{eq:nusmeft-d6} vanish for the flavor transitions with $i\neq j$, as we consider. The relevant off-diagonal terms in the running of the $\psi^4$ operators read~\cite{Ardu:2024tzb}
\begin{align}
\label{eq:nusmeft-rge-bis}
\dot{\bar{\mathcal{C}}}_{\substack{lNqd\\ \alpha N ij }}^{(1)} &= \dfrac{1}{2} \big{(}y_u y_u^\dagger\big{)}_{is}\,\bar{\mathcal{C}}_{\substack{lNqd\\ \alpha N sj}}^{(1)}+\dots\,,
\\[0.35em]
\dot{\bar{\mathcal{C}}}_{\substack{Nlqu\\N\alpha ij}} &= -2 \big{(}y_d\big{)}_{is} \big{(}y_u\big{)}_{tj}\, \bar{\mathcal{C}}_{\substack{lNqd\\ \alpha Nts}}^{(1)\,\ast} +\dots\,, \nonumber
\end{align}

\noindent where we have kept, once again, only the terms that can change quark flavor. Notice, also, that the vector operators $\bar{\mathcal{O}}_{Nd}$ and $\bar{\mathcal{O}}_{Nq}$ do not appear in the above equations, as they are chirality conserving, thus requiring an insertion of the neutrino Yukawa $y_N$ to mix into chirality-breaking operators. Notice, also, that the same flavor quark Yukawa combinations as in Sec.~\ref{ssec:smeft-nu-masses} appear in the above equations.

\subsubsection*{Leading-logarithm solution}

We now compute the RG contribution from the operators with fixed flavor indices $i\neq j$ to the off-diagonal element of the Dirac neutrino mass $m_D$, defined in Eq.~\eqref{eq:mD-tree},
\begin{align}
    m_D\equiv m_D^{(\text{0})}+\delta m_D\,.
\end{align}

\noindent By combining Eqs.~\eqref{eq:nusmeft-rge} and \eqref{eq:nusmeft-rge-bis}, we find that 
\begin{align}
\label{eq:mass-nusmeft-ij}
({\delta m_D})_{\alpha N} = -\frac{N_c\sqrt{2}}{(16\pi^2)^2}&\frac{v^3}{\Lambda^2} \log^2\left(\frac{\mu_{\text{ew}}}{\Lambda}\right)\Big{[}y_d^\dagger (y_u y_u^\dagger)^2-\dfrac{1}{4}y_d^\dagger y_d y_d^\dagger y_u y_u^\dagger\Big{]}_{ji}\bar{\mathcal{C}}^{(1)}_{\substack{l N qd\\\alpha N ij}}+\dots\,,
\end{align}
\noindent where the Yukawa-dependent prefactor is identical to the one obtained in Eq.~\eqref{eq:mass-smeft-ij} for the dimension-seven SMEFT operators. Indeed, these terms come from diagrams with the same fermionic loops, thus sharing similarities with the previous scenario, cf.~Fig.~\ref{fig:diags-mass-smeft} and \ref{fig:diags-mass-nusmeft}. In particular, we find in this case that the RG contributions to the neutrino Yukawa are canceled by the terms proportional to the Higgs quartic coupling $\lambda$, which are depicted by diagrams (a) and (b) in Fig.~\ref{fig:diags-mass-nusmeft}. Furthermore, the dominant contribution comes from the first term in Eq.~\eqref{eq:mass-nusmeft-ij}, which is induced by the two-step mixing, $\bar{\mathcal{C}}_{lNqd}^{(1)} \to \bar{\mathcal{C}}_{Nlqu}\to \big{\lbrace}y_N,\, \bar{\mathcal{C}}_{NH}\big{\rbrace} $\,.

\subsubsection{Numerical significance}

We now quantify Eq.~\eqref{eq:mass-nusmeft-ij} by fixing the quark flavors $i\neq j$. For the effective coefficients contributing to the $b\to s\nu N$ transition at tree-level,
 \begin{align}
    \label{eq:deltamD-num-B}
    (\delta m_D)_{\alpha N} \simeq -\bar{\mathcal{C}}_{\substack{lNqd\\\alpha N 23}}^{(1)}(\Lambda)\times \cfrac{14\,\mathrm{keV}}{(\Lambda/1\,\text{TeV})^2}\,-\bar{\mathcal{C}}_{\substack{lNqd\\\alpha N 32}}^{(1)}(\Lambda)\times \cfrac{0.3\,\mathrm{keV}}{(\Lambda/1\,\text{TeV})^2}\,+\,\dots\,,
\end{align}

\noindent where, once again, we have set $\Lambda=1~\mathrm{TeV}$ in the logarithm. The $\nu$SMEFT coefficients $\bar{\mathcal{C}}/\Lambda^2$ are constrained by the low-energy analysis in Sec.~\ref{sec:left}, which thus amount to at most $\delta m_D = \mathcal{O}(\mathrm{keV})$ for the couplings fixed by the excess in Belle-II data, cf.~Eq.~\eqref{eq:lambda-nusmeft-B-scalar}. Similarly, for the operators appearing in the kaon sector, we find that
 \begin{align}
    \label{eq:deltamD-num-K}
    (\delta m_D)_{\alpha N} \simeq -\bar{\mathcal{C}}_{\substack{lNqd\\\alpha N 12}}^{(1)}(\Lambda)\times \cfrac{2.7\,\mathrm{eV}}{(\Lambda/1\,\text{TeV})^2}\,-\bar{\mathcal{C}}_{\substack{lNqd\\\alpha N 21}}^{(1)}(\Lambda)\times \cfrac{0.13\,\mathrm{eV}}{(\Lambda/1\,\text{TeV})^2}\,+\,\dots\,,
\end{align}

\noindent in which we see a larger suppression, due to the relevant CKM elements and quark masses in this case. By replacing the lower limit on the effective scale from Eq.~\eqref{eq:lambda-nusmeft-K-scalar}, we find that the contribution to the Dirac mass is at most $\delta m_D = \mathcal{O}(10^{-3}~\mathrm{eV})$, which is in turn completely negligible. 

While the contributions to $\delta m_D$ seem large in Eq.~\eqref{eq:deltamD-num-B}, there is an important difference with respect to the previous scenario, as $m_D$ represents only the off-diagonal element of the neutrino-mass matrix in Eq.~\eqref{eq:nu-mass-matrix}, which must be diagonalized in order to quantify the effect on active neutrino masses, as we proceed in the following.

\subsubsection{Diagonalizing neutrino masses}

To obtain the RG-induced contribution to the active neutrino masses ($\delta m_\nu$), we now diagonalize the neutrino mass matrix $M_\nu$, defined in Eq.~\eqref{eq:nu-mass-matrix}, via a rotation $\nu_L^\prime \equiv U \,\nu_L$, where primed fields are in the mass basis, and we treat $N$ as a fourth neutrino. The physical masses of active neutrinos ($m_{\nu_\alpha}$) then become
\begin{align}
\label{eq:nu-mass-seesaw}
m_{\nu_\alpha} &= \dfrac{m_N}{2} \Big{[}\sqrt{1+4 m_D^2/m_N^2}-1\Big{]} \simeq \dfrac{m_D^2}{m_N}\,,
\end{align}

\noindent whereas the physical heavy-neutrino ($\nu_4$) has a mass,
\begin{align}
m_{\nu_4} &= \dfrac{m_N}{2} \Big{[}\sqrt{1+4 m_D^2/m_N^2}+1\Big{]} \simeq m_N \,,
\end{align}

\noindent where we assume $ m_N\gg m_D $ in the expansion, obtaining a seesaw-type relation. For simplicity, we have omitted the flavor indices in $m_D$ in the above equations.~\footnote{The above expression assumes a single sterile neutrino and operators with a fixed flavor $\alpha \in \lbrace e,\mu,\tau \rbrace$, but its generalization is straightforward. Furthermore, notice that we have absorbed a minus sign in the active neutrino mass via a field redefinition.}  By expanding Eq.~\eqref{eq:nu-mass-seesaw} in $\delta m_D$, we find that the RG-induced contributions to the active neutrino masses read
\begin{align}
\label{eq:deltam-nusmeft}
    \delta m_\nu \simeq 2 \,\dfrac{m_{D}^{(0)}\,\delta m_{D}}{m_N}+ \dfrac{\big{(}\delta m_D\big{)}^2}{m_N}\,,
\end{align}

\noindent where flavor indices have been omitted. By combining these expressions with Eqs.~\eqref{eq:deltamD-num-B} and \eqref{eq:deltamD-num-K}, we see that $\delta m_\nu$ remains small, provided that $m_\nu \ll \delta m_D \ll m_N$, i.e.~that the heavy neutrino mass $m_N$ is sufficiently large. Finally, $U_{\alpha N} \equiv U_{\alpha 4}$ needed to diagonalize $M_\nu$ is given by
\begin{align}
\label{eq:mix-nusmeft}
|U_{\alpha N}| = \Bigg{[} 1 + \dfrac{4m_D^2/m_N^2}{\Big{(}1-\sqrt{1+4 m_D^2/m_N^2}\Big{)}^2} \Bigg{]}^{-1/2}\simeq \bigg{|}\dfrac{m_D}{m_N}\bigg{|}\,, 
\end{align}

\noindent where we have once again expanded in $m_D^2/m_N^2 \ll 1$. In principle, the leptonic mixing $U_{\alpha N}$ could be tested experimentally as it modifies the charged-current interactions in the SM as follows,
\begin{align}
    \mathcal{L}_W \supset -\dfrac{g}{\sqrt{2}} \sum_{\alpha=e,\mu,\tau} \sum_{i=1}^{4} U_{\alpha i} \, \big{(}\bar{\ell}_{L\alpha} \gamma^\mu \nu_{Li}\big{)} \,W_\mu^- +\mathrm{h.c}\,,
\end{align}

\noindent so that $U_{\alpha N}$ could impact various low-energy observables~\cite{Atre:2009rg,Abada:2017jjx}. However, given our estimation of $m_D$ in Eqs.~\eqref{eq:deltamD-num-B} and~\eqref{eq:deltamD-num-K}, we find that $|U_{\alpha N}| \lesssim 3 \times 10^{-7}~\mathrm{GeV}/m_N$, which lies well below the seesaw line and the current experimental constraints, see e.g.~Ref.~\cite{Fernandez-Martinez:2023phj}. The above statements will be quantified more precisely in Sec.~\ref{ssec:nusmeft-numerical}.

\subsection{$0\nu\beta\beta$ decays}
\label{ssec:nusmeft-0nubb}

\begin{figure}
    \centering
    \includegraphics[width=0.88\linewidth]{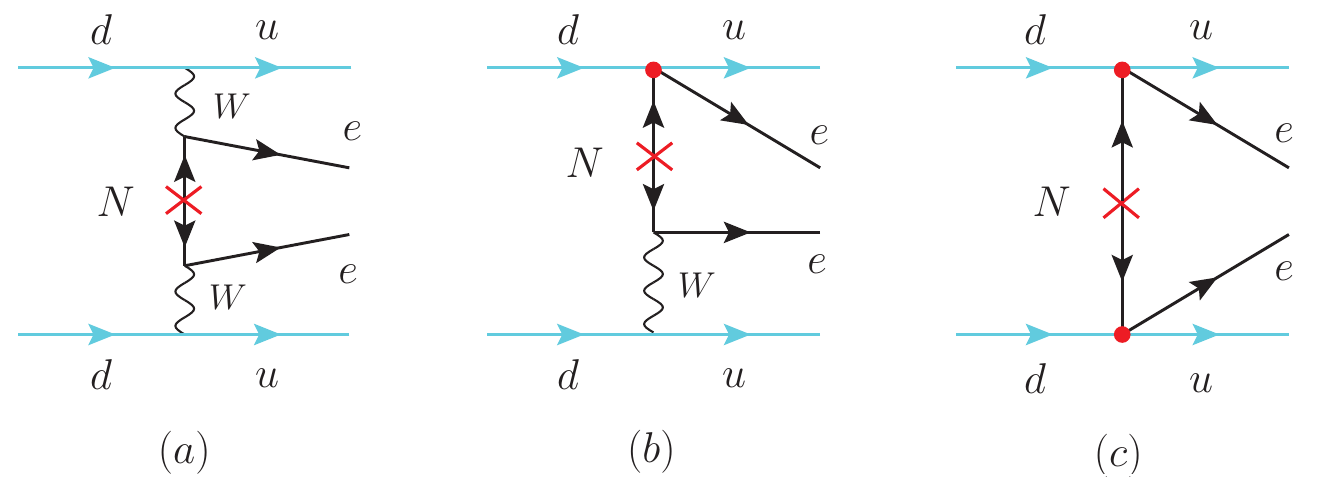}
    \caption{\label{fig:0vbb-LNC} \small \sl Contributions to $0\nu\beta\beta$  decays from the exchange of a sterile neutrino $N$ in our setup. The red dots represent the insertions of dimension-six $\nu$LEFT operators.}
\end{figure}

The considered $\nu$SMEFT operators can also induce $0\nu\beta\beta$ decay through heavy-neutrino exchange diagrams, which can be induced via the active-sterile mixing, and/or by insertions of $\nu$LEFT contact interactions with first-generation fermions, as depicted in Fig.~\ref{fig:0vbb-LNC}~\cite{Dekens:2020ttz,Dekens:2023iyc,Dekens:2024hlz}. From the above discussion, both the mixing and the contact terms can receive RG-induced contributions from the operators relevant to FCNC decays. While RG-induced contributions to $0\nu\beta\beta$ decays were highly suppressed in the SMEFT, these effects can be resonantly enhanced by the exchange of a Majorana neutrino, if its mass $m_{N}$ is close to the typical exchanged energy in these processes, namely $E\approx 100$ MeV.

After matching the $\nu$SMEFT operators onto the $\nu$LEFT, the procedure to compute contributions to the $0\nu\beta\beta$ decay half-life depends on the heavy-neutrino mass. If the heavy neutrino lies in the range $\Lambda_\chi<m_{N}\lesssim \mu_{\rm ew}$, where $\Lambda_\chi \simeq1$ GeV is the chiral-symmetry-breaking scale, the heavy neutrino must be integrated out at energy scales $\mu\simeq m_{N}$. Therefore, the LEFT Lagrangian describes physical interactions for energies $\Lambda_\chi<\mu\lesssim m_{N}$ and can then be matched onto the Chiral Lagrangian at $\Lambda_\chi$. The contributions to $0\nu\beta\beta$ decay are then computed as described in Sec.~\ref{ssec:smeft-0nubb}. 
If instead $m_{N}\lesssim\Lambda_\chi$, then $\nu_4$ remains a dynamical degree of freedom at the hadronic scale and must be included in the description of the hadronic interactions. Incorporating $\nu_4$ into the Chiral Lagrangian results in  LECs and NMEs that depend on the heavy-neutrino mass. This dependence is implemented in Ref.~\cite{Dekens:2020ttz}, where a master formula to compute the $0\nu\beta\beta$ decay half-life in these scenarios is also provided.

\subsubsection{Numerical significance}

In the following, we estimate these effects by focusing on the coefficients $\bar{\mathcal{C}}_{lNqd}^{(1)}$, which are associated with scalar contributions to FCNC processes. The lepton flavor index is set to the first-generation, i.e. $\alpha=1$, and we consider quark flavor indices $(i,j)$ that contribute to either the $b\to s$ or the $s\to d$ transition. We consider both the RG mixing of $\bar{\mathcal{C}}_{lNqd}^{(1)}$ into the $\nu$LEFT coefficients that enter $0\nu\beta\beta$ decays at tree level, as well as its RG-induced contribution to $|U_{eN}|$, which contribute to the diagrams (c) and (a) in Fig.~\ref{fig:0vbb-LNC}, respectively, in addition to the mixed contributions in panel (b).~\footnote{For simplicitly, we only consider the RG contribution to $|U_{eN}|$ via Eqs.~\eqref{eq:deltamD-num-B}--\eqref{eq:deltamD-num-K} and \eqref{eq:mix-nusmeft}.}

Starting with the $b\to s\nu N$ transition and assuming that contributions from $\bar{\mathcal{O}}_{lNqd}$ dominate over the standard ones, we find that the present limits on the $0\nu\beta\beta$ decay half-life imply the constraints~\cite{KamLAND-Zen:2024eml},
\begin{align}
    &\dfrac{1}{\Lambda^2}\,\bar{\mathcal{C}}_{\substack{lNqd\\1 N 23}}(\Lambda)  < (0.66\,{\rm TeV})^{-2}\,, & \dfrac{1}{\Lambda^2}\,\bar{\mathcal{C}}_{\substack{lNqd\\1 N 32}}(\Lambda)< (0.1\,{\rm TeV})^{-2} \,,
\end{align}

\noindent where we fix $m_{N}\sim 100$~MeV, for illustration, and we set $\Lambda = 1$ TeV in the logarithms. For this operator and $m_N$ value, the dominant contribution is given by diagram (a) in Fig.~\ref{fig:0vbb-LNC}, but these constraints turn out not to be competitive with fine-tuning considerations for neutrino masses. 

For the $s\to d\nu N$ transition, the dominant contributions to $0\nu\beta\beta$ decays from the operators with quark flavor indices $(i,j)=(2,1)$ are given by diagram (c) in Fig.~\ref{fig:0vbb-LNC}, which is induced at tree level in this case,
\begin{align}
    &\dfrac{1}{\Lambda^2}\,\bar{\mathcal{C}}_{\substack{lNqd\\1 N 21}}(\Lambda)  < (48\,{\rm TeV})^{-2}\,, 
\end{align}

\noindent where we take again $m_{N}\sim 100 $ MeV. This constraint is almost competitive with the other bounds considered in this study, as will be shown in the following. Differently, for flavor indices $(i,j)=(1,2)$, tree-level contributions are absent, and the leading effects are given by diagram (a). However, these contributions are too small due to CKM and quark-mass suppression, leading to constraints that are irrelevant for the following phenomenological comparison.

\subsection{Leptonic meson decays}
\label{ssec:nusmeft-leptonic}

Finally, we discuss the constraints arising from charged-current $P\to \ell+\mathrm{inv}$ processes, where $P$ denotes a pseudoscalar meson. Whenever these decays are kinematically allowed, they provide sensitivity to the scalar $\nu$SMEFT operator $\bar{\mathcal{O}}_{l N q d}$, which has both neutral- and charged-current components. Importantly, these charged-current channels can be useful for probing the lepton flavor of these operators, which cannot be accessed through $P\to P^\prime +\mathrm{inv}$ searches. 

Given the effective operators introduced in Eq.~\eqref{eq:nu-smeft}, the contributions to the $P\to \ell_\alpha N$ decay rate can be written as 
\begin{equation}
    \Gamma(P\to \ell_\alpha N) = \dfrac{G_F^2 f_P^2 m_P^3}{32\pi } \dfrac{\lambda^{1/2}(m_P^2,m_{\ell_\alpha}^2,m_N^2)}{{(m_{u_i}+m_{d_j})^2}}\bigg{(}1-\dfrac{m_\ell^2+m_N^2}{m_P^2}\bigg{)}\, \Big{|}\bar{C}_{\substack{S_{RR}\\ij\alpha N}}^{(cc)}\Big{|}^2 \,,
\end{equation}
where we assume $m_N<m_P-m_{\ell_\alpha}$, $\lambda(a^2,b^2,c^2)\equiv \big{(}a^2-(b-c)^2\big{)}\big{(}a^2-(b+c)^2\big{)}$ is the triangle function, $f_P$ denotes the $P$-meson decay constant~\cite{FlavourLatticeAveragingGroupFLAG:2024oxs}, and $m_{u_i}$ and $m_{d_j}$ stand for the relevant quark masses. The low-energy effective coefficient is given as follows at the electroweak scale,~\footnote{The corresponding low-energy operator is defined as $\bar{O}_{\substack{S_{RR}}}^{(\mathrm{cc})} =\big{(}\bar{u}_{L} d_{R}\big{)} \big{(} \bar{\ell}_{L} N\big{)}$, where we omit flavor indices. We note that other operators can be induced at dimension six in the $\nu$SMEFT, but they do not have a neutral-current counterpart.} 
\begin{equation}
    \bar{C}_{\substack{S_{RR}\\ij\alpha N}}^{(\mathrm{cc})} (\mu_{\mathrm{ew}}) = -\dfrac{v^2}{\Lambda^2} \sum_k V_{ik}\, \bar{\mathcal{C}}_{\substack{lNqd\\ \alpha N kj}}(\mu_{\mathrm{ew}})\,.
\end{equation}

\noindent which must be evolved down to the relevant low-energy scale. Similarly, the $\tau$-lepton flavor operators can also induce decays into light mesons for operators with light-quark flavors,
\begin{equation}
    \Gamma(\tau \to P N ) = { \dfrac{G_F^2 f_P^2 m_P^4}{64\pi m_\tau } \dfrac{\lambda^{1/2}(m_P^2,m_\tau^2,m_N^2)}{{(m_{u_i}+m_{d_j})^2}}\bigg{(}1+\dfrac{m_N^2-m_P^2}{m_\tau^2}\bigg{)}\, \Big{|}\bar{C}_{\substack{S_{RR}\\ij 3 N}}^{(cc)}\Big{|}^2 \,,}
\end{equation}
where we assume $m_N<m_\tau-m_P$.

In our numerical analysis, we compute the SM predictions following Ref.~\cite{Becirevic:2020rzi}, using the relevant decay constants reported in FLAG~\cite{FlavourLatticeAveragingGroupFLAG:2024oxs}, and the experimental averages from PDG~\cite{ParticleDataGroup:2024cfk}. By combining the relevant observables, we obtain the following constraints for the operators entering the $b\to s+\mathrm{inv}$ transition,
\begin{align}
    &\dfrac{1}{\Lambda^2}\,\bar{\mathcal{C}}_{\substack{lNqd\\\alpha N 23}}^{(1)}(\Lambda)  <\begin{cases} (10\,{\rm TeV})^{-2}\,,\\[0.3em](10\,{\rm TeV})^{-2}\,,\\[0.3em](4\,{\rm TeV})^{-2}\,,\end{cases}\quad & \dfrac{1}{\Lambda^2}\,\bar{\mathcal{C}}_{\substack{lNqd\\\alpha N 32}}^{(1)}(\Lambda)< \begin{cases} (7\,{\rm TeV})^{-2}\,,\,&\quad\alpha=e\\[0.3em](0.7\,{\rm TeV})^{-2}\,,\,&\quad\alpha=\mu\\[0.3em](0.1\,{\rm TeV})^{-2}\,,\,&\quad \alpha=\tau\end{cases} 
\end{align}

\noindent where we quote the bound for $m_N \approx 0$. Similarly, we find the following constraints for the operators relevant for the $s\to d+\mathrm{inv}$,
\begin{align}
    &\dfrac{1}{\Lambda^2}\,\bar{\mathcal{C}}_{\substack{lNqd\\\alpha N 12}}^{(1)}(\Lambda)  < \begin{cases} (110\,{\rm TeV})^{-2}\,,\\[0.3em](11\,{\rm TeV})^{-2}\,,\\[0.3em](2\,{\rm TeV})^{-2}\,,\end{cases} & \dfrac{1}{\Lambda^2}\,\bar{\mathcal{C}}_{\substack{lNqd\\\alpha N 21}}^{(1)}(\Lambda)< \begin{cases} (44\,{\rm TeV})^{-2}\,,\,&\alpha=e\\[0.3em](5\,{\rm TeV})^{-2}\,,&\alpha=\mu\\[0.3em](0.7\,{\rm TeV})^{-2}\,,\,&\alpha=\tau\end{cases}
\end{align}

\noindent In both cases, we find the most stringent constraints for the electron flavor, whereas the constraints are weaker for operators with $\tau$-leptons. These constraints will be compared to the other phenomenological constraints in the following section.

\subsection{Phenomenological analysis}
\label{ssec:nusmeft-numerical}
\noindent In the following, we confront the $B\to K\nu\nu$ and $K\to \pi\nu\nu$ decays discussed in Sec.~\ref{sec:left} with the indirect constraints from the absence of fine-tuning in neutrino masses and the non-observation of $0\nu\beta\beta$ decays, in the context of $\nu$SMEFT. 
In contrast to the SMEFT, we will show that fine-tuning constraints are fully satisfied in the regions allowed by current flavor data, in both the kaon and $B$-meson sectors.

\begin{figure}[!t]
    \begin{center}
    \includegraphics[width=0.48\textwidth]{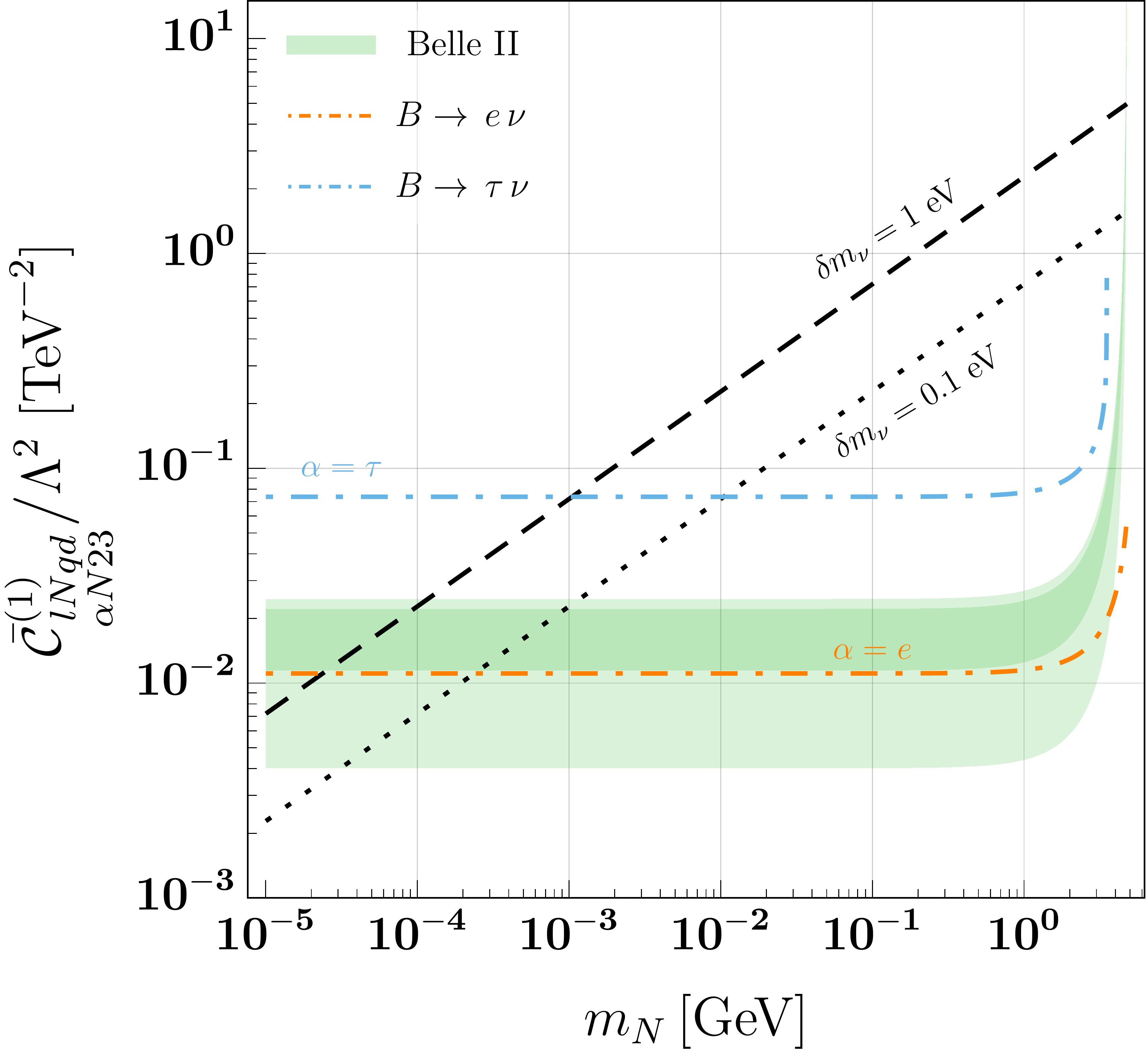}\hfill~\includegraphics[width=0.48\textwidth]{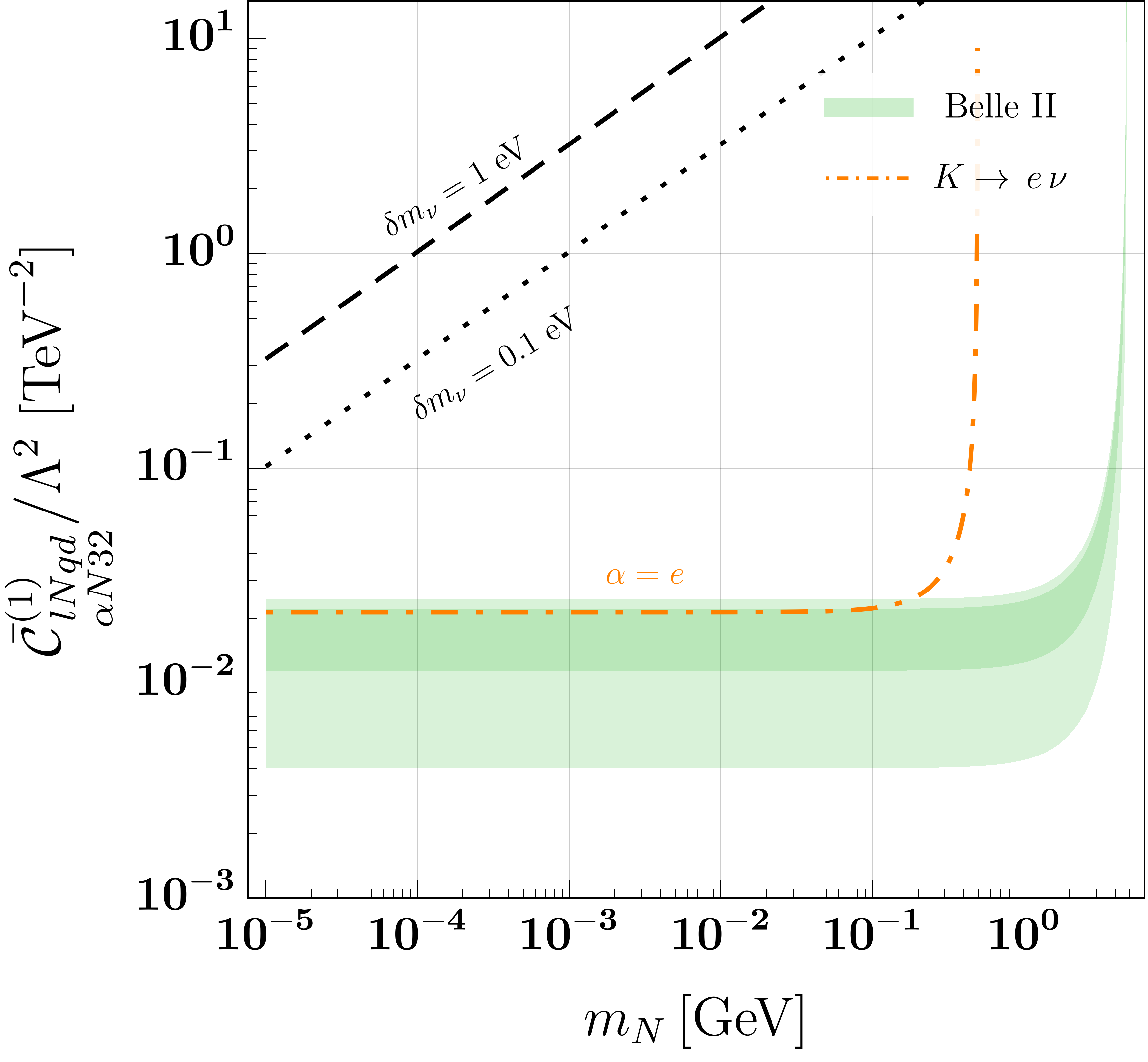}
    \caption{\label{fig:mass-nu-smeft-BK}\small \sl The constraints from $B\to K \nu\nu$ decays~\cite{Belle-II:2023esi} on the $\smash{\bar{C}_{lNqd}^{(1)}/\Lambda^2}$ effective coefficient, evaluated at the scale $\Lambda \simeq 1~\mathrm{TeV}$, as a function of the heavy neutrino mass $m_N$ are depicted by the light (dark) green regions to $1\sigma$ ($2\sigma$) accuracy. The flavor indices are fixed to $\alpha N23$ (left panel) and $\alpha N 32$ (right panel), which lead to tree-level contributions to these decays. These constraints are compared to the requirement that the RG contributions to neutrino masses $\delta m_\nu$ are smaller than $1$~eV (black dashed line) and $0.1$~eV (black dotted line). Constraints from charged-current meson decays are depicted by the dot-dashed orange and blue lines for $\alpha=e$ and $\alpha=\tau$, respectively, whereas constraints from $0\nu\beta\beta$ are not competitive and do not appear for this range of effective coefficients.}
    \end{center}
\end{figure}

\begin{figure}[!t]
    \begin{center}
    \includegraphics[width=0.48\textwidth]{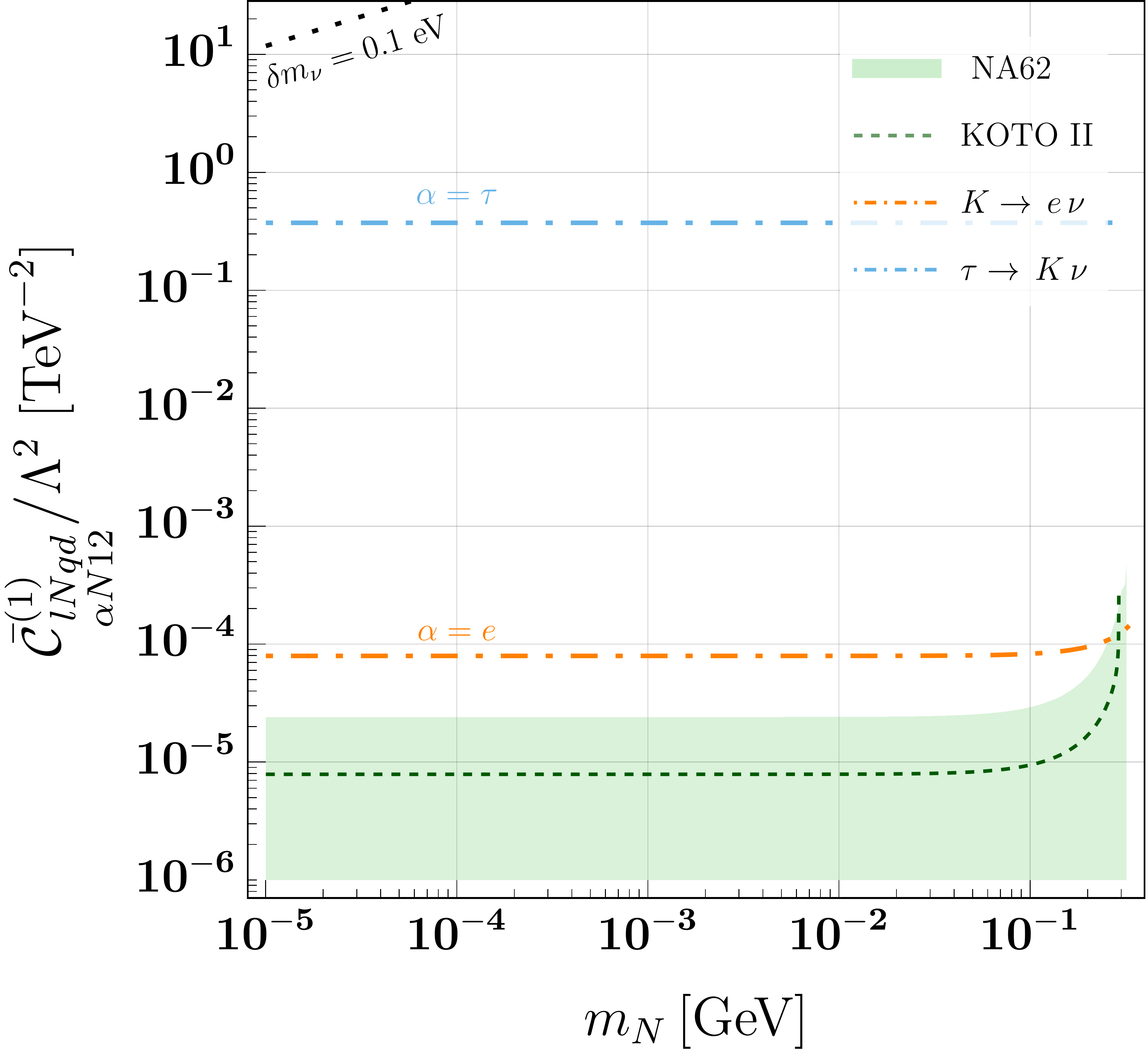}\hfill~\includegraphics[width=0.48\textwidth]{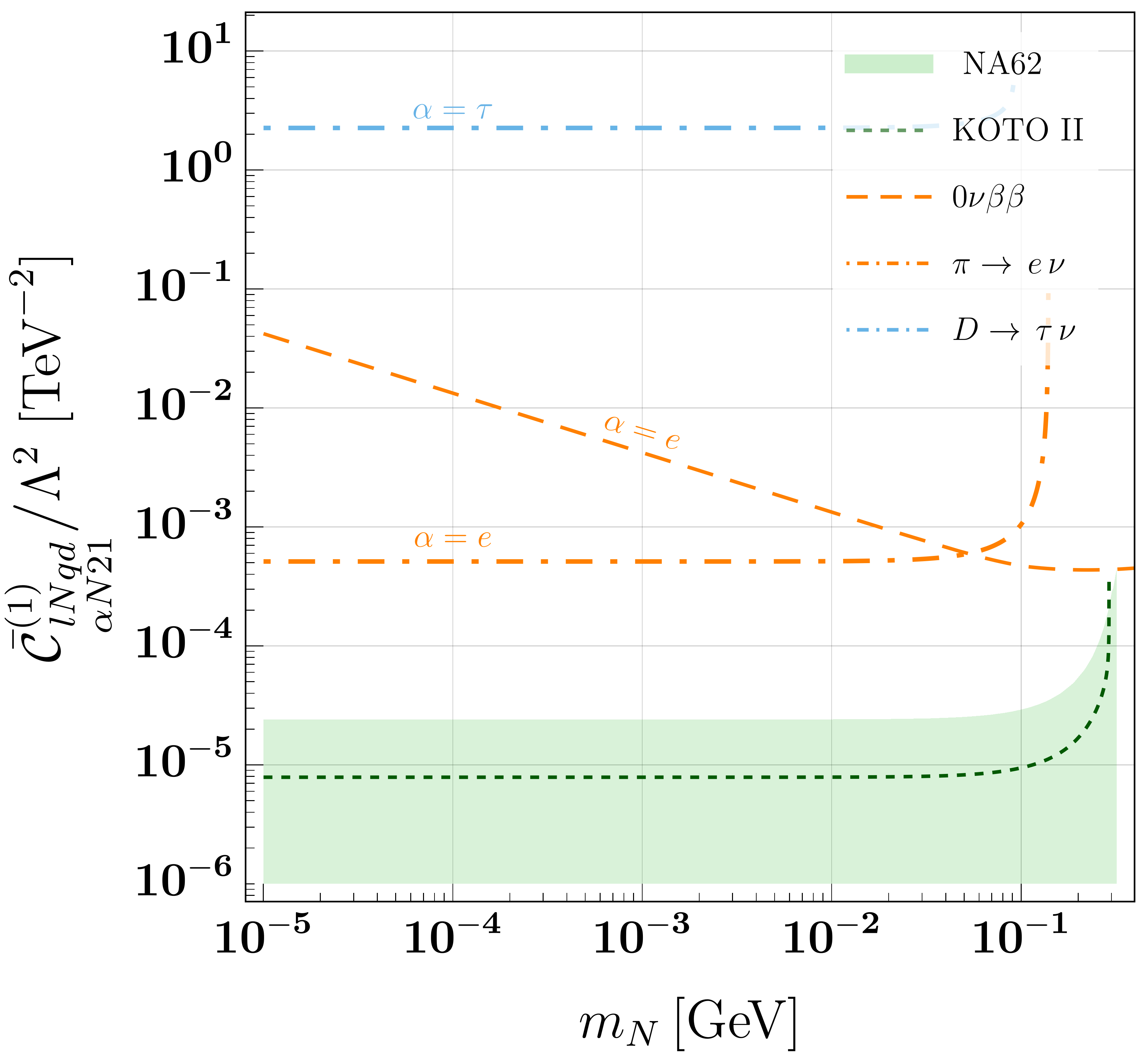
    }
    \caption{\label{fig:mass-nu-smeft-Kpi}\small \sl The constraints from $\mathcal{B}(K^+\to \pi^+\nu\bar{\nu})$ decays~\cite{Chang:2026vvx} on the $\smash{\bar{C}_{lNqd}^{(1)}/\Lambda^2}$ effective coefficient, evaluated at the scale $\Lambda \simeq 1~\mathrm{TeV}$, as a function of the heavy neutrino mass $m_N$ are depicted by the light (dark) green regions to $1\sigma$ ($2\sigma$) accuracy. The flavor indices are fixed to $\alpha N12$ (left panel) and $\alpha N 21$ (right panel), which lead to tree-level contributions to these decays. Future prospects for the KOTO-II experiment~\cite{KOTO:2025gvq}, assuming SM-like central values, are depicted by the green dashed lines. These constraints are compared to the requirement that the RG contributions to neutrino masses $\delta m_\nu$ are smaller than $0.1$~eV (black dotted line). 
    Constraints from $0\nu\beta\beta$ decays are depicted by the orange dashed lines for $\alpha=e$~\cite{KamLAND-Zen:2024eml}, whereas those derived from charged-current meson decays are displayed by the dot-dashed orange and blue lines for $\alpha=e$ and $\alpha=\tau$, respectively. }
    \end{center}
\end{figure}

\subsubsection*{\underline{$B\to K^{(\ast)}\nu\nu$}\,: }
In Fig.~\ref{fig:mass-nu-smeft-BK}, we depict the allowed regions from $B^+\to K^+ \nu\nu$~\cite{Belle-II:2023esi} and $B^0\to K^{\ast 0} \nu\nu$~\cite{Belle:2017oht} decays in the two-dimensional plane of $m_N$ vs.~$\mathcal{\bar{C}}_{lNqd}^{(1)}/\Lambda^{2}$. The Wilson coefficients are evaluated at the scale $\Lambda \simeq 1~\mathrm{TeV}$, and the quark flavor indices are fixed to $(i,j)=(2,3)$ and $(i,j)=(3,2)$ in the left and right panels, respectively, namely to those contributing at tree-level to these processes. For simplicity, we neglect the interference term in Eq.~\eqref{eq:deltam-nusmeft}, computing only the term proportional to $(\delta m_D)^2/m_N$ and checking that it is not spoiling the light neutrino masses.  From Fig.~\ref{fig:mass-nu-smeft-BK}, we find that the region favored by Belle-II is perfectly consistent with the fine-tuning constraints for $m_N \gtrsim 100~\mathrm{keV}$ for the coefficient in the left panel, whereas the coefficient in the right panel induces further suppressed contributions to neutrino masses. As mentioned above, this is due to the seesaw relation in Eq.~\eqref{eq:nu-mass-seesaw} that suppresses the RG-induced contribution to light-neutrino masses. 

The constraints from meson decays derived in Sec.~\ref{ssec:nusmeft-leptonic} are depicted by the dot-dashed lines in Fig.~\ref{fig:mass-nu-smeft-BK}. These constraints are particularly useful to probe the leptonic flavor of the operators. From Fig.~\ref{fig:mass-nu-smeft-BK}, we learn that operators with $\alpha=e$ that accommodate $B\to K+\mathrm{inv}$ data are already in tension with these constraints, and similar conclusions are valid for $\alpha=\mu$, which are not explicitly shown in this plot.~\footnote{Note that this tension increases when considering the reinterpretation from Ref.~\cite{Belle-II:2025lfq,Gartner:2024muk}, which would enhance the values of the Wilson coefficients by a a factor of $\approx 2$, as discussed in Sec.~\ref{sec:left}.} However, these constraints are weaker for operators with $\alpha=\tau$ flavor, which thus remain as the only viable possibility.

\subsubsection*{\underline{$K\to \pi\nu\nu$}\,: }

In Fig.~\ref{fig:mass-nu-smeft-Kpi}, we perform the same comparison for $K\to \pi\nu\nu$ observables, considering current data from NA62~\cite{Chang:2026vvx} and future prospects from KOTO-II~\cite{KOTO:2025gvq}. The two-dimensional constraints on the $m_N$ vs.~$\mathcal{\bar{C}}_{lNqd}^{(1)}/\Lambda^{2}$ plane are shown in the left and right panels, for the Wilson coefficients evaluated at the scale $\Lambda \simeq 1~\mathrm{TeV}$, with quark flavor indices $(i,j)=(1,2)$ and $(i,j)=(2,1)$, namely, those entering kaon decays at tree level.

Similarly to Sec.~\ref{ssec:smeft-numerical}, we find that the fine-tuning constraints are much weaker for the coefficients entering kaon decays, which are being competitive with low-energy constraints. We stress, once again, that this is due to the higher effective scales probed by kaon observables and by the suppressed RG-contributions from these operators to neutrino masses.  In this case, we find that $0\nu\beta\beta$ constraints are weaker than $K\to \pi +\mathrm{inv}$ bound, but become comparable near the endpoint of the spectrum,~$m_N\approx m_K-m_\pi$, for the coefficient shown in the right panel of Fig.~\ref{fig:mass-nu-smeft-Kpi}. Such constraints are absent in the left panel, as $0\nu\beta\beta$ decays are only induced at loop level for the coefficient there considered, cf.~Sec.~\ref{ssec:nusmeft-0nubb}. Finally, we find that charged-current meson decays provide weaker constraints than $K\to \pi +\mathrm{inv}$ decays for all lepton flavors.

\subsection{From EFT to concrete models}
\label{ssec:nusmeft-uv-completions}

From the above discussion, it is perfectly plausible to generate a distortion of the $q^2$ distribution of $B\rightarrow K +\text{inv}$ a
nd $K\to \pi +\text{inv}$ decays, using a dimension-six scalar operator built with sterile neutrinos. This operator should preferably couple to third-generation lepton doublets to avoid constraints from charged-current processes. The remaining question is whether the assumption that the scalar operator dominates over the vector operators can be realized in a minimal UV scenario, as we explore in this section.

We consider possible UV completions of the operators given in Eq.~\eqref{eq:smeft-d6} in weakly-coupled renormalizable scenarios. These operators can be induced by the exchange of a scalar or vector boson, as classified in Table~\ref{tab:nusmeft-mediators} in terms of their SM quantum numbers~\cite{Beltran:2023ymm}. The scenarios with the scalar leptoquarks $S_1 \sim ({\bf \bar{3}},{\bf 1},1/3)$ and $\widetilde{R}_2 \sim ({\bf 3},{\bf 2},{1/6})$ are of particular interest, as they induce scalar and tensor currents through Fierz transformations. The leptoquark interactions with fermions are described in these models by the following Lagrangians~\cite{Dorsner:2016wpm}~\footnote{Notice that we omit di-quark couplings for the $S_1$ model to avoid rapid proton decays.}
\begin{align}
    \mathcal{L}_{S_1} &\supset y_{1L}^{i\alpha}\, \overline{q^C_i}\,\epsilon\, l_\alpha \, S_1 + y_{1R}^{ij}\,\overline{u^C_i} e_j\, S_1 + \bar{y}_{1R}^{iN}\,\overline{d^C_i} N\, S_1 + \mathrm{h.c.}\,, \\[0.4em]
    \mathcal{L}_{\widetilde{R}_2} &\supset y_{2L}^{i\alpha}\, \overline{d_i}\, l_\alpha\,\epsilon\,\widetilde{R}_2 + \bar{y}_{2R}^{iN}\,\overline{q_i} \,\widetilde{R}_2 \,N +\mathrm{h.c.} \,, \nonumber
\end{align}

\noindent where the couplings to SM fermions are denoted by $y_{1L}$, $y_{1R}$ and $y_{2L}$, and those to right-handed neutrinos by $\bar{y}_{1R}$ and $\bar{y}_{2R}$. These Yukawa interactions induce the dimension-six SMEFT operators given in Eq.~\eqref{eq:smeft-lq-d6}, which we repeat below for convenience of the reader,
\begin{align}
  \label{eq:lq-eft-1}
  \dfrac{1}{\Lambda^2}\,\mathcal{C}_{\substack{lq\\ \alpha\beta ij}}^{(1)} &= -\dfrac{1}{\Lambda^2}\,\mathcal{C}_{\substack{lq\\ \alpha\beta ij}}^{(3)} = +\dfrac{y_{1L}^{j\beta} \,y_{1L}^{i\alpha\,\ast}}{4\,m_{S_1}^2}\,, &
  \dfrac{1}{\Lambda^2}\,\mathcal{C}_{\substack{ld\\ \alpha\beta ij}} &= -\dfrac{y_{2L}^{i\beta}\,y_{2L}^{j\alpha\,\ast}}{2\,m_{\widetilde{R}_2}^2}\,. 
\end{align}

\begin{table}[!t]
\renewcommand{\arraystretch}{1.6}
\centering
\begin{tabular}{c||c|c|cccc}
Mediator & $S$ & Quantum Numbers & $\bar{\mathcal{O}}_{Nd}$ & $\bar{\mathcal{O}}_{Nq}$ & $\bar{\mathcal{O}}_{lNqd}^{(1)}$  & $\bar{\mathcal{O}}_{lNqd}^{(3)}$ \\ \hline \hline
$\Phi$ & $0$ & $({\bf 1},{\bf 2},1/2)$ &  & & $\checkmark$  & \\
$S_1$ & $0$ & $({\bf \bar{3}},{\bf 1},1/3)$ & $\checkmark$ & & $\checkmark$  & $\checkmark$\\
$\widetilde{R}_2$ & $0$ & $({\bf 3},{\bf 2},1/6)$ & & $\checkmark$ & $\checkmark$ & $\checkmark$ \\ \hline
$Z^\prime $& $1$ & $({\bf 1},{\bf 1},0)$  & $\checkmark$ & $\checkmark$ & & \\
$\bar{U}_1$ & $1$ & $({\bf 3},{\bf 1},-1/3)$ & $\checkmark$ & & \\
$\widetilde{V}_2$ & $1$ & $({\bf \bar{ 3}},{\bf 2},-1/6)$ &  &$\checkmark$ & \\
\end{tabular}
\caption{\small \sl Mediators that induce the dimension-six $\nu$SMEFT operators, at tree-level, are classified in terms of their spin, $S$, and their SM quantum numbers, $\big{(}SU(3)_c,\,SU(2)_L,\,U(1)_Y\big{)}$. Notice that we use the operators $\bar{\mathcal{O}}_{lNqd}^{(1)}$ and $\bar{\mathcal{O}}_{lNqd}^{(3)}$ defined in Eq.~\eqref{eq:nu-smeft-redundant}, instead of the basis from Ref.~\cite{delAguila:2008ir}.}
\label{tab:nusmeft-mediators}
\end{table}

\noindent The leptoquarks interactions with sterile neutrinos $N$ amount to the following  operators,
\begin{align}
  \label{eq:lq-eft-2}
  \dfrac{1}{\Lambda^2}\,\bar{\mathcal{C}}_{\substack{lNqd\\\alpha N ij}}^{(3)} &= -\dfrac{\bar{y}_{1R}^{jN}\,y_{1L}^{i\alpha\,\ast}}{8\,m_{S_1}^2} + \dfrac{\bar{y}_{2R}^{i N}\,y_{2L}^{j\alpha\,\ast}}{8\,m_{\widetilde{R}_2}^2}\,, &   \dfrac{1}{\Lambda^2}\,\bar{\mathcal{C}}_{\substack{Nq\\ NNij}} &=  -\dfrac{\bar{y}_{2R}^{iN}\, \bar{y}_{2R}^{jN\,\ast}}{2\,m_{\widetilde{R}_2}^2}\,,\\[0.35em]
  \dfrac{1}{\Lambda^2}\,\bar{\mathcal{C}}_{\substack{lNqd\\ \alpha N ij}}^{(1)} &= +\dfrac{\bar{y}_{1R}^{jN}\,{y}_{1L}^{i\alpha\,\ast}}{2\,m_{S_1}^2}+ \dfrac{\bar{y}_{2R}^{i N}\,y_{2L}^{j\alpha\,\ast}}{2\,m_{\widetilde{R}_2}^2}\,, &
  \dfrac{1}{\Lambda^2}\,\bar{\mathcal{C}}_{\substack{Nd\\ NNij}} &= +\dfrac{\bar{y}_{1R}^{jN}\,\bar{y}_{1R}^{iN\,\ast}}{m_{S_1}^2}\,. \nonumber
\end{align}

\noindent From the above equations, we find that both models yield scalar and tensor currents via the $\smash{\bar{\mathcal{C}}_{lNqd}^{(1)}}$ and $\smash{\bar{\mathcal{C}}_{lNqd}^{(3)}}$ coefficients, respectively, cf.~Eq.~\eqref{eq:nusmeft-matching}. At tree-level, the scenarios with $S_1$ and $\widetilde{R}_2$ leptoquarks amount to $\bar{C}_{S_{LL}}=-4\, \bar{C}_T$ and $\bar{C}_{S_{LL}}=+4\, \bar{C}_T$, respectively. However, these relations are sizably affected by the QCD running from the matching scale, which we take to be $\mu\approx 1~\mathrm{TeV}$, down to the relevant low-energy scale. For $B$-meson decays, these effects amount to $\bar{C}_{S_{LL}}(m_b)\simeq \pm 8.5\, \bar{C}_T (m_b)$, whereas for kaon decays, $\bar{C}_{S_{LL}}(2~\mathrm{GeV})\simeq \pm 10.7\, \bar{C}_T (2~\mathrm{GeV})$, as considered in our low-energy analysis in Sec.~\ref{sec:left}.~\footnote{These relations are slightly modified by the electroweak mixing between scalar and tensor operators~\cite{Gonzalez-Alonso:2017iyc}.}

Finally, we note that the above leptoquark models generate both chirality-conserving and violating operators in Eqs.~\eqref{eq:lq-eft-1} and \eqref{eq:lq-eft-2}. The scalar and tensor coefficients can induce the dominant contributions to the $d_i\to d_j+\mathrm{inv}$ decay rates depending on the hierarchy of the leptoquark couplings. In such a scenario, there will be a distortion in the kinematical spectrum that can be studied experimentally, as performed, e.g., for $B\to K +\mathrm{inv}$ decays in Ref.~\cite{Belle-II:2025lfq}. Note, also, that vector-type contributions with two sterile neutrinos in the final state would also be potentially present in this case~\cite{Becirevic:2024iyi}. Complementary constraints can be derived from $\Delta F = 2$ processes~\cite{Aebischer:2020dsw}, as well as from monolepton production at the LHC~\cite{Allwicher:2022gkm,Allwicher:2022mcg}. Finally, we stress that larger contributions to neutrino masses can arise in these concrete scenarios, which could dynamically explain the experimentally observed neutrino masses~\cite{Chen:2025npb}.

\section{Summary and outlook}
\label{sec:conclusion}

In this paper, we have investigated the contributions of Lepton Number Violating (LNV) operators to Flavor Changing Neutral Current (FCNC) processes $B\to K^{(\ast)}\nu\nu$ and $K^+\to \pi^+ \nu\nu$, which are currently studied at the Belle-II~\cite{Belle-II:2023esi} and NA62 experiments~\cite{Chang:2026vvx}. These operators appear at dimension seven in the Standard Model Effective Field Theory (SMEFT) and are particularly relevant because they can modify the kinematic distributions of these decays, unlike the leading dimension-six contributions~\cite{Becirevic:2023aov}. The primary goal of this study was to explicitly determine, through a complete Renormalization Group (RG) analysis, whether such a scenario can be reconciled with constraints from neutrino physics and general ultraviolet (UV) consistency requirements.

By using the one-loop anomalous dimension matrix computed in Ref.~\cite{Zhang:2023kvw}, we have demonstrated that the LNV SMEFT operators contributing to $B\to K\nu\nu$ and $K\to \pi \nu\nu$ decays mix into the dimension-five and seven Weinberg operators through RG evolution, thus inducing potentially large contributions to the tiny active neutrino masses. We have shown that the leading effects appear only at one-loop squared order, with a GIM-like suppression factor. These effects are particularly large for the operators entering $B$-meson decays, which would imply a large degree of fine-tuning to reproduce viable neutrino masses, $m_\nu \lesssim 0.1~\mathrm{eV}$~\cite{Planck:2018vyg}. Consequently, we found that Belle-II can only probe LNV contributions within the SMEFT in a highly fine-tuned configuration. 

In the kaon sector, we have shown that the fine-tuning constraints are weaker due to a stronger suppression of the corresponding RG-induced contributions, thus being fully compatible with the effects currently probed by NA62~\cite{Chang:2026vvx} and by the proposed KOTO-II experiment~\cite{KOTO:2025gvq}. These conclusions agree with the findings from Ref.~\cite{Endo:2026qof} in a similar context. Moreover, meson decays are complementary to neutrinoless double-beta decay ($0\nu\beta\beta$) searches~\cite{KamLAND-Zen:2024eml}, as previously discussed in Ref.~\cite{Deppisch:2020oyx} and references therein. However, as we have argued in this paper, the assumption that dimension-seven operators dominate $K\to \pi\nu\nu$ decay rates over dimension-six ones can only be realized in rather specific UV completions.

We have proposed a minimal extension of the SMEFT that can circumvent the limitations outlined above by introducing a light right-handed Majorana neutrino, $N$, into the spectrum~\cite{Rosauro-Alcaraz:2024mvx}, i.e., the so-called $\nu$SMEFT. We have shown that the RG-induced contributions to the renormalization of the Dirac neutrino mass, $m_D$, do not spoil light neutrino masses for $m_\nu \ll m_N$, for the Wilson coefficients allowed by Belle-II data, due to a seesaw-like relation, $m_\nu \simeq m_D^2/m_N$. Furthermore, these operators can be generated at dimension six above the electroweak scale, thus not posing a problem for the EFT power counting. Therefore, the scenarios with a light right-handed neutrino provide a theoretically robust framework for generating EFTs with sizable FCNC scalar operators in the neutrino sector. We emphasize, once again, that a clear prediction of this setup is a modified shape of the $B\to K+ \mathrm{inv}$ differential distribution, which can be probed at Belle-II~\cite{Belle-II:2025lfq}.

Finally, several directions remain to be explored. We have shown that $0\nu\beta\beta$ decays are almost as competitive as FCNC processes in the viable scenario discussed above, due to resonantly enhanced contributions, for $m_N$ in the $\mathcal{O}(100~\mathrm{MeV})$ range. Similar resonant effects can be explored in other processes~\cite{DeVries:2020jbs,Beltran:2022ast}, in particular in the context of future experiments such as SHiP~\cite{SHiP:2015vad}, which can probe long-lived sterile neutrinos produced in heavy-meson decays.  Sterile neutrinos with masses above a few MeV can also alter the thermal history
of the early universe, even if they do not decay during Big Bang Nucleosynthesis. These constraints offer a powerful and complementary probe to collider, beam-dump, and $0\nu\beta\beta$ decay searches for sterile neutrino masses $m_N$
above $100$ MeV~\cite{Braat:2026wgm}.  We leave a detailed investigation of these effects to future work.

\section*{Acknowledgments}

We warmly thank C.~Cornella, S.~Urrea and S.~Rosauro-Alcaraz for discussions, as well as L.~Gärtner, S.~Stefkova and T.~Wuerthner for providing useful inputs on Ref.~\cite{Belle-II:2025lfq}. This project has received support from the Agence Nationale de la Recherche under the contract ANR25-CE31-2504, from the European Union’s Horizon 2020 research and innovation programme under the Marie Skłodowska-Curie grant agreement N$^\circ$~101086085-ASYMMETRY, from the IN2P3 (CNRS) Master Projects HighPTflavor and ``Hunting for Heavy Neutral Leptons" (12-PH-0100), from the SPRINT/CNRS agreement supported by FAPESP under Contract No.~2023/00643-0 and from the USP-COFECUB project Uc Ph194-25/25. R.Z.F. also thanks IJCLab for their hospitality during part of this work, as well as CNPq for partial financial support. A.A.~acknowledges support from the Institut Universitaire de France (IUF).

\appendix

\section{Notation and conventions}
\label{app:conventions}

In this Appendix, we briefly summarize our notation and conventions. The SM Lagrangian is expressed as 
\begin{align}
    \mathcal{L}_{\mathrm{SM}} &= -\dfrac{1}{4}G_{\mu\nu}^A G^{A\,\mu\nu} -\dfrac{1}{4}W_{\mu\nu}^I W^{I\,\mu\nu} -\dfrac{1}{4}B_{\mu\nu} B^{\mu\nu}\\[0.3em]
    &+\big{(}D_\mu H^\dagger\big{)} \big{(}D_\mu H\big{)}-\mu_H^2\,H^\dagger H - \lambda \big{(}H^\dagger H\big{)}^2+\sum_\psi \bar{\psi} i \slashed{D} \psi+ \mathcal{L}_{\mathrm{yuk}}\,, \nonumber
\end{align}

\noindent where $\psi\in \lbrace q,l,u,d,e\rbrace$ and the covariant derivative is defined with the convention $D_\mu = \partial_\mu+i g_3 T^A G_\mu^A+i g_2 t^I W_\mu^I+i g_1 y B_\mu$, where $T^A$ are the $SU(3)$ generators, $t^I$ are the $SU(2)$ generators and $y$ is the $U(1)$ hypercharge generator. We remind that the Yukawa Lagrangian reads
\begin{align}
\mathcal{L}_{\mathrm{yuk}}=-y_u\,\bar{q} \widetilde{H} u-y_d\,\bar{q}{H}d-y_\ell\,\bar{l}{H}e + \mathrm{h.c.}\,,
\end{align}

\noindent where flavor indices are implicit.  We work in the flavor basis with diagonal down-type quark Yukawas, so that the CKM matrix appears in the upper component of the quark doublet, i.e.~$q_i = [(V^\dagger u_L)_i\,,~d_{Li}]^T$. Throughout this manuscript, flavor indices for quarks and leptons are denoted by Latin and Greek letters, respectively.

\section{Expressions for rare-meson decays}
\label{app:formulas}

In this Appendix, we collect the expressions for the $P\to P^\prime$ and $P\to V$ hadronic matrix elements in terms of form-factors, and we provide the expressions for the differential distributions of $B\to K^{(\ast)}\nu\nu$ and $K\to\pi \nu\nu$ decays in terms of the LEFT and $\nu$LEFT Wilson coefficients. For convenience, we define in the first case,
\begin{align}
    \label{eq:app-repl}
    C_{VL} \equiv C_{V_{RL}} + C_{V_{LL}}\,, && C_{AL} & \equiv C_{V_{RL}} -  C_{V_{LL}}\,, \\[0.3em]
    C_{SL} \equiv C_{S_{RL}} + C_{S_{LL}}\,, && C_{PL} & \equiv C_{S_{RL}} -  C_{S_{LL}}\,, \nonumber
\end{align}

\noindent where flavor indices are omitted. We will use similar definitions for the $\nu$LEFT, i.e., through the replacement $C_J \to \bar{C}_J$ in the above expressions. In the latter case, there will be coefficients with the opposite neutrino chirality, which can be obtained by replacing $L\to R$ in Eq.~\eqref{eq:app-repl}.

\subsection{$P\to P^\prime + \mathrm{inv}$}

For $P\to P^\prime\nu\nu$ decays based on the $d_i\to d_j\nu\nu$ transition, the hadronic matrix-elements can be generally expressed as
\begin{align}
\label{eq:app-PPp-FF}
\langle P^\prime (k)|\bar{d}_j\gamma^\mu d_i|P(p)\rangle &= \left[ (p+k)^\mu - \frac{M^2 - m^2}{q^2} q^\mu \right] f_+(q^2)  + \frac{M^2 - m^2}{q^2} q^\mu f_0(q^2)\,, \\[0.35em]
\langle P^\prime (k) | \bar{d_j} \sigma^{\mu\nu} d_i | P(p) \rangle &= -i \, (p^\mu k^\nu - p^\nu k^\mu)\frac{2\,f_T(q^2, \mu)}{M + m}\,, \nonumber
\end{align}

\noindent where $q^2=(p-k)^2$, $M$ ($m$) denote the initial (final) state meson masses, and $f_+$, $f_0$ and $f_T$ stand for the vector, scalar and tensor $P\to P^\prime$ form-factors, respectively. These form factors are constrained by the relation $f_+(0)=f_0(0)$. The scalar matrix element can be derived from Eq.~\eqref{eq:app-PPp-FF} by using the Ward identities.

\subsubsection*{LEFT} By using the Lagrangian defined in Eq.~\eqref{eq:left} and the hadronic matrix-elements defined above, the $P\to P^\prime \nu_\alpha \nu_\beta$ differential decay-rate can be expressed in the LEFT as follows, 
\begin{align}
    \dfrac{\mathrm{d}\Gamma}{\mathrm{d}q^2}(P\to P^\prime \nu_\alpha \nu_\beta) &= \frac{\sqrt{\lambda_{P}} }{(4 \pi)^3 M^3 v^4\left(1+\delta_{\alpha \beta}\right)}\left[\frac{\lambda_{P}}{24 }\left|f_{+}(q^2)\right|^2\bigg{(}\big{|}C_{\substack{VL\\ s b\alpha \beta }}\big{|}^2 +\big{|}C_{\substack{VL\\ s b\beta \alpha }}\big{|}^2\bigg{)}\right. \\[0.4em]
    & \qquad+\frac{q^2\left(M^2-m^2\right)^2}{4\left(m_{d_i}-m_{d_j}\right)^2}\left|f_0(q^2)\right|^2\bigg{(}\big{|}C_{\substack{{SL}\\ s b\alpha \beta }}\big{|}^2+\big{|}C_{\substack{{SL}\\ bs\alpha \beta }}\big{|}^2\bigg{)} \nonumber \\[0.4em]
    & \qquad \left.+\frac{4 q^2 \lambda_{P} }{3\left(M+m\right)^2}\left|f_T(q^2)\right|^2\bigg{(}\big{|}C_{\substack{T_{L}\\ s b\alpha \beta }}\big{|}^2+\big{|}C_{\substack{T_{L}\\ bs\alpha \beta }}\big{|}^2\bigg{)}\right]\,,  \nonumber
\end{align}

\noindent where $\lambda_P\equiv \lambda (q^2,m^2,M^2)$, and we remind that $\lambda(a^2,b^2,c^2)\equiv (a^2-(b-c)^2)(a^2-(b+c)^2)$. Notice, in particular, that we have used the symmetry properties for scalar and tensor coefficients under $\alpha \leftrightarrow\beta$ to simplify the above equation. This expression agrees with the one provided in Ref.~\cite{Felkl:2023ayn}.

\subsubsection*{$\nu$LEFT} In the case of the $\nu$LEFT, there are two decays that can be possible depending on the sterile neutrino mass $m_N$. For $m_N < M - m$, 
\begin{align}
    \dfrac{\mathrm{d}\Gamma}{\mathrm{d}q^2}(P\to P^\prime \nu_\alpha N) &= \frac{\sqrt{\lambda_{P}}}{(4 \pi)^3 M^3 v^4} \bigg{(}1-\dfrac{m_N^2}{q^2}\bigg{)}^2
    \left[\frac{q^2\,(M^2-m^2)^2}{16(m_{d_i}-m_{d_j})^2}\left|f_0(q^2)\right|^2   \left(\big{|}\bar{C}_{\substack{{SL}\\ s b\alpha N }}\big{|}^2+\big{|}\bar{C}_{\substack{{SL}\\ bs\alpha N }}\big{|}^2\right) \right. \nonumber\\[0.4em]
    &\left.\qquad +\frac{q^2\,\lambda_{P}}{3(M+m)^2} \bigg{(}1+\dfrac{2 m_N^2}{ q^2}\bigg{)}\left|f_T(q^2)\right|^2\bigg{(}\big{|}\bar{C}_{\substack{T_{L}\\ s b\alpha N }}\big{|}^2+\big{|}\bar{C}_{\substack{T_{L}\\ bs\alpha N }}\big{|}^2\bigg{)}\right]\,.
\end{align}

\noindent Similarly, for $2m_N < M- m$, 
\begin{align}
\frac{\mathrm{d}\Gamma}{\mathrm{d}q^2}(P\to P'NN)
=\frac{\sqrt{\lambda_P\,\lambda_N}}
{24(4\pi)^3\,M^3\,q^4\,v^4}&\Big{|}\bar{C}_{\substack{VR\\ s bNN }}+\bar{C}_{\substack{VL\\ s bNN }}\Big{|}^2\\\nonumber
\times&\left[
6\,|f_0(q^2)|^2\,m_N^2\left(M^2-m^2\right)^2
+
\frac{|f_+(q^2)|^2\,\lambda_P\,\lambda_N}{q^2}
\right]
\,.
\end{align}

\noindent where we define $\lambda_N \equiv  \lambda (q^2,m_N^2,m_N^2)$. 

\subsection{$P\to V + \mathrm{inv}$}

For $P\to V\nu\nu$ decays, the vector hadronic matrix-elements read,
\begin{align}
\langle V(k,\varepsilon)|\bar{d_j}\gamma^\mu d_i|P(p)\rangle &= \varepsilon_{\mu\nu\rho\sigma} \,\varepsilon^{\nu\ast}p^\rho k^\sigma \dfrac{2V(q^2)}{M+m}\,,\\[0.4em]
\langle V(k,\varepsilon)|\bar{d_j}\gamma^\mu \gamma_5 d_i| P(p)\rangle &= i \varepsilon_\mu^\ast  (m+M) A_1(q^2)-i(p+k)_\mu (\varepsilon^\ast \cdot q) \dfrac{A_2(q^2)}{m+M} \nonumber \\[0.4em]
&\hspace{9.5em}-i q_\mu (\varepsilon^\ast \cdot q) \dfrac{2m}{q^2} \big{[}A_3(q^2)-A_0(q^2)\big{]}\,,  \nonumber
\end{align}

\noindent where $q^2=(p-k)^2$ and $M$ ($m$) denote the initial (final) state meson masses, as before. The the $V$-meson polarization is denoted by $\varepsilon$, and $V$ and $A_{1,2,3}$ stand for the $P\to V$ form-factors. The $A_3$ form-factor is not independent, but related to the others through the relation, $2m A_3(q^2)\equiv (M+m)\,A_1(q^2)- (M-m) \,A_2(q^2)$, and it satisfies $A_3(0)=A_0(0)$.~\footnote{The pseudoscalar matrix element can be determined through the Ward identities.} For convenience, we also define the following  combination of form factors,
\begin{align}
 A_{12}(q^2) \equiv \dfrac{(M+m)(M^2-m^2-q^2)\, A_1(q^2)}{16 M m^2}- \dfrac{\lambda_P\, A_2(q^2)}{16 M m^2(M+m)}\,,
\end{align}

\noindent where we remind that $\lambda_P\equiv \lambda (q^2,m^2,M^2)$. Finally, the tensor matrix-element can also be decomposed in full generality as follows,
\begin{align}
\langle {V}(k,\varepsilon)|\bar{d_j}\sigma^{\mu\nu} d_i|P(p)\rangle &= \varepsilon_{\mu\nu\alpha\beta} \bigg{\lbrace}-\varepsilon^{\ast \alpha} (p+k)^\beta \, T_1(q^2)+\varepsilon^{\ast \alpha} q^\beta \dfrac{M^2-m^2}{q^2}\Big{[}T_1(q^2)-T_2(q^2)\Big{]}\nonumber\\[0.35em]
&+(\varepsilon^\ast\cdot q) p^\alpha k^\beta \dfrac{2}{q^2}\bigg{[}T_1(q^2)-T_2(q^2)-\dfrac{q^2}{M^2-m^2}\,T_3(q^2)\bigg{]} \bigg{\rbrace}\,,
\end{align}

\noindent where $T_{1,2,3}$ are the tensor form-factors, which satisfy $T_1(0)=T_2(0)$. It is also convenient to define,
\begin{align}
    T_{23}(q^2) \equiv \dfrac{(M+m)(M^2+3m^2-q^2)\,T_2(q^2)}{8Mm^2}  -\dfrac{\lambda_{P}\,T_3(q^2)}{8Mm^2(M-m)}\,.
\end{align}

\subsubsection*{LEFT} By using the Lagrangian defined in Eq.~\eqref{eq:nuleft} and the hadronic matrix-elements defined above, the $P\to V \nu_\alpha \nu_\beta$ differential decay-rate can be expressed in the LEFT as follows, 
\begin{align}
    \dfrac{\mathrm{d}\Gamma}{\mathrm{d}q^2}(P\to V \nu_\alpha \nu_\beta) &= \frac{\sqrt{\lambda_P} }{12(4 \pi)^3 M^3 v^4\left(1+\delta_{\alpha \beta}\right)}\left[\frac{\lambda_P\,q^2\,|V(q^2)|^2}{\left(M+m\right)^2}\left(\big{|}C_{\substack{VL\\ s b\alpha \beta }}\big{|}^2+\big{|}C_{\substack{VL\\ s b \beta \alpha}}\big{|}^2\right)\right. \nonumber \\[0.4em]
& +\left({32 M^2 m^2\,\left|A_{12}(q^2)\right|^2}+q^2\left(M+m\right)^2\left|A_1(q^2)\right|^2\right)\left(\big{|}C_{\substack{AL\\ s b\alpha \beta }}\big{|}^2+\big{|}C_{\substack{AL\\ s b\beta \alpha }}\big{|}^2\right) \nonumber \\[0.4em]
& +\frac{3\lambda_P\,q^2\,\left|A_0(q^2)\right|^2}{\left(m_{d_i}+m_{d_j}\right)^2}\left(\big{|}C_{\substack{PL\\ s b\alpha \beta }}\big{|}^2+\big{|}C_{\substack{PL\\ bs\alpha \beta }}\big{|}^2\right) \nonumber \\[0.4em]
&\left. +32\,\left({ \lambda_{P}}\left|T_1(q^2)\right|^2+{\left(M^2-m^2\right)^2}\left|T_2(q^2)\right|^2+\frac{8 M^2 m^2 q^2\,\left|T_{23}(q^2)\right|^2}{\left(M+m\right)^2}\right)\right.\nonumber\\[0.4em]
&\left.\times\left(\big{|}C_{\substack{T_{L}\\ s b\alpha \beta }}\big{|}^2+\big{|}C_{\substack{T_{L}\\ bs\alpha \beta }}\big{|}^2\right)\right]\,,
\end{align}

\noindent which is in agreement with the expression provided in Ref.~\cite{Felkl:2023ayn}.

\subsubsection*{$\nu$LEFT}

Once again, in the case of the $\nu$SMEFT, there are two possible decay modes. Firstly, for $m_N<M-m$,
\begin{align}
    \dfrac{\mathrm{d}\Gamma}{\mathrm{d}q^2}(P\to V &\nu_\alpha N) =\frac{\sqrt{\lambda_P}}{(4\pi)^3  M^3 v^4}  \bigg{(}1-\dfrac{m_N^2}{q^2}\bigg{)}^2 \left[ \frac{ \lambda_P \, q^2\,|A_0(q^2)|^2}{16(m_{d_i} + m_{d_j})^2}  \left(\big{|} \bar{C}_{\substack{PL \\ sbN\alpha }} \big{|}^2 + \big{|} \bar{C}_{\substack{PL \\ bs N\alpha}}\big{|}^2\right)\right.\nonumber \\[0.4em]
&\left.+ \dfrac{2}{3}\left(1+\dfrac{2m_N^2}{q^2} \right) \left( \big{|} \bar{C}_{\substack{T_{L} \\ sbN\alpha }} \big{|}^2 + \big{|} \bar{C}_{\substack{T_{L} \\ bs N\alpha}} \big{|}^2 \right)\right.\nonumber \\[0.4em]
&\left.\times \left( {\lambda_P} \, |T_1(q^2)|^2 +{(M^2-m^2)^2}\, |T_2(q^2)|^2+ \frac{8 M^2 m^2 q^2\, |T_{23}(q^2)|^2}{(M+m)^2}\right)\right]\,.
\end{align}
For $2m_N<M-m$, the following decay is also possible,
\begin{align}
    \dfrac{\mathrm{d}\Gamma}{\mathrm{d}q^2}(P &\to V N N) =\frac{\sqrt{\lambda_P\,\lambda_{N}}}{12 (4\pi)^3 M^3 v^4}\Bigg{\lbrace}   \frac{\lambda_P\,\lambda_N \,|V(q^2)|^2}{(M + m)^2\,q^4} \, \big{|} \bar{C}_{\substack{VR \\ sbNN}} \big{|}^2 \\[0.4em]
    &+ \Bigg{[}\frac{3 m_N^2\lambda_P\,|A_0(q^2)|^2}{q^4}+\dfrac{\lambda_N}{q^4}\bigg{(} (M+m)^2\,|A_1(q^2)|^2 +\dfrac{32M^2 m^2\,|A_{12}(q^2)|^2}{q^2}\bigg{)} \Bigg{]}\,\big{|} \bar{C}_{\substack{AR \\ sb N N}}   \big{|}^2\Bigg{\rbrace}
\,. \nonumber    
\end{align}


\end{document}